\documentclass[12pt,reqno]{amsart}

\usepackage{fullpage}
\usepackage{amsmath,amssymb,amsfonts}
\usepackage{graphicx}
\usepackage{booktabs}
\usepackage{natbib}
\usepackage{algorithm}
\usepackage{algpseudocode}
\usepackage{xcolor}
\usepackage[colorlinks=true,linkcolor=black,citecolor=purple,urlcolor=blue]{hyperref}

\graphicspath{{Images/}}

\newcommand{\bs}{\boldsymbol}

\theoremstyle{plain}

\theoremstyle{definition}

\theoremstyle{remark}

\title{Spatial Disease Mapping and Disparity Detection Using Generative AI: An Amortized Bayesian Learning Framework}

\author{Luca Aiello*\textsuperscript{1}}
\author{Sudipto Banerjee*}
\thanks{*Department of Biostatistics, University of California, Los Angeles, Los Angeles, CA, USA}
\thanks{\textsuperscript{1}Email: \href{mailto@g.ucla.edu}{laiello@g.ucla.edu}.}

\date{\today}

\begin{document}

\begin{abstract}
We introduce an amortized Bayesian framework for spatial boundary detection that generalizes posterior inference across areal graphs with varying numbers of regions and diverse adjacency structures. The underlying model couples a Poisson count likelihood with a covariate-driven rule to interrupt smoothing across dissimilar neighboring areas, utilizing a directed acyclic graph autoregressive (DAGAR) prior to capture residual spatial dependence. To approximate the target posterior distribution, a neural engine is trained on simulated maps: a permutation-invariant summary network encodes graph-aware representations of the observed counts, offsets, covariates, and adjacency matrices, while a conditional normalizing flow generates the approximate posterior draws. Simulation studies demonstrate accurate parameter recovery, near-nominal interval coverage, well-calibrated posterior predictive behavior, and informative posterior boundary probabilities. Benchmarking against Markov chain Monte Carlo (MCMC) confirms close agreement regarding primary boundary evidence, and an ablation study validates the inclusion of model-guided graph summaries. Finally, applications to Glasgow respiratory disease and California lung cancer data demonstrate that a single trained neural engine can be seamlessly deployed across real-world maps with distinct graph structures, yielding boundary conclusions consistent with established localized smoothing analyses.
\end{abstract} 


\maketitle

\section{Introduction}

%Bayesian spatial models are usually fitted one map at a time. In disease mapping, this means that every new region, administrative partition, adjacency graph, outcome, or sensitivity analysis typically requires a new posterior computation. This is natural from the perspective of standard Bayesian computation, but it becomes restrictive in real-world evidence settings where similar spatial analyses may need to be repeated across many maps, often within a short period of time. The difficulty is not only that spatial posteriors can be computationally demanding; it is that the input itself changes. A new map may contain a different number of areal units, a different adjacency structure, and a different pattern of covariate dissimilarities. Thus, a central statistical question is whether one can train a single inferential system that learns Bayesian inference for a class of spatial graphs, rather than for one fixed graph, while retaining calibrated uncertainty quantification. This is an AI-assisted inference problem, but with a statistical constraint: the learned system must generalize across heterogeneous real-world spatial inputs while retaining calibrated posterior uncertainty.
Disease mapping refers to the task of producing spatial maps exhibiting statistical estimates of counts or rates (e.g., incidence or mortality) to better understand the geographic variation of diseases \citep{koch2005cartographies, lawson2016handbook}. Spatial dependence is introduced using stochastic models on graphs, where the nodes correspond to regions, and an edge between two nodes relates them as neighbors. %This question is especially relevant for spatial boundary detection. 
Areal disease-mapping models are widely used in spatial epidemiology, public-health surveillance, and real-world evidence generation, where disease counts are observed over administrative or geographical units together with expected counts, exposures, risk factors, and adjacency information \citep{wakefield2007disease,lawson2018bayesian}. Standard Bayesian models typically borrow strength across neighboring areas through conditional autoregressive or related Gaussian Markov random field priors \citep{besag1974spatial,besag1991bayesian,leroux2000estimation}. These priors stabilize noisy small-area estimates and produce interpretable smoothed risk surfaces. However, %the 
same %smoothing that makes these models useful 
can also be misleading when neighboring areas lie on opposite sides of a socioeconomic, environmental, administrative, or infrastructural boundary. In such cases, smoothing across the border may attenuate meaningful contrasts and obscure localized high-risk areas.

A substantial literature has, therefore, considered boundary detection and localized smoothing in disease mapping. Early approaches identified boundaries post hoc from large posterior differences in fitted risks across adjacent areas \citep{lu2005bayesian}, while later models introduced edge-specific indicators or neighbor-specific weights to weaken or remove spatial dependence across selected borders \citep{lu2007bayesian,ma2007bayesian,ma2010hierarchical}. These formulations are flexible but can introduce many edge-level latent quantities, complicating posterior learning and prior specification \citep{li2015bayesian}. More structured approaches link boundary formation to observed dissimilarities in explanatory variables, allowing smoothing to be interrupted across edges where neighboring areas differ sufficiently with respect to a relevant covariate \citep{lee2012boundary,lee2014bayesian,rushworth2017adaptive,lee2021improved,gao2023spatial,aiello2023detecting,wu2025assessing}.

In this article, we develop a reusable, amortized posterior approximation for heterogeneous areal graphs for spatial boundary detection. The statistical model combines a Poisson observation model for areal counts, a covariate-driven mechanism that modifies the observed adjacency graph, and a directed acyclic graph autoregressive \citep[DAGAR;][]{datta2019spatial} prior for latent spatial effects on the resulting graph. By estimating the boundary parameter jointly with the DAGAR residual dependence parameter, the model separates local boundary formation from residual effects. The posterior targets are therefore not only node-level disease risks but also edge-level posterior boundary probabilities and residual spatial dependence. This makes the model a useful test case for AI-assisted Bayesian inference: the inferential output is a full posterior distribution over scientifically interpretable spatial quantities, not a point prediction or a fixed risk score.

The posterior distribution under this model is nonstandard because the graph itself depends on an unknown boundary parameter. A dataset-specific Markov chain Monte Carlo (MCMC) algorithm can be constructed, but it must be rerun for each new map. Amortized Bayesian inference \citep[ABI;][]{radev2020bayesflow,sainsbury2024likelihood,zammit2025neural} offers an alternate strategy: train a neural inference system on simulated parameter-data pairs generated from the model, then reuse the trained system to approximate posterior inference for new datasets from the same deployment regime. In this sense, the neural network acts as a reusable posterior-computation device for real-world spatial analysis in which the same scientific question may recur across different regions, outcomes, or administrative partitions. This approach is closely related to simulation-based inference and neural posterior estimation with normalizing flows \citep{rezende2015variational,papamakarios2017masked,durkan2019neural,papamakarios2021normalizing}. 

For Bayesian workflows, however, amortization is useful only if the learned posterior approximation is validated through recovery, calibration, posterior predictive checks, and comparison with model-matched posterior computation \citep{cook2006validation}. In our setting, this means treating the neural network as an inferential approximation whose uncertainty quantification, calibration, and range of reliable deployment must be assessed, rather than as a black-box prediction device.

The key methodological challenge is that real-world maps do not have a fixed input dimension. Different applications may involve different numbers of regions, adjacency structures, exposure patterns, and covariate dissimilarities. To address this, each simulated or observed areal dataset is represented as an unordered collection of node-specific, graph-aware feature vectors constructed from the observed counts, offsets, covariates, and adjacency matrix. The number of feature vectors is the number of areas in the map and can vary across datasets. A ``SetTransformer'' summary network maps this variable-size unordered set to a fixed-dimensional representation while preserving permutation invariance \citep{zaheer2017deep,lee2019set}. A conditional normalizing flow then uses this representation to approximate the posterior distribution of the model parameters. The resulting network learns a posterior operator over a class of maps and adjacency graphs, rather than a posterior approximation tied to a single spatial layout.

We aim for three specific contributions here. First, we develop an amortized Bayesian inference workflow that accepts varying-size areal graphs as input and returns approximate posterior draws for a spatial boundary-detection model. Second, we formulate the boundary model by combining covariate-driven edge modification with a DAGAR prior, yielding joint posterior inference for local discontinuities and residual spatial dependence. Third, we validate the reusable posterior approximation through parameter recovery, posterior calibration, posterior boundary probabilities, posterior predictive behavior, agreement with a model-matched MCMC-DAGAR benchmark, and deployment on Glasgow respiratory disease and California lung cancer data with a comparison to the \texttt{CARBayes} package \citep{CARBayes2013}. Section~\ref{sec:model} introduces the model, Section~\ref{sec:ABI} describes the varying-size graph ABI workflow, Section~\ref{sec:simulation_experiments} presents simulation validation, Section~\ref{sec:real_data_analysis} reports the empirical analyses, and Section~\ref{sec:discussion} concludes.

\section{Model}\label{sec:model}

We develop a hierarchical spatial model for areal count data with three components: an observation model for the counts, a covariate-driven mechanism that modifies the adjacency graph to accommodate local boundaries, and a DAGAR prior that induces residual spatial dependence on the resulting graph. This construction separates two distinct features of spatial variation: local discontinuities in smoothing and broader residual spatial dependence.

Our inferential targets include the area-specific relative risks, posterior probabilities that geographic edges are ``difference boundaries'', and the residual spatial dependence that remains after such boundaries have been accommodated. Thus, the model is designed to produce uncertainty-quantified spatial evidence at two levels: node-level disease risk and edge-level discontinuity structure. This distinction is important in repeated disease-mapping applications, where public-health interpretation depends both on where elevated risks occur and on whether adjacent areas should be smoothed together.

Let $i=1,\dots,N$ index the areal units, $y_i$ denote the observed count for region $i$, $e_i$ the expected count, and $x_i$ an area-level covariate used to characterize local boundary information specified below. Conditional on a latent spatial effect $w_i$, we specify
\begin{equation}\label{eq:y_model}
    y_i \mid w_i, \beta_0 \sim \mathrm{Poisson}(\lambda_i),
    \qquad
    \log \lambda_i = \log e_i + \beta_0 + w_i .
\end{equation}
Equivalently, $\lambda_i = e_i \exp(\beta_0 + w_i)$, so that $\exp(\beta_0 + w_i)$ has the interpretation of a relative risk for area $i$ with respect to the expected count $e_i$. Under \eqref{eq:y_model}, $\beta_0$ is a global log-risk intercept, while $w_i$ captures residual spatial variation after accounting for the offset. This specification is standard in disease mapping and spatial epidemiology; see, for example, \cite{wakefield2007disease} and \cite{lawson2018bayesian}.

\subsection{Covariate-driven boundary mechanism}

We now specify the covariate-driven boundary mechanism underlying the model. Our construction is motivated by the literature on boundary detection in disease mapping, and in particular, by the formulation of \citet{lee2012boundary} where adjacency relationships between neighboring areas are allowed to vary according to observed dissimilarity measures. In that framework, latent spatial effects are modeled using a conditional autoregressive \citep[CAR;][]{besag1991bayesian} prior of the type proposed by \citet{leroux2000estimation}. We adopt the same covariate-informed view of boundary formation, but replace the CAR with a DAGAR prior \citep{datta2019spatial} for the spatial random effects. For details on explicitly incorporating adjacency information within a DAGAR formulation, see \citet{aiello2023detecting}.

A substantial body of work has addressed spatial boundary detection under CAR-based disease-mapping models. Early approaches often identified boundaries post hoc through large pairwise differences in fitted disease risks across neighboring areas, yielding so-called boundary likelihood values \citep{lu2005bayesian}. Subsequent formulations introduced latent edge indicators or neighbor-specific weights, so that boundaries correspond to conditional independence or weakened dependence between adjacent random effects \citep{lu2007bayesian,ma2007bayesian,ma2010hierarchical}. Although flexible, such approaches can introduce many edge-specific quantities, making posterior learning and prior specification more challenging \citep{li2015bayesian}. More structured alternatives link boundary formation to observed dissimilarities, allowing boundaries to be driven by covariates rather than estimated independently edge by edge \citep{lee2012boundary,lee2014bayesian,rushworth2017adaptive}. Recent work has further developed graph-based and multiple-outcome formulations, as well as Bayesian approaches to spatial disparity detection \citep{lee2021improved,gao2023spatial,wu2025assessing}.

Our model follows the covariate-driven perspective of \citet{lee2012boundary}, but differs in two ways. First, the residual spatial effects are assigned a DAGAR prior rather than a CAR prior. Second, the residual spatial dependence parameter is estimated jointly with the boundary parameter. This yields a formulation in which local boundary formation and residual spatial smoothing are modeled by distinct components of the hierarchy.

Let $\bs{A}=(a_{ij})$ denote the binary adjacency matrix of the areal graph, where $a_{ij}=1$ if areas $i$ and $j$ are neighbors and $a_{ij}=0$ otherwise. Following \citet{lee2012boundary}, we quantify the covariate dissimilarity between neighboring areas $i$ and $j$ by $z_{ij}=|x_i-x_j|$, where $x_i$ and $x_j$ are the values of the boundary-driving covariate in the two areas. To make the boundary parameter comparable across covariate scales and graphs, we rescale these dissimilarities using their empirical median over neighboring pairs, i.e., $Z_{0.5}=\mathrm{median}\{z_{ij}:a_{ij}=1\}$. The elements of the effective adjacency matrix are then defined by
\begin{equation}\label{eq:boundary_model}
    a^{\ast}_{ij} =
    \begin{cases}
        1, & \text{if } a_{ij}=1 \text{ and } \exp(-\eta z_{ij}) \geq \tfrac{1}{2}, \\
        0, & \text{otherwise}.
    \end{cases}
\end{equation}
Equivalently, an observed edge is retained when $\eta z_{ij}\leq \log 2$ and removed when $\eta z_{ij}>\log 2$. Therefore, neighboring areas with sufficiently large covariate dissimilarity are prevented from smoothing across their shared border. The parameter $\eta$ therefore controls the degree of local boundary formation: larger values imply stronger penalization of dissimilar neighboring pairs, leading to the removal of more edges and sharper local discontinuities. Since $\eta$ is inferred from the data, the boundary rule in \eqref{eq:boundary_model} induces posterior boundary probabilities for each observed edge, $\Pr(a_{ij}^{\ast}=0\mid \bs{y})$,
%\begin{equation*}
%    \Pr(a_{ij}^{\ast}=0\mid \bs{y}),
%\end{equation*}
which can be used directly for uncertainty-aware boundary assessment or translated into a binary boundary map through a chosen decision rule.

\subsection{DAGAR prior and parameter specification}

Given the modified adjacency structure $\bs{A}^\ast=(a_{ij}^\ast)$ through \eqref{eq:boundary_model}, we assign a DAGAR prior to the latent spatial effects. Let $\pi=(\pi_1,\dots,\pi_N)$ denote a fixed ordering of the nodes, and define $N_{\pi}(\pi_k)=\{\pi_\ell:\ell<k,\; a^\ast_{\pi_k,\pi_\ell}=1\}$
% \begin{equation*}
%     N_{\pi}(\pi_k)=\{\pi_\ell:\ell<k,\; a^\ast_{\pi_k,\pi_\ell}=1\},
% \end{equation*}
to be the set of predecessors of node $\pi_k$ under this ordering. The DAGAR model specifies
\begin{equation}\label{eq:w_model}
    \bs{w}\sim\mathcal{N}\!\left(\bs{0},\sigma_w^2\bs{Q}(\rho)^{-1}\right),
    \qquad
    \bs{Q}(\rho)=(\bs{I}-\bs{B})^\top \bs{\Lambda}(\bs{I}-\bs{B}),
\end{equation}
where $\bs{w} = (w_1,\dots,w_N)$, $\bs{B} = (b_{ij})$ is strictly lower triangular and $\bs{\Lambda}=\mathrm{diag}(\lambda_1,\dots,\lambda_N)$. The nonzero entries of $\bs{B}$ and the diagonal elements of $\bs{\Lambda}$ are given by
\begin{equation*}
\begin{aligned}
    b_{ij}
    &=
    \begin{cases}
        \dfrac{\rho}{1+(n_i-1)\rho^2}, & \text{if } \pi_j\in N_{\pi}(\pi_i),\\
        0, & \text{otherwise},
    \end{cases}
    &\qquad \text{and}\qquad
    \lambda_i
    &=
    \frac{1+(n_i-1)\rho^2}{1-\rho^2}
\end{aligned}\;,
\end{equation*}
respectively, where $n_i=|N_{\pi}(\pi_i)|$. Under \eqref{eq:w_model}, the parameter $\sigma_w^2>0$ controls the marginal variability of the spatial random effects, whereas $\rho\in(0,1)$ governs the strength of residual spatial dependence on the graph induced by $\bs{A}^\ast$. Unlike many CAR formulations, the DAGAR dependence parameter has a direct correlation-like interpretation, which facilitates prior specification and posterior interpretation.

A key distinction from the boundary-detection framework of \citet{lee2012boundary} is that $\rho$ is treated as an unknown parameter and estimated from the data. In contrast, Lee and Mitchell fix the residual spatial dependence close to one in their CAR formulation so that smoothing is driven primarily by the local edge weights induced by the boundary mechanism. By learning $\rho$ jointly with $\eta$, the proposed model separates local boundary formation from residual spatial persistence: $\eta$ determines which edges are removed on the basis of covariate dissimilarity, whereas $\rho$ determines the strength of residual smoothing over the retained graph. This separation is the main modeling distinction between the proposed DAGAR formulation and CAR-based localized smoothing models in which boundary formation and residual smoothing are more tightly coupled.

The full parameter vector is $ \bs{\theta} = (\beta_0,\sigma_w^2,\eta,\rho)$
%\begin{equation*}
%    \bs{\theta} = (\beta_0,\sigma_w^2,\eta,\rho),
%\end{equation*}
and we assign the priors
\begin{equation}\label{eq:priors}
    \beta_0 \sim \mathcal{N}(0,\sigma_\beta^2),
    \qquad
    \sigma_w^2 \sim \mathrm{HalfNormal}(0.5),
    \qquad
    \eta \sim \mathrm{Unif}(0,M),
    \qquad
    \rho \sim \mathrm{Unif}(0,1).
\end{equation}
In \eqref{eq:priors}, the prior on $\eta$ is scaled by $M=\log(2)/Z_{0.5}$, where $Z_{0.5}$ is the median covariate dissimilarity over neighboring pairs. This scaling makes the boundary parameter comparable across graphs and covariate scales: $\eta=0$ retains the original adjacency graph, while values near $M$ allow edges with above-median dissimilarity to be removed. Thus, the prior on $\eta$ is specified on a graph-adapted scale governing the amount of covariate-driven boundary formation.

The half-normal prior on $\sigma_w^2$ mildly regularizes the marginal variance of the spatial random effects toward smaller values while retaining support for substantial residual heterogeneity. This regularization is useful because the magnitude of residual spatial variation, the residual dependence parameter $\rho$, and the boundary parameter $\eta$ are learned jointly; the prior therefore helps stabilize the separation between overall spatial variability, global spatial persistence, and local boundary formation.

Taken together, the proposed hierarchy \eqref{eq:y_model}--\eqref{eq:priors} defines a spatial Poisson model in which local boundaries are induced by covariate dissimilarity, while residual spatial association is modeled through a DAGAR prior on the resulting graph. The model, therefore, provides joint posterior inference on node-level risk, edge-level boundary probabilities, and residual spatial dependence, which are the primary inferential targets carried forward into the amortized posterior approximation.

\section{Amortized Bayesian Inference}\label{sec:ABI}

Posterior inference is carried out using ABI. In statistical terms, ABI replaces repeated dataset-specific posterior computation by a supervised conditional-density-estimation problem. We simulate many parameter-data pairs from the Bayesian model and train a neural conditional density estimator to approximate the map from an observed dataset to the corresponding posterior distribution. Once trained, the same estimator can be reused for new datasets generated from the same deployment regime.

In the present setting, the central challenge is that each dataset is observed on an areal graph whose size and adjacency structure may differ from one application to another. ABI is therefore used not only to avoid repeated dataset-specific posterior computation, but to learn uncertainty-quantified inference across a class of spatial graphs generated from the proposed hierarchical model in Section~\ref{sec:model}. The neural component does not replace the Bayesian model; it approximates posterior inference under the specified generative model.

The ABI workflow used here has three components. First, a simulator generates synthetic areal datasets from the hierarchical model presented in Section~\ref{sec:model}. Second, a summary network maps each variable-size areal graph, represented through observed counts, offsets, covariates, and adjacency information, to a fixed-dimensional representation. Third, an inference network uses this representation to approximate the posterior distribution of the model parameters. The reliability of this approximation must then be assessed through parameter recovery, posterior calibration, posterior predictive checking, boundary-probability assessment, and benchmark comparisons.

\subsection{Training data generated from the hierarchical model}

Let $\bs{\theta}=(\beta_0,\sigma_w^2,\eta,\rho)$ denote the model parameters. Training is based on synthetic parameter-data pairs $(\bs{\theta}^{(g)},\mathcal{D}^{(g)})$, for $g=1,\ldots,G$, where $g$ indexes simulated training datasets. For each $g$, $\bs{\theta}^{(g)}\sim p(\bs{\theta})$, and $\mathcal{D}^{(g)}\sim p(\mathcal{D}\mid\bs{\theta}^{(g)})$ denotes a synthetic areal dataset generated from the hierarchical model \eqref{eq:y_model}--\eqref{eq:priors}. Here $\mathcal{D}$ denotes the observable quantities available in an application: counts, offsets, covariates, and the observed adjacency matrix.

Each synthetic dataset is generated as follows. A graph size $N$ is sampled from the training design. Conditional on $N$, planar coordinates are generated, and a base adjacency matrix $\bs{A}$ is obtained from the Delaunay triangulation. A covariate vector $\bs{x}$ is then generated, neighboring dissimilarities $z_{ij}=|x_i-x_j|$ are computed, and the graph-specific scaling constant $M=\log 2/Z_{0.5}$ is calculated from the median neighboring dissimilarity $Z_{0.5}$. The boundary parameter $\eta\in(0,M)$ determines the modified adjacency matrix $\bs{A}^{\ast}$ through the thresholding rule in \eqref{eq:boundary_model}. Given $\bs{A}^{\ast}$, latent spatial effects are simulated from the DAGAR prior, and counts are generated from the Poisson observation model.

This simulator defines the joint distribution $p(\bs{\theta},\mathcal{D})=p(\bs{\theta})p(\mathcal{D}\mid\bs{\theta})$ used to train the amortized posterior approximation. Importantly, the training distribution is not tied to a single fixed map: graph size, topology, covariate surface, boundary structure, and residual dependence vary across simulations. The simulator therefore defines the deployment regime over which the amortized posterior approximator learns inference and over which its reliability is later evaluated.

\subsection{Summary network for variable-size areal datasets}
\label{sec:summary_network}

A key feature of the proposed amortized workflow is that the input map is not assumed to have a fixed size. Different simulated and observed datasets may contain different numbers of areal units and different adjacency structures. We therefore represent each dataset $\mathcal{D}$ as an unordered collection of node-specific feature vectors, $\mathcal{S}(\mathcal{D}) = \{\bs{s}_1,\ldots,\bs{s}_N\}$, where $N$ is the number of areal units in the corresponding map. The length of this set is allowed to vary across datasets. A summary network then maps the variable-size set to a fixed-dimensional representation,
\begin{equation}\label{eq:summary_network}
    \bs{z}=h_{\bs{\psi}}(\mathcal{S}(\mathcal{D})),
\end{equation}
where $h_{\bs{\psi}}$ has parameters $\bs{\psi}$. This step is what allows the inference network to operate on maps with different numbers of regions: the posterior approximation is conditioned on the fixed-dimensional summary $\bs{z}$, while the raw graph-level input remains variable-size.

The node-level features $\bs{s}_i$ are constructed from observed quantities only: the count $y_i$, offset $e_i$, covariate value $x_i$, and the observed adjacency matrix $\bs{A}$. The latent filtered graph $\bs{A}^{\ast}$ and latent spatial effects $\bs{w}$ are not used as inputs, because they are unavailable in empirical applications and would not be available at inference time.

The representation is model-guided. Its purpose is not to replace the spatial model, but to provide the amortized posterior approximator with observable information aligned with the inferential roles of $\beta_0$, $\sigma_w^2$, $\eta$, and $\rho$. The summaries retain information about four main aspects of the data. Marginal count and offset summaries, including $y_i$, $e_i$, $\log(1+y_i)$, $\log e_i$, and an offset-adjusted residual proxy, primarily inform the global log-risk level $\beta_0$ and the overall magnitude of residual variation $\sigma_w^2$. Neighborhood residual summaries, such as local averages and local roughness over observed neighbors, describe agreement or disagreement among adjacent areas and are informative about $\rho$ and $\sigma_w^2$. Dissimilarity-based summaries, including local covariate dissimilarities and dissimilarity-binned edge summaries, capture the relationship between residual jumps and covariate differences across neighboring areas and are designed to inform the boundary parameter $\eta$. Finally, graph-level autocorrelation, lag-concordance, and roughness summaries capture residual spatial dependence over the observed graph and are primarily informative about $\rho$.

Full definitions of all node-level, edge-binned, and graph-level summaries are provided in Appendix~\ref{supp:observed_data_representation}. The main point is that the summaries are constructed from observable graph-indexed quantities, while the summary network learns how to aggregate them across maps of different sizes.

In our implementation, the summary network $h_{\bs{\psi}}$ in \eqref{eq:summary_network} is a SetTransformer \citep{lee2019set}. This choice is natural because the numerical labels assigned to areal units should not carry inferential meaning, and because the number of areal units can vary across datasets. The SetTransformer aggregates the unordered collection $\mathcal{S}(\mathcal{D})$ into the fixed-dimensional representation $\bs{z}$ while preserving permutation invariance. Thus, the architecture separates two tasks: model-guided graph summaries encode information relevant to the spatial boundary model, and the SetTransformer learns how to aggregate this information across varying-size areal graphs.

\subsection{Inference network and posterior approximation}

The Bayesian inferential target is the posterior distribution of $\bs{\theta}=(\beta_0,\sigma_w^2,\eta,\rho)$ given an observed areal dataset $\mathcal{D}$. Since $\mathcal{D}$ may be observed on a graph with a different number of areas from other datasets, we first map it to the fixed-dimensional representation $\bs{z}$ through \eqref{eq:summary_network}. The inference network then approximates posterior inference through a conditional density
\begin{equation}\label{eq:inference_network}
    q_{\bs{\phi}}(\bs{\theta}\mid\bs{z})
    =
    q_{\bs{\phi}}
    \left(
    \bs{\theta}
    \mid
    h_{\bs{\psi}}(\mathcal{S}(\mathcal{D}))
    \right).
\end{equation}
Thus, the learned summary $\bs{z}$ provides a common conditioning object for datasets whose original graph-indexed representations have different sizes.

We use a conditional normalizing flow for $q_{\bs{\phi}}$ in \eqref{eq:inference_network}. The flow defines an invertible transformation, conditional on $\bs{z}$, between the parameter vector $\bs{\theta}$ and a reference Gaussian variable $\bs{u}$:
\begin{equation*}
    \bs{u}=f_{\bs{\phi}}(\bs{\theta};\bs{z}),
    \qquad
    \bs{\theta}=f_{\bs{\phi}}^{-1}(\bs{u};\bs{z}),
    \qquad
    \bs{u}\sim\mathcal{N}(\bs{0},\bs{I}).
\end{equation*}
The conditional density in \eqref{eq:inference_network} is the density induced by this transformation through the change-of-variables formula,
\begin{equation}\label{eq:change_of_variable_inference}
    q_{\bs{\phi}}(\bs{\theta}\mid\bs{z})
    =
    p_{\bs{u}}
    (
    f_{\bs{\phi}}(\bs{\theta};\bs{z})
    )
    \left|
    \det
    \frac{\partial f_{\bs{\phi}}(\bs{\theta};\bs{z})}
         {\partial \bs{\theta}}
    \right|,
\end{equation}
where $p_{\bs{u}}$ is the standard Gaussian density. Training therefore uses the forward map $f_{\bs{\phi}}$ to evaluate the density assigned to simulated parameters, whereas posterior sampling uses the inverse map $f_{\bs{\phi}}^{-1}$.

The training objective can be understood as a conditional-density projection. The ideal target is to make the learned conditional density  $q_{\bs{\phi}}(\bs{\theta}\mid \bs{z})$ close to the conditional distribution of the simulator parameters given the learned representation. This corresponds to minimizing
\begin{equation}\label{eq:KL_loss}
    \mathbb{E}_{p(\bs{z})}
    \left[
    \mathbb{KL}
    \left\{
    p(\bs{\theta}\mid \bs{z})
    \,\|\, 
    q_{\bs{\phi}}(\bs{\theta}\mid \bs{z})
    \right\}
    \right].
\end{equation}
For fixed $\bs{z}$, the entropy term in the KL decomposition does not depend on $q_{\bs{\phi}}$. Therefore, as in likelihood-based training of neural posterior estimators and normalizing flows, minimizing the KL criterion in \eqref{eq:KL_loss} is equivalent to minimizing the expected negative log-density assigned by the flow to parameters generated by the simulator \citep{radev2020bayesflow}. The summary network and inference network are therefore trained jointly on simulated parameter-data pairs using the objective
\begin{equation}\label{eq:abi_objective}
    (\hat{\bs{\phi}},\hat{\bs{\psi}})
    =
    \arg\min_{\bs{\phi},\bs{\psi}}
    \left\{
    -
    \mathbb{E}_{p(\bs{\theta},\mathcal{D})}
    \left[
    \log q_{\bs{\phi}}
    \left(
    \bs{\theta}
    \mid
    h_{\bs{\psi}}(\mathcal{S}(\mathcal{D}))
    \right)
    \right]
    \right\}.
\end{equation}
Thus, in the idealized limit of unlimited simulations, sufficient network capacity, and successful optimization, the learned flow targets the conditional distribution of the parameters given the learned representation $\bs{z}$. In practice, the approximation may differ from the exact posterior $p(\bs{\theta}\mid\mathcal{D})$ because $\bs{z}$ may not be sufficient, the network class is finite, the simulator is sampled finitely, and numerical optimization is imperfect. This is why the amortized posterior approximation must be evaluated empirically on held-out datasets through recovery, calibration, posterior predictive checks, boundary-probability diagnostics, and benchmark comparison.

% For an observed dataset $\mathcal{D}_{\mathrm{obs}}$, we first compute
% \begin{equation*}
%     \bs{z}_{\mathrm{obs}}
%     =
%     h_{\hat{\bs{\psi}}}
%     \{
%     \mathcal{S}(\mathcal{D}_{\mathrm{obs}})
%     \}.
% \end{equation*}
% Posterior samples are then obtained by drawing
% \begin{equation*}
%     \bs{u}^{(\ell)}\sim\mathcal{N}(\bs{0},\bs{I}),
%     \qquad \ell=1,\ldots,L,
% \end{equation*}
% and transforming back to parameter space:
% \begin{equation*}
%     \bs{\theta}^{(\ell)}
%     =
%     f_{\hat{\bs{\phi}}}^{-1}
%     \left(
%     \bs{u}^{(\ell)};\bs{z}_{\mathrm{obs}}
%     \right).
% \end{equation*}
% The resulting draws
% $\{\bs{\theta}^{(\ell)}\}_{\ell=1}^L$
% are samples from the learned approximation
% \begin{equation*}
%     q_{\hat{\bs{\phi}}}
%     (\bs{\theta}\mid \bs{z}_{\mathrm{obs}}).
% \end{equation*}
For an observed dataset $\mathcal{D}_{\mathrm{obs}}$, we compute $\bs{z}_{\mathrm{obs}}=h_{\hat{\bs{\psi}}}(\mathcal{S}(\mathcal{D}_{\mathrm{obs}}))$. Posterior samples are obtained by drawing $\bs{u}^{(\ell)}\sim\mathcal{N}(\bs{0},\bs{I})$ and setting $\bs{\theta}^{(\ell)}=f_{\hat{\bs{\phi}}}^{-1}(\bs{u}^{(\ell)};\bs{z}_{\mathrm{obs}})$, so that the resulting draws approximate $q_{\hat{\bs{\phi}}}(\bs{\theta}\mid\bs{z}_{\mathrm{obs}})$.

Algorithm~\ref{alg:abi_dagar} summarizes the training and deployment workflow corresponding to the objective in \eqref{eq:abi_objective}. During training, simulated datasets are transformed into observed-data representations, summarized by $h_{\bs{\psi}}$, and used to train the conditional flow. During deployment, a new observed map is passed through the fitted summary network and inverse flow to obtain approximate posterior draws.

\begin{algorithm}[t]
\caption{Amortized Bayesian inference for the DAGAR boundary-detection model}
\label{alg:abi_dagar}
%\fontsize{9}{11}\selectfont
\begin{algorithmic}[1]
\Require Simulator for $(\bs{\theta},\mathcal{D})$, feature map $\mathcal{S}(\cdot)$, summary network $h_{\bs{\psi}}$, conditional flow $f_{\bs{\phi}}$, batch size $B$, number of posterior draws $L$
\Statex

\State \textbf{Training phase}
\Repeat
    \For{$b=1,\ldots,B$}
        \State Simulate parameters and data $(\bs{\theta}^{(b)},\mathcal{D}^{(b)})\sim p(\bs{\theta},\mathcal{D})$
        \State Construct the observed-data representation $\mathcal{S}^{(b)}=\mathcal{S}(\mathcal{D}^{(b)})$
        \State Compute the summary representation $\bs{z}^{(b)}=h_{\bs{\psi}}(\mathcal{S}^{(b)})$
        \State Evaluate the flow density $q_{\bs{\phi}}(\bs{\theta}^{(b)}\mid\bs{z}^{(b)})$
    \EndFor
    \State Update $(\bs{\phi},\bs{\psi})$ by minimizing the minibatch negative log-density loss
\Until{convergence}

\Statex
\State \textbf{Deployment phase for an observed dataset $\mathcal{D}_{\mathrm{obs}}$}
\State Construct $\mathcal{S}_{\mathrm{obs}}=\mathcal{S}(\mathcal{D}_{\mathrm{obs}})$
\State Compute $\bs{z}_{\mathrm{obs}}=h_{\hat{\bs{\psi}}}(\mathcal{S}_{\mathrm{obs}})$
\For{$\ell=1,\ldots,L$}
    \State Draw $\bs{u}^{(\ell)}\sim\mathcal{N}(\bs{0},\bs{I})$
    \State Set $\bs{\theta}^{(\ell)}=f_{\hat{\bs{\phi}}}^{-1}(\bs{u}^{(\ell)};\bs{z}_{\mathrm{obs}})$
\EndFor
\State Return $\{\bs{\theta}^{(\ell)}\}_{\ell=1}^L$ as approximate posterior draws from $q_{\hat{\bs{\phi}}}(\bs{\theta}\mid\bs{z}_{\mathrm{obs}})$
\end{algorithmic}
\end{algorithm}

\subsection{Approximation and validation}

The resulting posterior draws are samples from the learned approximation $q_{\hat{\bs{\phi}}}\!(\bs{\theta}\mid h_{\hat{\bs{\psi}}}\{\mathcal{S}(\mathcal{D}_{\mathrm{obs}})\})$, 
% \begin{equation*}
%     q_{\hat{\bs{\phi}}}\!\left(
%     \bs{\theta}\mid
%     h_{\hat{\bs{\psi}}}\{\mathcal{S}(\mathcal{D}_{\mathrm{obs}})\}
%     \right),
% \end{equation*}
rather than exact draws from the Bayesian posterior. The preceding conditional-density-estimation argument clarifies the population target of training, but it does not by itself provide finite-sample or out-of-regime guarantees. Approximation error can arise from finite simulation, limitations of the observed-data representation, finite network capacity, and numerical optimization. This distinction is especially important here because posterior uncertainty is used not only for scalar model parameters, but also for edge-level boundary probabilities, decision rules, and public-health interpretation.

For this reason, validation of the amortized posterior approximation is an essential part of the proposed workflow. The central inferential question is not simply whether posterior samples can be generated efficiently for a new map, but whether the learned approximation provides reliable uncertainty quantification for the scientific targets of interest across the class of graphs on which it is intended to be deployed. In the validation analyses below, we therefore assess parameter recovery, posterior calibration, posterior predictive behavior, boundary-probability quality, and agreement with benchmark Bayesian analyses.

\section{Simulation experiments}\label{sec:simulation_experiments}

The simulation study was designed to evaluate whether a single amortized posterior approximator can learn reliable Bayesian inference across areal graphs of varying size and topology. All datasets were generated from the proposed Poisson areal count model with covariate-driven boundaries and DAGAR residual spatial dependence. The validation therefore targets both components of the contribution: the statistical boundary-detection model and the learned posterior operator used to approximate inference for new maps. We assessed parameter recovery, posterior calibration, posterior predictive behavior, boundary-probability quality, agreement with benchmark Bayesian analyses, and practical deployment after the one-time training stage.

\subsection{Simulation design}

The training and validation design was intentionally heterogeneous. For each simulated dataset, the number of areas was drawn uniformly between 40 and 300, so that the amortized posterior approximator was required to learn across maps with different numbers of regions and different graph topologies rather than being tuned to a single spatial configuration. For each realized graph and covariate configuration, the graph-specific scaling factor $M$ was computed from the median neighboring dissimilarity, and parameters were generated according to the prior distributions specified in \eqref{eq:priors} with $\sigma_\beta = 0.5$.
% \begin{equation*}
% \beta_0 \sim \mathcal{N}(0,0.5^2), \qquad
% \sigma_w^2 \sim \mathrm{HalfNormal}(0.5), \qquad
% \eta \sim \mathrm{Unif}(0,M), \qquad
% \rho \sim \mathrm{Unif}(0,1).
% \end{equation*}

Here, $\beta_0$ controls the global log-risk level in the observation model, $\sigma_w^2$ governs the magnitude of latent spatial variation, $\eta$ controls covariate-driven edge deletion, and $\rho$ controls residual spatial dependence under the DAGAR prior; see \eqref{eq:y_model}--\eqref{eq:w_model}. This design defines the deployment regime for the amortized posterior approximation: graph size, graph topology, covariate dissimilarity structure, boundary configuration, and residual dependence vary simultaneously.

\begin{figure}[t]
\centering
\includegraphics[width=\textwidth]{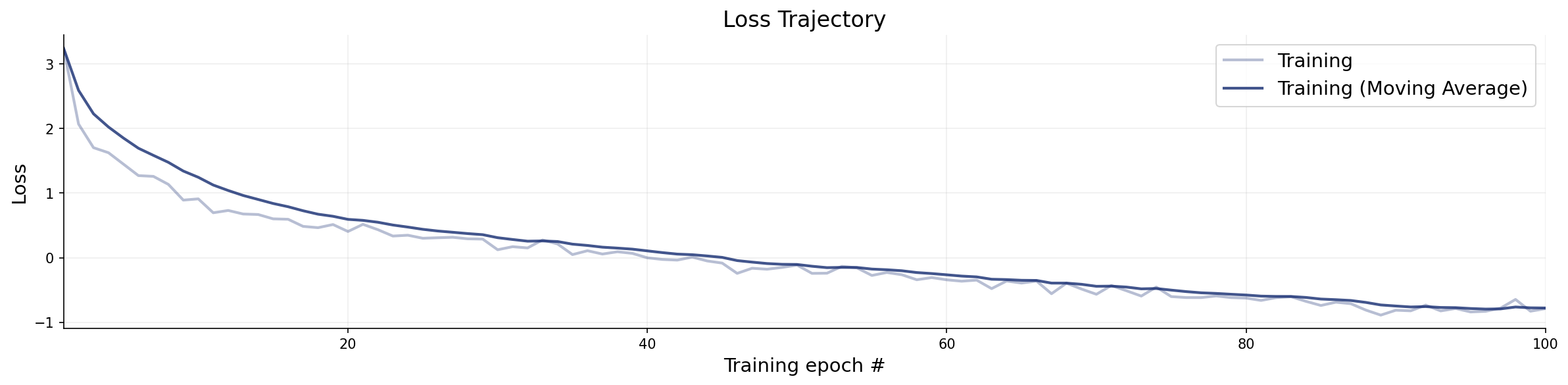}
\caption{Training loss over 100 epochs for the amortized posterior approximation.}
\label{fig:history}
\end{figure}

Training was performed online for 100 epochs with batch size 64 and 200 batches per epoch, so that the amortized posterior approximator was exposed to $100 \times 200 \times 64 = 1{,}280{,}000$ synthetic datasets generated under the model. As shown in Fig.~\ref{fig:history}, optimization was stable throughout: the loss decreased from 3.259 at the first epoch to $-0.710$ at the final epoch, attained a minimum of $-0.793$, and averaged $-0.700$ over the last five epochs. This trajectory indicates stable optimization of a posterior approximation trained over substantial heterogeneity in graph size, graph topology, and parameter configuration.

\subsection{Parameter recovery and posterior calibration}

For validation, we used 200 held-out simulated datasets and obtained 10{,}000 posterior draws per dataset. In the realized validation sample, the number of areas ranged from 40 to 299, with mean 160.615 and standard deviation 72.840. Thus, validation was performed on maps spanning essentially the full graph-size range used during training. All posterior draws respected the required support constraints, with $\sigma_w^2>0$, $\eta\in(0,M)$, and $\rho\in(0,1)$ in every case.

\begin{figure}[t]
\centering
\includegraphics[width=\textwidth]{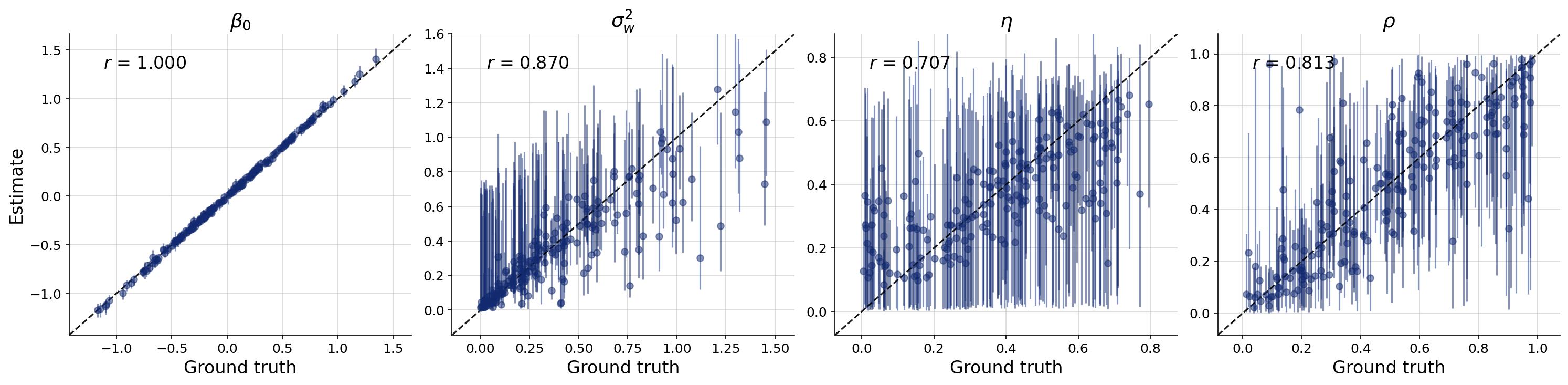}
\caption{Parameter recovery on 200 held-out simulated datasets.}
\label{fig:recovery}
\end{figure}

Fig.~\ref{fig:recovery} and Table~\ref{tab:sim_recovery} summarize parameter recovery on the held-out simulated datasets. Here, $r$ denotes the Pearson correlation between posterior mean estimates and true generating values across datasets, while $R^2$ is the corresponding coefficient of determination. Recovery was strongest for the global intercept $\beta_0$, for which both bias and RMSE were negligible and the posterior mean was almost perfectly aligned with the truth ($r=1.000$, $R^2=0.999$). Recovery was also strong for $\sigma_w^2$ ($r=0.870$, $R^2=0.752$) and informative for both $\eta$ ($r=0.707$, $R^2=0.491$) and $\rho$ ($r=0.813$, $R^2=0.678$). This ordering is substantively reasonable: $\beta_0$ is identified through the global mean structure, whereas $\sigma_w^2$, $\eta$, and $\rho$ must be learned indirectly through the interaction of latent spatial variation, boundary formation, and graph topology. Importantly, biases were small for all parameters and empirical 95\% interval coverage remained close to nominal, indicating that the amortized posterior approximation preserves the main uncertainty structure of the generative model across heterogeneous held-out graphs.

\begin{table}[t]
\centering
\caption{Parameter recovery on 200 held-out simulated datasets. RMSE is based on posterior means; $r$ is the Pearson correlation between posterior means and true generating values across datasets; $R^2$ is the corresponding coefficient of determination; coverage is empirical 95\% interval coverage.}
\label{tab:sim_recovery}
\begin{tabular}{lccccc}
\hline
\textbf{Parameter} & \textbf{Bias} & \textbf{RMSE} & $\bs{r}$ & $\bs{R^2}$ & \textbf{95\% coverage} \\
\hline
$\beta_0$    & 0.003    & 0.019 & 1.000 & 0.999 & 0.975 \\
$\sigma_w^2$ & $-$0.008 & 0.164 & 0.870 & 0.752 & 0.945 \\
$\eta$       & $-$0.020 & 0.150 & 0.707 & 0.491 & 0.955 \\
$\rho$       & $-$0.011 & 0.151 & 0.813 & 0.678 & 0.955 \\
\hline
\end{tabular}
\end{table}

The width of the posterior intervals in Fig.~\ref{fig:recovery} also illustrates an important feature of the approximation. Uncertainty is much tighter for $\beta_0$, which is informed directly by the global count level, whereas intervals are wider for $\sigma_w^2$, $\eta$, and $\rho$. This is expected because these parameters are learned indirectly through residual variation, graph topology, covariate-driven edge contrasts, and spatial dependence. In particular, different combinations of boundary strength, residual dependence, and latent variance can generate similar spatial count patterns, making these components intrinsically harder to separate. Thus, the validation results suggest that the amortized posterior is most precise for the mean component and more diffuse for the latent-structure components, while still maintaining small bias and close-to-nominal empirical coverage.

\begin{figure}[t]
\centering
\includegraphics[width=\textwidth]{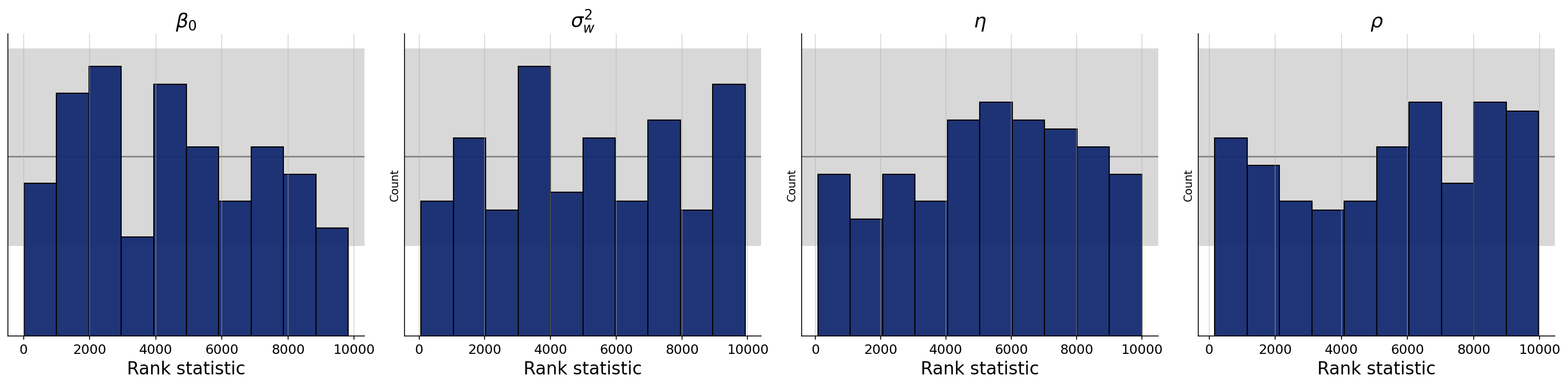}
\caption{Simulation-based calibration rank histograms on 200 held-out simulated datasets.}
\label{fig:sbc_hist}
\end{figure}

Posterior calibration was assessed using simulation-based calibration (SBC), empirical interval coverage, and graph-size-stratified posterior discrepancy summaries. The SBC rank histograms in Fig.~\ref{fig:sbc_hist} are broadly consistent with calibration. Across parameters, the histograms show only mild irregularities and do not exhibit pronounced U-shapes, edge pile-up, or other patterns suggesting severe systematic miscalibration. This visual impression is consistent with the empirical coverage results: 95\% interval coverage was 0.975 for $\beta_0$, 0.945 for $\sigma_w^2$, and 0.955 for both $\eta$ and $\rho$. Overall, these results indicate that the amortized posterior approximation provides reasonably calibrated uncertainty quantification on held-out maps with varying sizes and graph structures. The corresponding ECDF-difference plots, graph-size-stratified posterior $z$-score diagnostics, and regime-specific error heatmaps are reported in Appendix~\ref{sec:supp_sim_recovery_calibration}.

\subsection{Boundary-probability quality and decision performance}

A central feature of the proposed approach is that it returns posterior boundary probabilities for neighboring pairs, rather than only a thresholded boundary map. We therefore evaluated these probabilities directly on the pooled edge set across all held-out datasets. The posterior boundary probabilities achieved an area under the receiver operating characteristic curve (AUROC) of 0.970, an average precision (AP) of 0.882, and a Brier score of 0.057. These summaries assess complementary aspects of performance: ranking of true boundaries against non-boundaries, precision-recall behavior in the presence of potentially rare boundaries, and the accuracy of the predicted probabilities themselves. Taken together, the results indicate that the amortized posterior approximation produces boundary probabilities that are both highly discriminative and informative beyond any single thresholding rule.

We next examined two pre-specified decision rules applied to the posterior boundary probabilities. Because sensitivity is undefined for datasets containing no true boundaries, sensitivity summaries were computed over the 161 validation datasets containing at least one true boundary, whereas specificity remained well defined and was computed over all 200 validation datasets. The FDR-controlling rule employed by \cite{li2015bayesian} with $\alpha=0.05$ produced a highly conservative operating point. Across datasets containing true boundaries, its mean sensitivity was 0.169 (s.d. 0.329), while its mean specificity across all datasets was 0.995 (s.d. 0.016). Thus, the rule selected only the strongest boundary signals and yielded very few false-positive selections, but at the cost of missing many true boundaries. This low-recovery behavior should be interpreted as a consequence of the stringent FDR decision rule, rather than as evidence that the posterior boundary probabilities are uninformative.

\begin{figure}[t]
\centering
\includegraphics[width=\textwidth]{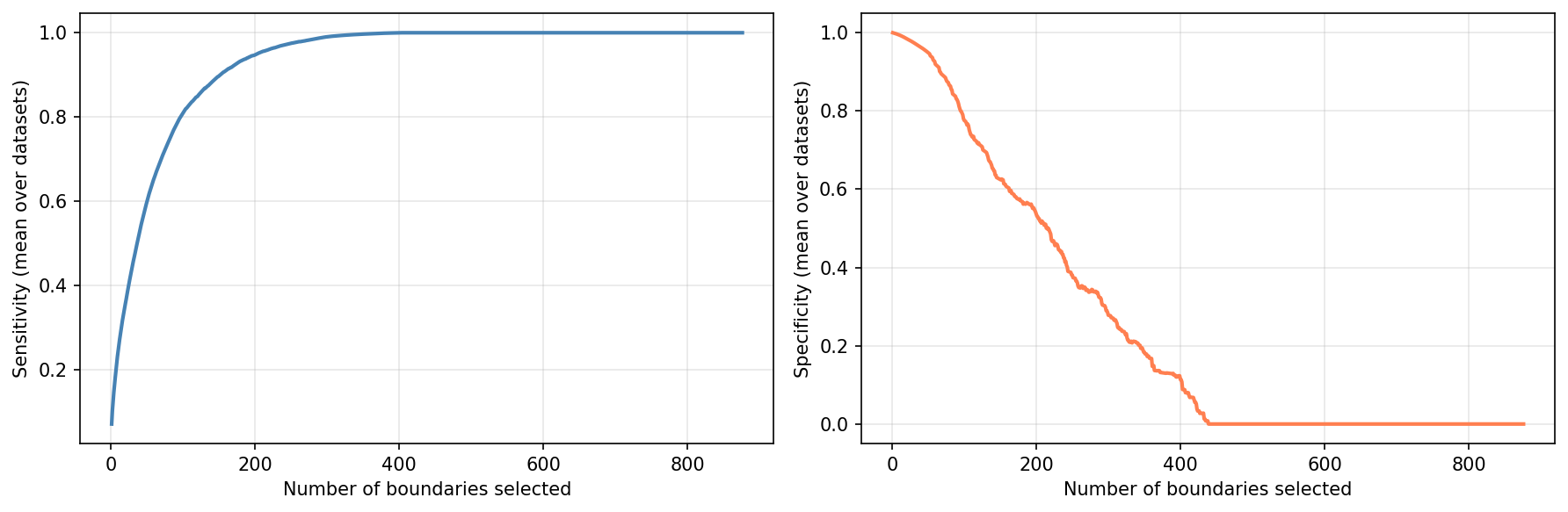}
\caption{Ranking diagnostic for posterior boundary probabilities. Mean sensitivity (left) and specificity (right) are shown as functions of the number of selected boundaries. Sensitivity excludes datasets containing no true boundaries; at each selection count, averages use datasets for which the corresponding metric and selection count are defined.}
\label{fig:boundary_curve}
\end{figure}

Fig.~\ref{fig:boundary_curve} places this operating point in a broader ranking perspective by plotting sensitivity and specificity as functions of the number of selected boundaries. The posterior boundary probabilities induce a useful ordering of candidate edges: sensitivity increases rapidly as the highest-ranked edges are selected, while specificity remains high over a substantial range of less conservative operating points before declining as increasingly marginal edges are added. This shows that the posterior probabilities contain strong ranking information even though the FDR rule itself is too stringent for high boundary recovery. At each selection count, averages are computed over datasets for which the corresponding selection count and performance metric are defined.

Under the median-probability rule, the amortized posterior selected on average 80.74 boundaries, compared with a mean true count of 91.24. Across the 161 datasets containing at least one true boundary, mean sensitivity was 0.719 (s.d. 0.322). Mean specificity across all 200 datasets was 0.962 (s.d. 0.059). Thus, the median-probability rule recovered a large fraction of true boundaries while retaining high specificity, giving a substantially less conservative and more balanced operating point than the FDR-controlling rule.

\begin{figure}[t]
\centering
\includegraphics[width=\textwidth]{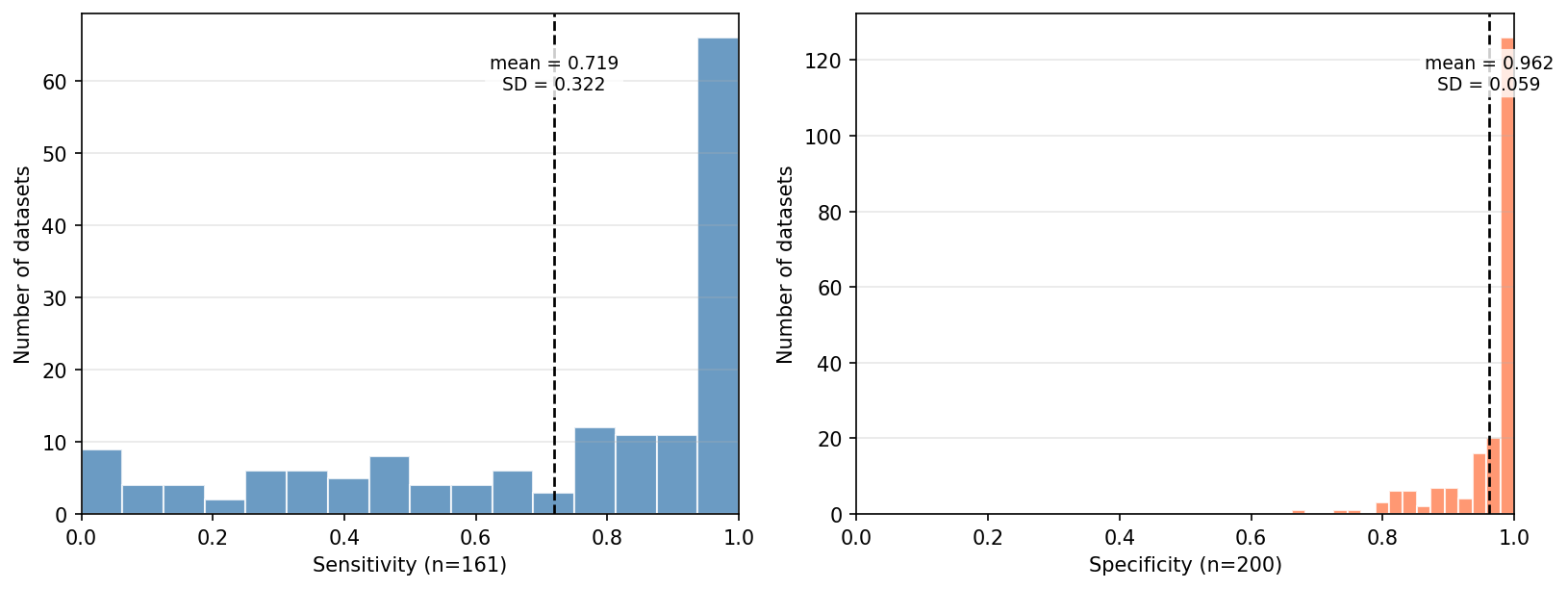}
\caption{Sensitivity (left) and specificity (right) under the median-probability rule. Sensitivity is shown for the 161 datasets containing at least one true boundary, while specificity is shown for all 200 held-out datasets. Dashed lines indicate the corresponding means.}
\label{fig:boundary_mpm_hist}
\end{figure}

Fig.~\ref{fig:boundary_mpm_hist} shows that sensitivity is frequently high but heterogeneous across datasets, whereas specificity is strongly concentrated near one. This pattern indicates that the median-probability rule controls false positives well across most simulated graphs, while the difficulty of recovering all true boundaries varies more strongly with the number, strength, and spatial configuration of generated boundaries. The zero-boundary datasets are omitted from the sensitivity panel because sensitivity is undefined for them, but they remain included in the specificity panel.

The posterior distribution of the total number of boundaries was also well calibrated. The true number of boundaries was contained in the corresponding 95\% posterior interval in 0.995 of held-out datasets, and the posterior mass below the true count averaged 0.553. Overall, these results show that the amortized posterior approximation provides useful probabilistic summaries of boundary evidence, induces sensible ranking behavior across operating points, and captures the aggregate level of boundary uncertainty on held-out graphs of varying size and topology. Additional dataset-level summaries, reliability plots, and dissimilarity-profile diagnostics are reported in Appendix~\ref{sec:supp_sim_boundary}.

\subsection{Posterior predictive checks}\label{sec:ppc}

Posterior predictive checks were conducted on 40 representative held-out datasets spanning the graph-size range, using 100 posterior predictive replicates per dataset. For each dataset and posterior draw, replicated counts were generated from the posterior predictive distribution induced by the amortized posterior approximation. We evaluated three classes of summaries: node-level count coverage, Moran-type residual spatial dependence, and edge-level residual contrasts stratified by covariate dissimilarity. For the latter two classes, posterior predictive $p$-values were computed as posterior predictive tail probabilities for the corresponding observed summaries.

Node-level count coverage was strong: the observed counts fell within the 95\% posterior predictive interval for a mean of 0.94 of areas per dataset, close to the nominal 0.95 benchmark. This suggests that the amortized posterior predictive distribution reproduces the marginal count scale well across held-out maps.

Moran-type summaries were also well reproduced. The observed Moran-type statistic was covered by the corresponding 95\% posterior predictive interval in 0.975 of datasets, and the associated posterior predictive $p$-values were centered near 0.5, with mean 0.486. This indicates that the amortized posterior approximation captures residual spatial dependence without systematic under- or over-smoothing.

Edge-contrast summaries stratified by dissimilarity bin behaved similarly, with mean posterior predictive $p$-values 0.535, 0.489, and 0.561 in the low-, medium-, and high-dissimilarity bins, respectively. These values indicate no systematic mismatch in the local contrast structure, either among neighboring areas with small dissimilarities or among those with larger dissimilarities where boundaries are more likely to arise.

Overall, these posterior predictive diagnostics support adequate fit at three levels: the marginal count scale, the global residual dependence structure, and the local edge-level contrast structure induced by the boundary mechanism. Thus, the amortized approximation reproduces not only scalar parameter uncertainty and boundary probabilities, but also key data-generating features on held-out graphs of varying size. Full posterior predictive plots and distributional summaries are reported in Appendix~\ref{sec:supp_sim_ppc}.

\subsection{Benchmark against model-matched MCMC-DAGAR}\label{sec:benchmark_mcmc_dagar}

As a second validation exercise, distinct from the 200-dataset truth-recovery study above, we compared ABI-DAGAR with an MCMC implementation of the same thresholded Poisson-DAGAR model, denoted MCMC-DAGAR. The MCMC sampler was constructed to follow the localized updating logic of the \texttt{CARBayes} framework as closely as possible while replacing the latent CAR prior with the DAGAR prior used in the proposed model. We applied both methods to an additional bank of 100 held-out simulated datasets and obtained 10{,}000 posterior draws per dataset with each method.

\begin{figure}[t]
\centering
\includegraphics[width=\textwidth]{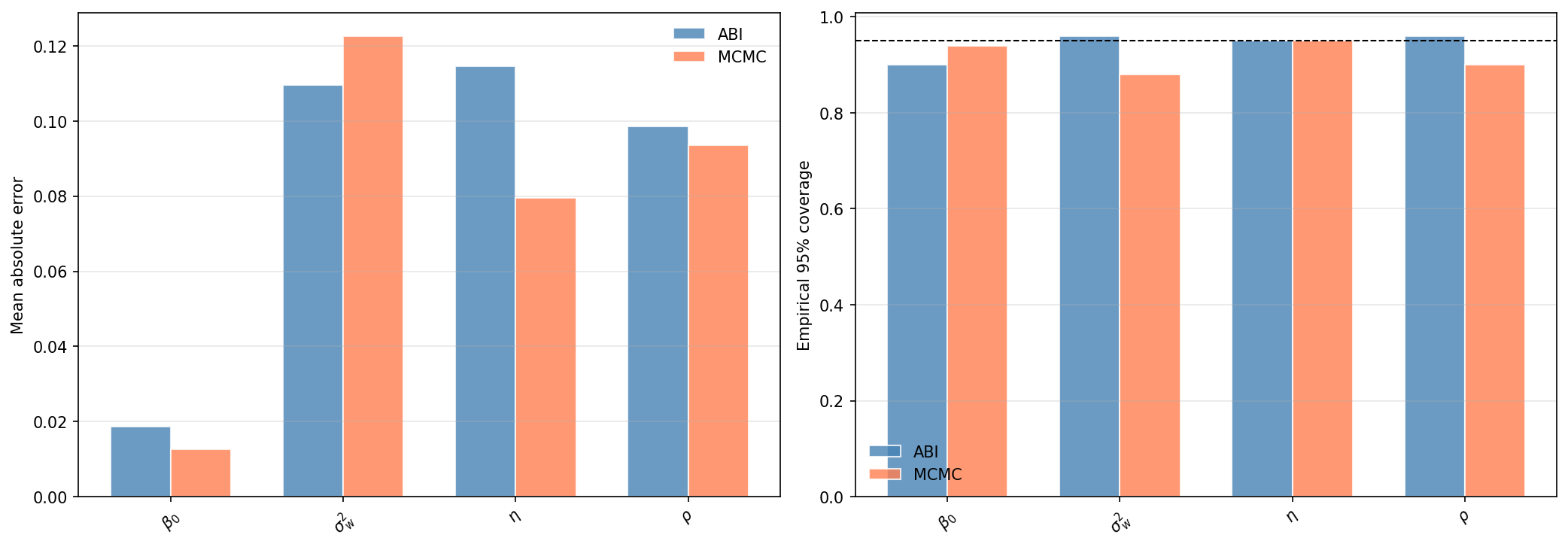}
\caption{Comparison between ABI-DAGAR and model-matched MCMC-DAGAR on 100 held-out simulated datasets. Left: mean absolute error by parameter. Right: empirical 95\% interval coverage.}
\label{fig:sim_abi_mcmc_recovery_main}
\end{figure}

The goal of this comparison is to assess whether the amortized approximation reproduces the posterior behavior of a dataset-specific Bayesian sampler for the same statistical model. Fig.~\ref{fig:sim_abi_mcmc_recovery_main} summarizes the parameter-level comparison through mean absolute error and empirical 95\% interval coverage. Agreement is strongest for the intercept $\beta_0$, for which both methods recover the truth extremely well. For the spatial parameters, differences are more visible but remain moderate: MCMC-DAGAR yields smaller mean absolute error for $\eta$ and $\rho$, whereas ABI-DAGAR performs slightly better for $\sigma_w^2$. Empirical 95\% interval coverage is mixed rather than uniformly favoring one method: MCMC-DAGAR attains higher coverage for $\beta_0$, ABI-DAGAR attains higher coverage for $\sigma_w^2$ and $\rho$, and the two methods coincide for $\eta$. Thus, at the scalar-parameter level, the amortized posterior approximation broadly tracks the model-matched MCMC-DAGAR benchmark, with the largest discrepancies occurring for the latent-structure components that are also most difficult to recover in the truth-based validation study.

\begin{figure}[t]
\centering
\includegraphics[width=\textwidth]{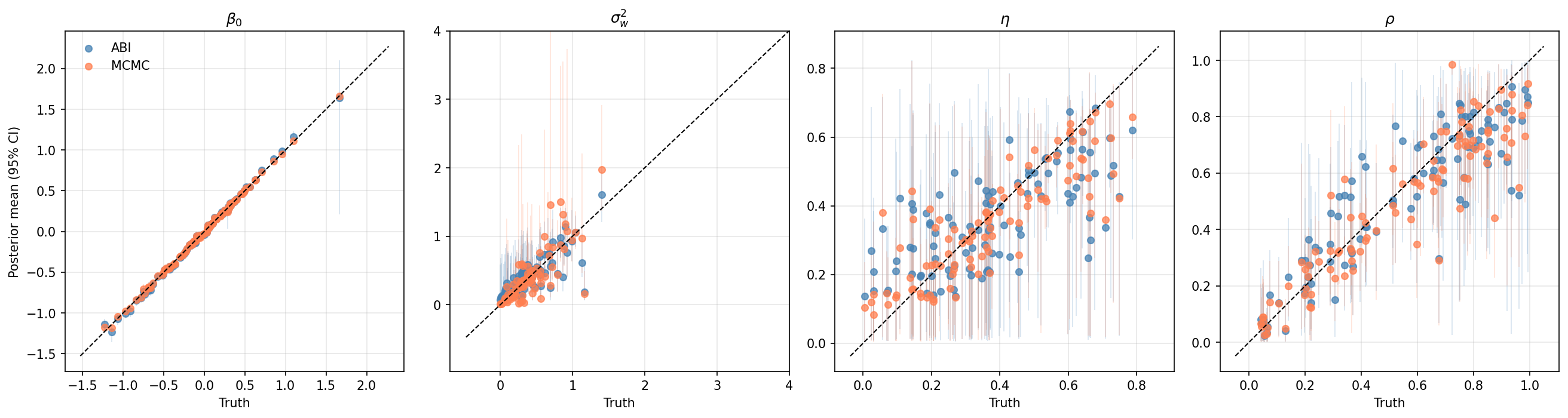}
\includegraphics[width=\textwidth]{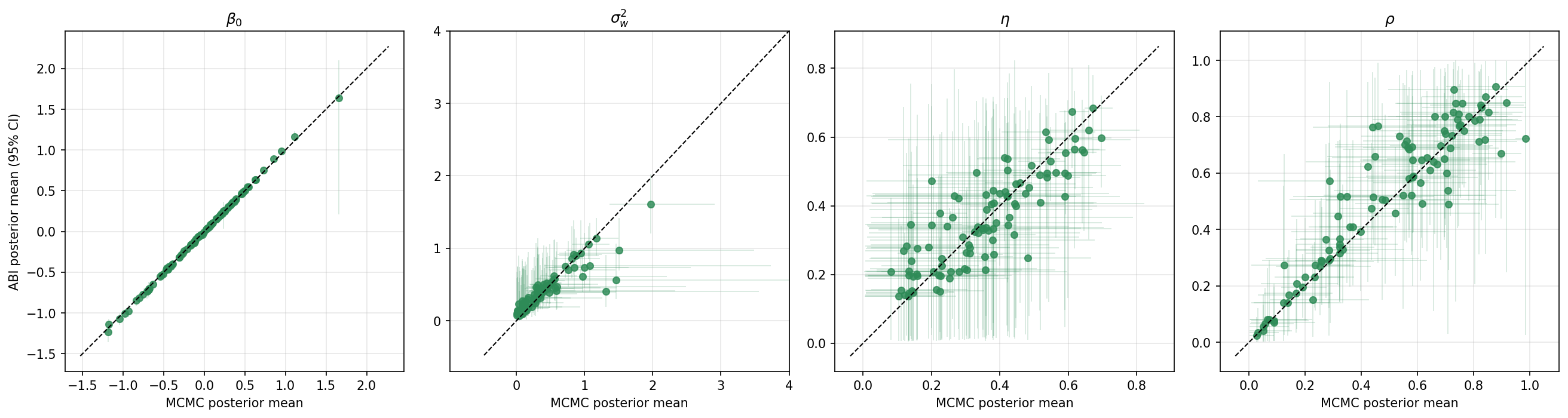}
\includegraphics[width=\textwidth]{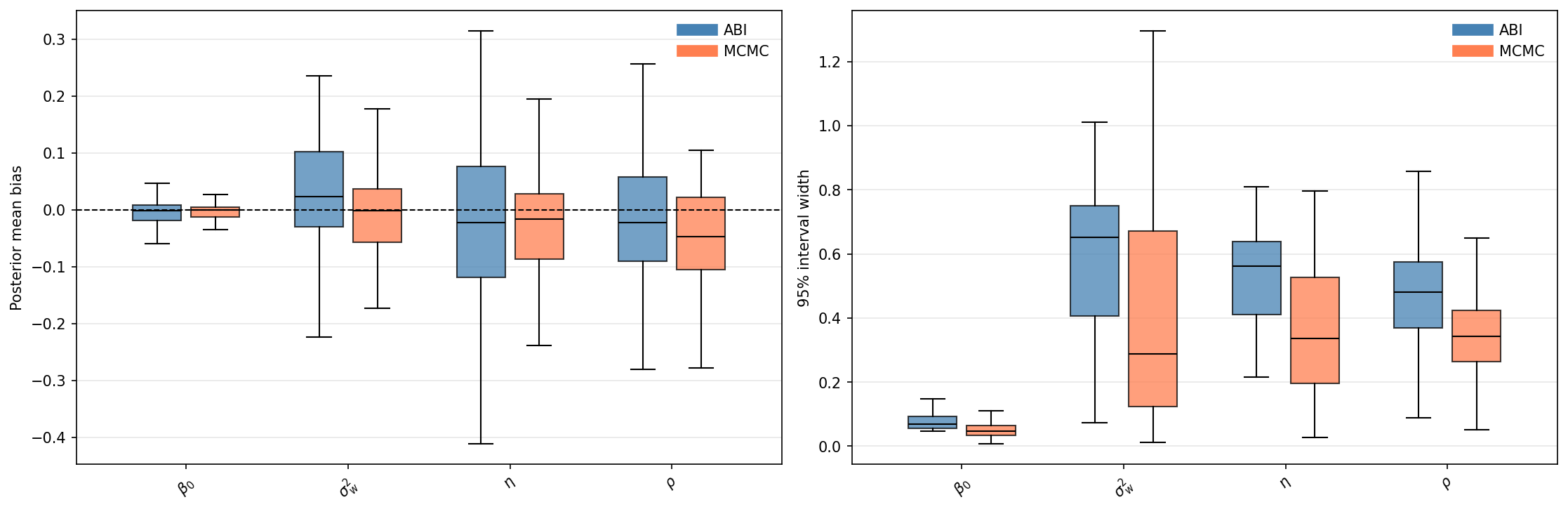}
\caption{Detailed parameter-level comparison between ABI-DAGAR and model-matched MCMC-DAGAR on 100 held-out simulated datasets. Top: truth versus posterior mean, and ABI-DAGAR versus MCMC-DAGAR posterior mean, for each parameter. Bottom: distributions of posterior mean bias and 95\% interval width across datasets.}
\label{fig:sim_abi_mcmc_param_main}
\end{figure}

Fig.~\ref{fig:sim_abi_mcmc_param_main} provides a more detailed view of the same benchmark. The scatterplots show that posterior means from ABI-DAGAR and MCMC-DAGAR are strongly aligned across datasets, indicating that the amortized approximation recovers the main posterior location. The boxplots show that the larger differences occur primarily in posterior spread for the latent-structure parameters, rather than in systematic shifts of posterior means. This distinction is important: the ABI approximation is not simply biased away from the MCMC-DAGAR benchmark, but tends to be more conservative for parameters whose information is carried indirectly through residual spatial dependence, graph topology, and covariate-driven edge contrasts.

The ABI-DAGAR versus MCMC-DAGAR comparison also clarifies the nature of the remaining approximation error. As shown in Fig.~\ref{fig:sim_abi_mcmc_param_main}, the larger differences occur primarily in posterior spread for $\sigma_w^2$, $\eta$, and $\rho$, rather than in posterior location. This alignment is strongest for $\beta_0$ and remains substantial for the spatial parameters, although the agreement is weaker for the latent-structure components. The larger differences occur primarily in posterior spread for $\sigma_w^2$, $\eta$, and $\rho$. This behavior is consistent with the conditional-density-estimation target of ABI: the learned posterior is conditioned on a fixed-dimensional representation $\bs{z}=h_{\hat{\bs{\psi}}}\{\mathcal{S}(\mathcal{D})\}$, rather than on the full dataset through a dataset-specific sampler. If this representation does not fully distinguish weakly identified posterior regimes, or if the finite-capacity flow smooths over ridges or weak multimodality in $(\sigma_w^2,\eta,\rho)$, the learned approximation may cover these regimes with a broader distribution. In this sense, the wider ABI-DAGAR intervals for $\eta$ and $\rho$ can be interpreted as conservative uncertainty for latent-structure parameters, while posterior location and edge-level boundary evidence remain broadly aligned with the model-matched MCMC-DAGAR benchmark.

The edge-level comparison provides the most direct assessment of posterior agreement for the boundary-detection target. Posterior boundary probabilities from ABI-DAGAR and MCMC-DAGAR showed a mean correlation of 0.920, a median correlation of 0.957, and a mean absolute difference of 0.066. Under the median-probability rule, ABI-DAGAR selected 94.48 boundaries on average and MCMC-DAGAR selected 94.47, with mean Jaccard overlap 0.593. This indicates strong agreement in the underlying posterior boundary probabilities, with more moderate agreement after converting probabilities into binary boundary sets. Boundary discrimination was extremely strong for both methods: dataset-level mean AUROC and mean AP were numerically equal to 1.000 at the reported precision.

ABI-DAGAR required 0.78 seconds per dataset on average to produce 10{,}000 posterior draws, compared with 2.84 seconds for MCMC-DAGAR, corresponding to a deployment speedup of about 3.64 after training. These timings characterize the deployment cost of the validated amortized approximation. The main inferential conclusion is that the reusable posterior approximator closely reproduces the model-matched Bayesian benchmark, especially for the edge-level posterior boundary probabilities that are central to the proposed boundary-detection workflow. Additional boundary summaries and runtime diagnostics are reported in Appendix~\ref{sec:supp_sim_abi_mcmc}.

\subsection{Ablation study of summary statistics}

Because the neural component relies on a learned representation of graph-indexed data, we conducted an ablation study to assess whether the engineered summaries provide statistically meaningful information for posterior inference, rather than serving only as a generic high-dimensional input representation. The baseline model uses the full summary design described in Section~\ref{sec:summary_network}. We compared this baseline with six alternatives: removing the core observation features, graph-topology features, dissimilarity features, local spatial features, or global graph features one block at a time, and a core-observation-only representation. For each representation, we retrained a separate amortized posterior approximator from scratch using the same network architecture and training protocol. Thus, the ablation compares posterior approximations learned under different summary designs, rather than post hoc masking of inputs in a single trained network. All runs were evaluated on the same validation datasets; only the summary inputs were changed. Validation was performed on 4050 held-out datasets, consisting of 50 datasets for each graph size $N=40,\ldots,120$. Additional implementation details and full ablation summaries are reported in Appendix~\ref{sec:supp_ablation}.

\begin{figure}[t]
\centering
\includegraphics[width=\textwidth]{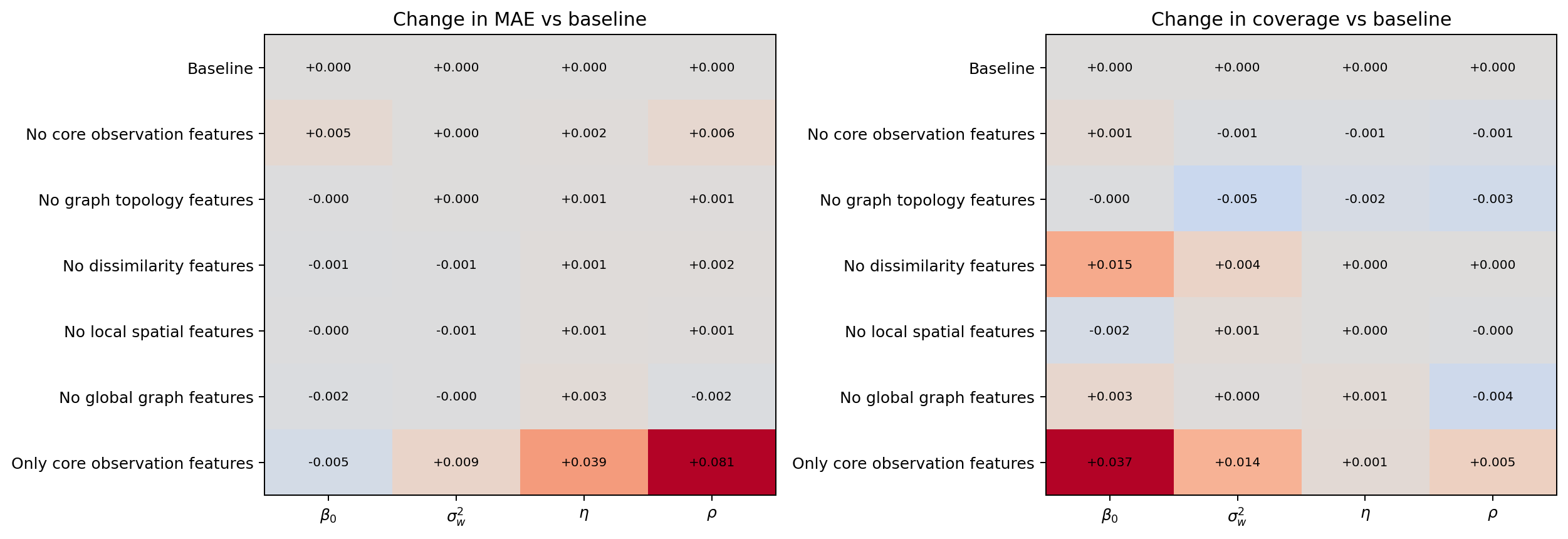}
\caption{Differences in scalar recovery metrics relative to the full-summary baseline.}
\label{fig:ablation_delta_main}
\end{figure}

The most informative comparison is relative to the full-summary baseline, since the methodological question is which components of the model-guided representation are needed to preserve posterior quality. Fig.~\ref{fig:ablation_delta_main} shows the changes in scalar recovery metrics, mean absolute error and coverage, when each summary block is removed. The full representation gave the strongest recovery of the boundary parameter $\eta$, which is the central parameter for the boundary-detection component of the model. Removing any single non-core block produced only modest changes, suggesting some redundancy and robustness in the engineered representation. By contrast, the core-observation-only representation showed clear deterioration, especially for $\eta$ and $\rho$, indicating that the core observation features alone are not sufficient for the full spatial boundary-detection task. Removing the core observation block also produced the largest degradation among the one-block ablations and substantially worsened the training and validation losses, indicating that these features remain important for stable learning.

\begin{figure}[t]
\centering
\includegraphics[width=\textwidth]{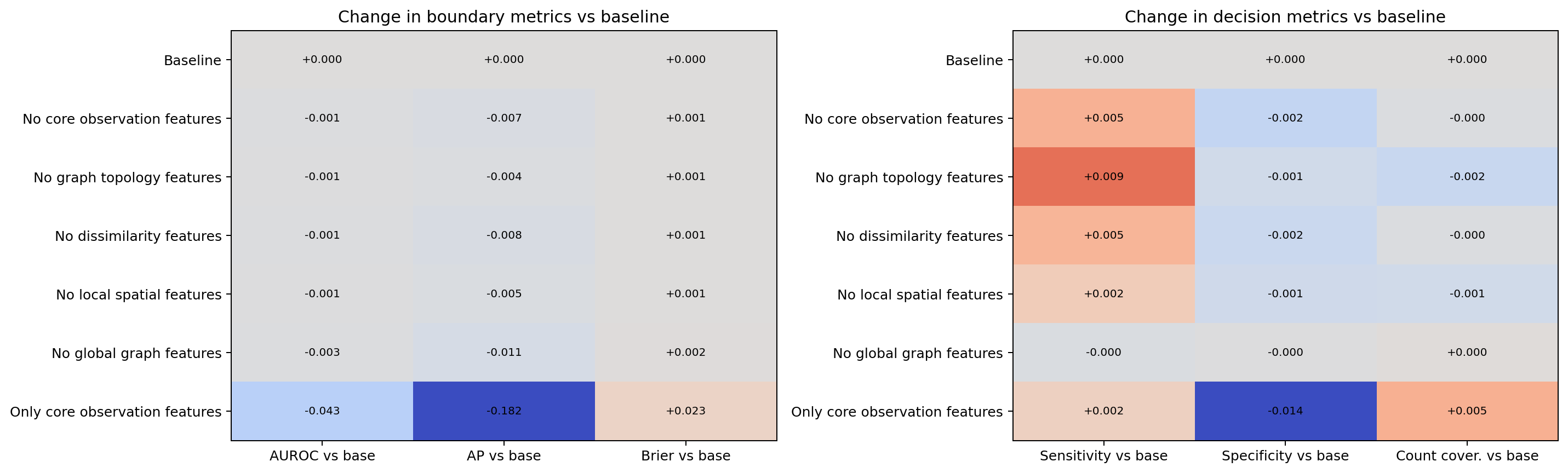}
\caption{Differences in boundary-detection metrics relative to the full-summary baseline.}
\label{fig:ablation_boundary_delta_main}
\end{figure}

The same conclusion holds at the edge level. Fig.~\ref{fig:ablation_boundary_delta_main} shows the corresponding changes in boundary-detection performance relative to the baseline. The core-observation-only representation exhibited the clearest deterioration, with a large reduction in pooled AP and AUROC and a marked increase in the Brier score, demonstrating that the richer graph-aware and dissimilarity-aware summaries are important for producing high-quality posterior boundary probabilities. By contrast, the one-block non-core ablations remained comparatively close to the baseline, indicating that these blocks contribute mainly through refinement and robustness rather than as isolated carriers of information.

Taken together, the ablation study supports the use of the full model-guided summary design. The representation contains some redundancy, but retraining under reduced summary sets shows that the full design provides the most stable balance across global mean recovery, spatial dependence learning, boundary-parameter inference, and posterior boundary-probability quality.

\subsection{Computational performance}

We report computational cost separately for the one-time amortization stage and posterior inference on new datasets. On an Intel(R) Core(TM) i7-10750H CPU, the one-time training stage required 5 hours and 45 minutes. After training, generating 10{,}000 posterior draws for each of the 200 held-out validation datasets required 153.59 seconds in total, corresponding to 0.77 seconds per dataset, with a validation memory footprint of 1.168 GB. The posterior predictive diagnostic block described in Section~\ref{sec:ppc}, based on 40 datasets and 100 posterior predictive replicates per dataset, required an additional 23.52 seconds. Additional computational details are reported in Appendix~\ref{sec:supp_sim_computation}.

These timings characterize the deployment profile of the validated amortized posterior approximation, rather than serving as the primary measure of methodological success. The main inferential question is whether the trained approximation remains reliable when applied to new areal graphs from the deployment regime.

All experiments used fixed random seeds for graph simulation, parameter generation, neural-network initialization, minibatch generation, and posterior sampling. The seeds are reported in the accompanying code repository.

Overall, the simulation experiments support the proposed ABI-DAGAR framework as a reusable posterior approximation for heterogeneous areal graphs. Across held-out maps of varying size and topology, the method recovered the model parameters, provided reasonably calibrated posterior uncertainty, produced informative posterior boundary probabilities beyond any single thresholding rule, reproduced key posterior predictive features of the data-generating model, and remained practically deployable after the one-time training stage.

\section{Real-data deployment and benchmark comparison}
\label{sec:real_data_analysis}

We applied the trained amortized posterior approximation to two empirical areal health datasets with different graph sizes and spatial structures: the Glasgow respiratory disease dataset originally analyzed by \citet{lee2012boundary}, obtained through the \texttt{R} package \texttt{CARBayes} \citep{CARBayes2013}, and the California lung cancer dataset considered by \citet{gao2023spatial}, extracted from the SEER$^*$Stat database using the SEER$^*$Stat software \citep{seer}. In both applications, the same trained amortized posterior approximator was used without dataset-specific retraining, and posterior inference was carried out jointly for $\bs{\theta}=(\beta_0,\sigma_w^2,\eta,\rho)$.
% \begin{equation*}
%     \bs{\theta}=(\beta_0,\sigma_w^2,\eta,\rho).
% \end{equation*}

The empirical analyses have two aims. First, they illustrate deployment of a reusable posterior approximation on real areal health datasets with different numbers of regions, covariate structures, and boundary patterns. Second, they provide an external benchmark by comparing the resulting boundary conclusions with those from a localized smoothing analysis implemented in \texttt{CARBayes}. This comparison is important because the goal of the amortized approach is not to redefine the scientific notion of a spatial boundary, but to assess whether a trained posterior approximation can reproduce the main uncertainty-quantified boundary conclusions of an established model-based Bayesian workflow on real areal health data. For each application, posterior summaries were based on 10{,}000 draws from the amortized posterior approximation. Posterior predictive checks for the empirical count data are reported in Appendix~\ref{sec:additional_details_ppc}. Additional fitted-risk comparison plots, edge-probability-versus-dissimilarity plots, and a model-matched real-data comparison with the MCMC-DAGAR implementation are reported in Appendices~\ref{sec:additional_details_risk}--\ref{sec:additional_details_mcmc_dagar}.

\subsection{Glasgow respiratory disease data}

We first considered the Glasgow respiratory disease dataset analyzed by \citet{lee2012boundary}, using the standardized income-deprivation covariate to define edge dissimilarities. The dataset contains $N=134$ areas and 360 adjacent edges, with an average of 5.37 neighbors per area. The empirical median standardized edge dissimilarity was 0.597, corresponding to a graph-specific scaling factor $M=1.162$.

The ABI-DAGAR posterior suggests a spatial regime in which local boundary structure coexists with strong residual dependence. The posterior median for the effective boundary parameter $\eta$ was 0.830, with 95\% credible interval $(0.114,\,1.137)$, while the posterior median for $\rho$ was 0.878, with 95\% credible interval $(0.455,\,0.975)$. Thus, the posterior places mass away from zero for both the boundary and dependence components. This indicates that the data are not well described by either a globally smooth risk surface or a completely fragmented one. Instead, the fitted model supports persistent residual spatial organization together with localized interruptions in smoothing.

Using the median-probability rule, under which an edge is selected as a boundary whenever its posterior boundary probability exceeds 0.5, ABI-DAGAR classified 134 of the 360 adjacent edges as boundaries. This corresponds to a dense boundary configuration, consistent with the posterior evidence for an active boundary mechanism. The Glasgow analysis therefore illustrates a real-data setting in which the trained amortized posterior approximation identifies both broad residual spatial dependence and numerous local discontinuities in the risk surface.

\subsection{California lung cancer data}

We next considered the California lung cancer dataset discussed by \citet{gao2023spatial}, using standardized smoking prevalence as the dissimilarity covariate. The dataset contains $N=58$ areas and 139 adjacent edges, with an average of 4.79 neighbors per area. The empirical median standardized edge dissimilarity was 0.698, yielding a graph-specific scaling factor $M=0.993$.

Relative to Glasgow, the California analysis suggests a weaker and more localized boundary signal. The ABI-DAGAR posterior median for $\eta$ was 0.471, with 95\% credible interval $(0.017,\,0.964)$, while the posterior median for $\rho$ was 0.816, with 95\% credible interval $(0.050,\,0.990)$. The posterior therefore supports residual spatial dependence and some boundary formation, but with greater uncertainty about the strength of both components than in the Glasgow application. This pattern is consistent with a spatial regime in which smoothing remains broadly persistent, while local discontinuities are less widespread.

Under the same median-probability rule, ABI-DAGAR classified 33 of the 139 adjacent edges as boundaries. The selected boundary set is substantially sparser than in the Glasgow analysis, indicating that the California data are characterized by fewer sharp local discontinuities. Taken together, the two empirical analyses show that the same trained amortized posterior approximation can be deployed without dataset-specific retraining on areal datasets with different graph sizes, covariate structures, and spatial regimes, while still providing joint uncertainty quantification for local boundary formation and residual spatial dependence.

\subsection{Comparison with \texttt{CARBayes}}
\label{sec:carbayes-comparison}

A natural empirical benchmark for the proposed approach is the localized CAR model implemented by the \texttt{S.CARdissimilarity} function in the \texttt{CARBayes} package \citep{CARBayes2013}, which builds directly on the framework of \citet{lee2012boundary}. We therefore compared ABI-DAGAR with \texttt{S.CARdissimilarity} on both empirical datasets, using the same observed graph, the same standardized dissimilarity covariate, and the same median-probability rule for boundary selection.

We treat this comparison as an external empirical validation exercise for the trained amortized posterior approximation. Since the two methods use different latent spatial priors, exact agreement in all posterior summaries is neither expected nor required. The most relevant comparison is at the level of the boundary mechanism: posterior boundary probabilities, selected boundary sets, and the spatial configuration of detected discontinuities. In the present single-dissimilarity setting, the effective boundary parameters are directly comparable at the edge-mechanism level. By contrast, the variance components play analogous roles but arise under different latent spatial priors, and should therefore be interpreted more cautiously.

At the parameter level, the strongest agreement occurs for quantities directly tied to the mean structure and boundary mechanism. As shown in Table~\ref{tab:carbayes-parameter-comparison}, posterior summaries for the intercept and the effective boundary parameter are similar in both datasets. In Glasgow, the ABI-DAGAR posterior median for $\eta$ is 0.830, compared with 0.673 under \texttt{CARBayes}; in California, the corresponding medians are 0.471 and 0.443. The intercept estimates are also closely aligned. The variance components differ more noticeably, especially in Glasgow, as expected given the different spatial prior constructions. A distinctive feature of ABI-DAGAR is that it also provides posterior inference for the residual dependence parameter $\rho$, for which there is no directly comparable free parameter in the benchmark analysis.

\begin{table}[t]
\centering
\caption{Posterior medians with 95\% credible intervals for ABI-DAGAR and \texttt{CARBayes}.}
\label{tab:carbayes-parameter-comparison}
\begin{tabular}{lccc}
\hline
\textbf{Dataset} & \textbf{Parameter} & \textbf{ABI-DAGAR} & \texttt{CARBayes} \\
\hline
Glasgow & $\beta_0$    & $-0.239$ $(-0.309,\,-0.171)$ & $-0.220$ $(-0.242,\,-0.197)$ \\
Glasgow & $\sigma_w^2$ & $0.336$ $(0.079,\,1.100)$    & $0.136$ $(0.096,\,0.192)$    \\
Glasgow & $\eta$       & $0.830$ $(0.114,\,1.137)$    & $0.673$ $(0.624,\,0.688)$    \\
Glasgow & $\rho$       & $0.878$ $(0.455,\,0.975)$    & ---                          \\
\hline
California & $\beta_0$    & $0.086$ $(0.014,\,0.152)$ & $0.095$ $(0.071,\,0.117)$ \\
California & $\sigma_w^2$ & $0.106$ $(0.011,\,0.847)$ & $0.033$ $(0.017,\,0.059)$ \\
California & $\eta$       & $0.471$ $(0.017,\,0.964)$ & $0.443$ $(0.310,\,0.716)$ \\
California & $\rho$       & $0.816$ $(0.050,\,0.990)$ & ---                       \\
\hline
\end{tabular}
\end{table}

The ABI-DAGAR credible intervals in Table~\ref{tab:carbayes-parameter-comparison} are wider than those from \texttt{CARBayes}, especially for the latent-structure parameters. This difference should be interpreted cautiously. First, \texttt{CARBayes} and ABI-DAGAR use different latent spatial priors and different dependence parameterizations, and $\rho$ has no directly comparable free parameter in the \texttt{CARBayes} benchmark. Second, the model-matched MCMC-DAGAR comparison reported in Section \ref{sec:benchmark_mcmc_dagar} and Appendix~\ref{sec:additional_details_mcmc_dagar} shows that part of the interval-width difference is already present at the model-class level: DAGAR-based MCMC can yield wider uncertainty than the localized CAR benchmark for key spatial parameters. Relative to MCMC-DAGAR, the amortized approximation remains somewhat more diffuse, especially for $\eta$, but posterior centers and edge-level boundary summaries remain well aligned. We therefore emphasize posterior boundary probabilities and boundary-set agreement as the primary empirical comparison, while treating the scalar ABI intervals as conservative uncertainty assessments.

The edge-level comparison provides the most direct benchmark for boundary detection. Table~\ref{tab:carbayes-agreement} shows that posterior boundary probabilities from ABI-DAGAR and \texttt{CARBayes} are highly correlated in both applications, with correlations of 0.862 in Glasgow and 0.863 in California. Under the median-probability rule, ABI-DAGAR selects more boundaries than \texttt{CARBayes}: 134 versus 99 in Glasgow and 33 versus 24 in California. The overlap is nevertheless substantial. In Glasgow, all 99 \texttt{CARBayes}-selected boundaries are also selected by ABI-DAGAR, while 73.9\% of ABI-DAGAR-selected boundaries are selected by \texttt{CARBayes}. In California, all 24 \texttt{CARBayes}-selected boundaries are recovered by ABI-DAGAR, while 72.7\% of ABI-DAGAR-selected boundaries are selected by \texttt{CARBayes}.

\begin{table}[t]
\centering
\caption{Agreement between ABI-DAGAR and \texttt{CARBayes} (CB) under the median-probability rule. 
% ``ABI in CB'' denotes the percentage of ABI-DAGAR selected boundaries also selected by \texttt{CARBayes}; ``CB in ABI'' denotes the percentage of \texttt{CARBayes} selected boundaries also selected by ABI-DAGAR.
}
\label{tab:carbayes-agreement}
\resizebox{\textwidth}{!}{%
\begin{tabular}{lccccccc}
\hline
\textbf{Dataset} & \textbf{Risk corr.} & \textbf{Bound. corr.} & \textbf{ABI sel.} & \textbf{CB sel.} & \textbf{Shared} & \textbf{ABI in CB} & \textbf{CB in ABI} \\
\hline
Glasgow & 0.767 & 0.862 & 134 & 99 & 99 & 73.9 \% & 100.0 \% \\
California & 0.843 & 0.863 & 33 & 24 & 24 & 72.7 \% & 100.0 \%\\
\hline
\end{tabular}%
}
\end{table}

\begin{figure}[t]
\centering
\includegraphics[width=0.495\textwidth]{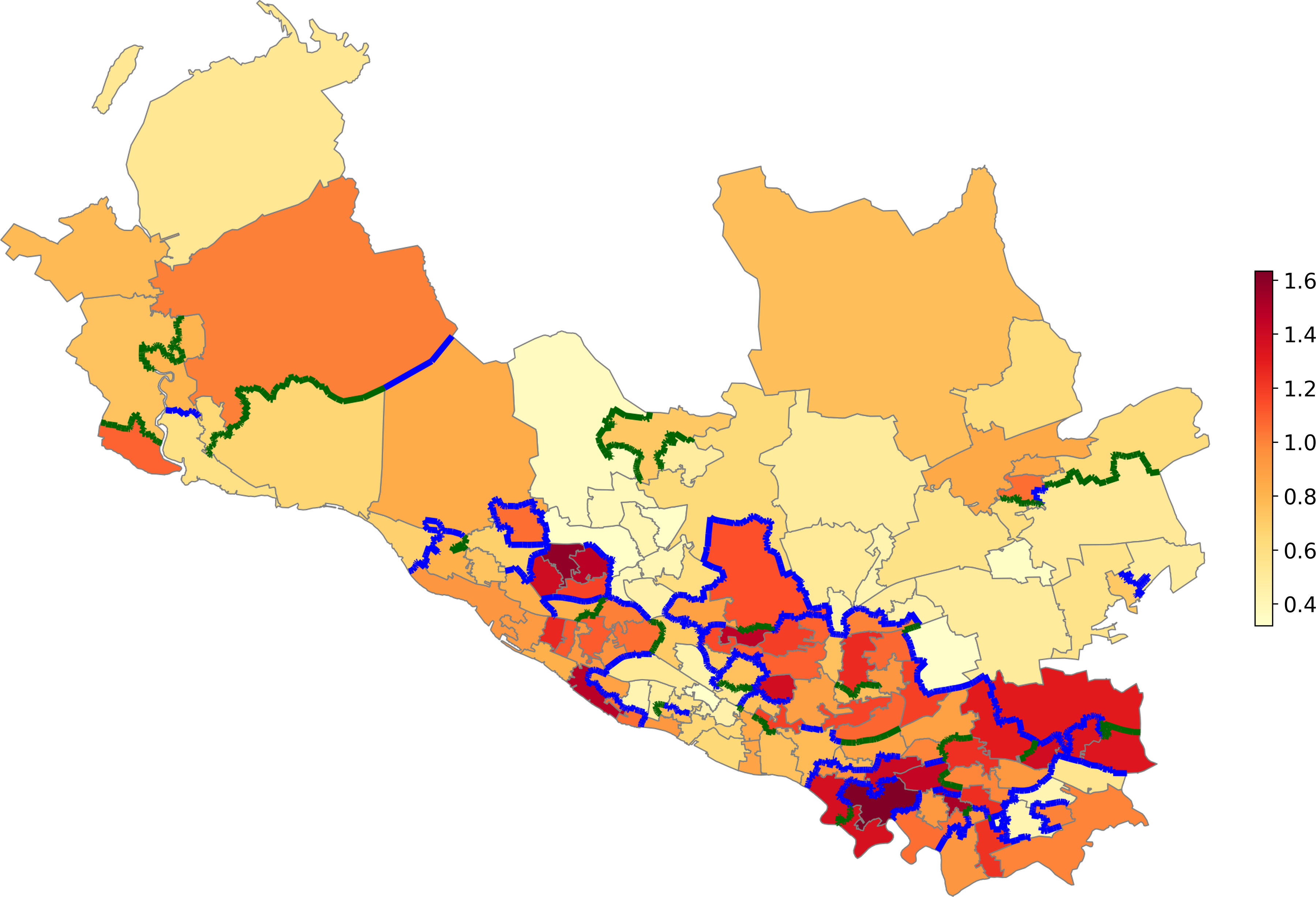}
\includegraphics[width=0.495\textwidth]{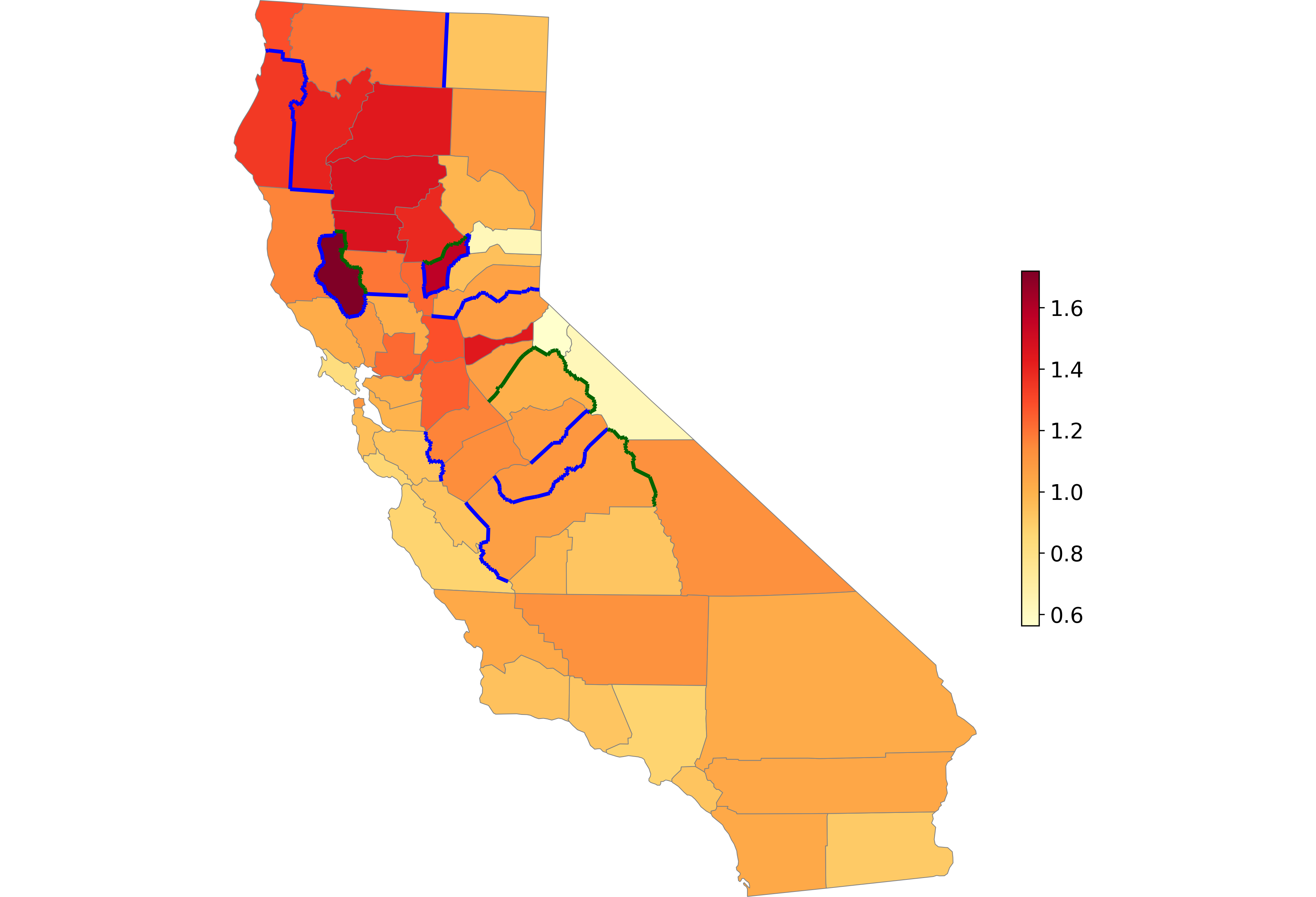}
\caption{Glasgow (left) and California (right) boundary agreement between ABI-DAGAR and \texttt{CARBayes}. Blue indicates boundaries selected by both methods, and green ABI-DAGAR-only edges.}
\label{fig:glasgow_california_real}
\end{figure}

Fig.~\ref{fig:glasgow_california_real} displays the spatial agreement between the two boundary-selection procedures. Blue edges are selected by both methods and green edges are selected by ABI-DAGAR but not by \texttt{CARBayes}. In both datasets, all benchmark-selected boundaries are recovered by ABI-DAGAR. The additional ABI-DAGAR-only edges appear mainly as local expansions around the same dominant discontinuity structure, rather than as qualitatively different boundary patterns.

The fitted risk surfaces are also positively associated, with correlations of 0.767 in Glasgow and 0.843 in California. This agreement is weaker than for the boundary probabilities, which is expected because the two approaches combine boundary formation with different latent spatial priors. For this reason, the fitted-risk comparison should be viewed as a secondary diagnostic, whereas the edge-level agreement provides the more direct evidence that the two methods lead to similar substantive boundary-detection conclusions. Additional fitted-risk comparison plots and edge-probability-versus-dissimilarity plots are reported in Appendices~\ref{sec:additional_details_risk} and~\ref{sec:additional_details_edge}.

For all \texttt{CARBayes} posterior summaries and boundary comparisons reported above, we used the longer MCMC configuration with 300{,}000 iterations, 100{,}000 burn-in iterations, and thinning by 20, yielding 10{,}000 retained draws. Under this setup, \texttt{CARBayes} required 615 seconds for Glasgow and 395 seconds for California. By comparison, generating 10{,}000 draws from the trained ABI-DAGAR posterior approximation required approximately 1 second per dataset. Including the one-time ABI-DAGAR training cost of 5 hours and 45 minutes, the corresponding break-even points are approximately 34 Glasgow-sized analyses or 53 California-sized analyses. As a computational sensitivity check only, we also recorded shorter \texttt{CARBayes} runs with 20{,}000 iterations, 10{,}000 burn-in iterations, and no thinning; these required 45 seconds for Glasgow and 30 seconds for California, corresponding to break-even points of approximately 471 and 714 analyses, respectively. Thus, the empirical results shown in Tables~\ref{tab:carbayes-parameter-comparison}--\ref{tab:carbayes-agreement} and Fig.~\ref{fig:glasgow_california_real} are based on the longer \texttt{CARBayes} runs, while the shorter runs are reported only to contextualize deployment cost. The timing details are reported in Appendix~\ref{sec:additional_details_runtime}. Neural-network implementation details, software versions, training configuration, and reproducibility information are reported in Appendix~\ref{sec:supp_network_reproducibility}.

Overall, the empirical comparison shows that ABI-DAGAR reproduces the main substantive boundary conclusions of a localized smoothing benchmark in both applications. This is a central validation result for deployment: the same trained amortized posterior approximation, applied without dataset-specific retraining, preserves the main edge-level posterior evidence for boundary formation while substantially reducing the cost of repeated posterior sampling relative to the \texttt{CARBayes} benchmark. The contribution of the proposed approach is therefore to provide a reusable posterior approximation for the same scientific question, with uncertainty quantification retained at the level of model parameters and posterior boundary probabilities. At the same time, the DAGAR formulation extends the benchmark comparison by providing joint posterior inference for residual spatial dependence through $\rho$, rather than requiring the dependence component to be fixed in advance.

\section{Discussion}\label{sec:discussion}

This paper shows that amortized Bayesian inference can be used to learn posterior inference across a family of spatial graphs, rather than for a single fixed map. The contribution is therefore not a predictive AI model, but a validated neural approximation to Bayesian inference for heterogeneous real-world spatial datasets. In many spatial health applications, the map is not a fixed object: new analyses may involve different regions, administrative resolutions, adjacency structures, outcomes, or numbers of areal units. Standard Bayesian workflows treat each such dataset as a new posterior computation problem. We instead train a single neural posterior approximator over heterogeneous simulated maps, so that posterior inference can be reused on new areal graphs within a validated deployment regime. In this sense, the neural network acts as an AI-assisted posterior-computation device: it supports model-based Bayesian inference without replacing the statistical model or its uncertainty quantification.

The boundary-detection model provides a demanding test case for this idea. The posterior is not only a distribution over global parameters, but also determines edge-level posterior boundary probabilities through a graph that depends on an unknown boundary parameter. The model combines a covariate-driven adjacency modification, motivated by localized smoothing approaches in disease mapping \citep{lee2012boundary,lee2014bayesian,rushworth2017adaptive}, with a DAGAR prior for residual spatial dependence \citep{datta2019spatial}. This separates local discontinuities from residual spatial persistence and yields uncertainty-quantified inference at both the node level and the edge level. Thus, the amortized approximation is not merely learning a smoothed risk predictor; it is learning a posterior distribution for a structured spatial model whose inferential targets include boundary probabilities.

The ability to handle varying-size input graphs is what makes the approach reusable. Each map is represented as an unordered set of node-specific, graph-aware summaries constructed from the observed counts, offsets, covariates, and adjacency matrix. The SetTransformer summary network maps this variable-size set to a fixed-dimensional representation while preserving permutation invariance \citep{zaheer2017deep,lee2019set}, and the conditional normalizing flow then produces approximate posterior draws. During training, graph size, topology, covariate surface, boundary configuration, and residual dependence are all varied. The trained network therefore learns an inferential operator over a class of areal datasets, rather than a posterior approximation tied to one graph. This is the mechanism by which the method addresses a central source of heterogeneity in real-world spatial evidence generation: different studies may involve different maps.

The validation study is essential to this claim. Across held-out graphs with 40 to 300 areas, the amortized posterior approximation recovered the model parameters with small bias, achieved empirical coverage close to nominal, and showed no severe systematic miscalibration under simulation-based diagnostics \citep{cook2006validation}. Boundary-specific results were also encouraging: posterior boundary probabilities discriminated well between true boundaries and non-boundaries, showed good probabilistic behavior, and gave calibrated uncertainty for the total number of boundaries. Posterior predictive checks indicated that the approximation reproduced the marginal count scale, residual spatial dependence, and local edge-contrast structure induced by the boundary mechanism. These diagnostics are not secondary computational checks; they are what make the learned neural approximation a statistically credible inferential tool.

The model-matched MCMC-DAGAR benchmark provides a second, complementary form of validation. Whereas the truth-based simulations assess recovery under known parameters, the MCMC comparison asks whether the amortized posterior reproduces the behavior of a dataset-specific Bayesian sampler for the same thresholded Poisson-DAGAR model. For the main scientific target of the paper, ABI-DAGAR and MCMC-DAGAR produced strongly correlated posterior boundary probabilities. This suggests that the reusable neural approximation preserves the main edge-level posterior evidence for boundary formation, rather than merely producing a computationally convenient surrogate.

A recurring pattern across the benchmark comparisons is that ABI-DAGAR reproduces posterior location and boundary-probability structure more closely than it reproduces scalar posterior spread for the latent-structure parameters. This distinction is important for interpreting the empirical analyses. The narrower intervals from \texttt{CARBayes} should not be treated as a model-matched uncertainty standard for ABI-DAGAR, because the two methods use different spatial priors and different dependence parameterizations. The more relevant comparison is with dataset-specific MCMC-DAGAR, where posterior centers and edge-level boundary summaries are broadly aligned, while the amortized approximation remains somewhat more diffuse, especially for $\eta$. This behavior is consistent with a conservative amortized approximation for weakly identified latent-structure parameters, whose information is carried indirectly through residual spatial dependence, graph topology, and covariate-driven edge contrasts. It also suggests a clear direction for future work: richer graph summaries, higher-capacity flows, or graph-specific recalibration may sharpen scalar uncertainty while preserving the reusable posterior approximation.

The empirical analysis reinforces the deployment argument. The same trained network was applied to the Glasgow respiratory disease and California lung cancer datasets, which differ in graph size, covariate structure, and boundary pattern. In both cases, ABI-DAGAR produced boundary conclusions broadly consistent with the localized \texttt{CARBayes} benchmark \citep{CARBayes2013}. Exact agreement is not expected because the latent spatial priors differ, but the high correlation in posterior boundary probabilities and the substantial overlap in selected boundary sets indicate that the amortized approximation captures the dominant boundary evidence in both applications. These examples illustrate the intended use of the method: once trained and validated, the same posterior approximator can be deployed on new maps without rebuilding the inferential machinery from scratch.

The approach should nevertheless be interpreted within its training deployment regime. In this study, that regime consists of Poisson areal count data with offsets, sparse locally connected adjacency graphs, graph sizes comparable to those used during training, a single covariate-driven boundary mechanism, and DAGAR residual spatial dependence. Applications with substantially different graph topology, outcome distribution, count scale, covariate behavior, overdispersion, or multiple boundary-driving covariates would require renewed validation, recalibration, or retraining. More broadly, the deployment regime could be expanded by training the amortized posterior approximator on a richer simulation design covering a wider range of graph structures, outcome distributions, or covariate mechanisms, although doing so would increase the up-front training cost because the network would need to learn across a larger and more heterogeneous set of scenarios. The present paper provides empirical validation rather than formal theoretical assurances for the amortized estimator; developing asymptotic theory for neural posterior approximations on varying-size graphs is an important direction for future work. This is not a weakness specific to ABI-DAGAR, but a basic requirement of simulation-based amortized inference: the learned posterior approximation is only as relevant as the simulator and training distribution used to define the inferential task \citep{radev2020bayesflow,sainsbury2024likelihood,zammit2025neural}.

Several extensions follow naturally. The current model uses a single boundary-driving covariate, but many applications may involve multiple candidate drivers of discontinuity, such as deprivation, environmental exposure, urbanicity, demographic composition, or access to care. A more ambitious direction would encode the set of covariates through an additional permutation-invariant network and return posterior uncertainty for covariate-specific boundary relevance. Future work should also examine robustness to covariate measurement error and misspecification, and compare the present model-guided summaries with graph neural encoders or hybrid summary architectures in richer spatial settings \citep{sainsbury2025neural,wikle2023statistical}. Amortized learning for complex multivariate spatial dependencies across regions using areal models \citep[i.e., joint modeling of multiple dependent health outcomes; see Chapter~11 in][and references therein]{banerjee2025book} and subsequent boundary detection for health disparities \citep{aiello2023detecting} has yet to be explored and will comprise future research. 

It is also worth noting that the transfer learning framework we investigated here is based on unsupervised learning. Supervised learning frameworks can be built by training the network using desired outputs. Here, we foresee future explorations by adapting the DeepRV framework of \cite{navottEtAl2026} to train the  nonparametric models in \cite{aiello2023detecting}. Alternatively, for simpler models such as in our current work, we can supervise the training using output from posterior analysis without computationally expensive MCMC or other iterative algorithms. Bayesian predictive stacking \citep{zhang2025jasa, panEtAl2025ba}, for example, uses conjugate posterior distributions to generate rapid inference and has recently been used effectively for amortized inference \citep{presicce_bayesian_2024}. Such methods can be adapted for spatial boundary detection.

Overall, ABI-DAGAR provides evidence that neural amortization can make Bayesian spatial inference reusable across varying-size maps while retaining an explicit generative model and posterior uncertainty. The key message is not simply that inference becomes faster after training. Rather, the paper demonstrates that a single validated posterior approximator can learn how to perform Bayesian boundary detection on new areal graphs, producing uncertainty-quantified evidence about both residual spatial dependence and local discontinuities. This suggests a broader role for AI-assisted statistical inference in real-world evidence generation: neural networks can support, rather than replace, model-based reasoning by making carefully validated Bayesian workflows portable across repeated analysis with heterogeneous spatial structures.

% \section*{Acknowledgments}

% The authors acknowledge federal grants from the National Institute of Health NIH-NIGMS R01GM148761-04 and the National Science Foundation NSF 2515898.


% The authors declare no competing interests.

\section*{Data availability}

The empirical datasets used in this article are publicly available from the sources described in Section~\ref{sec:real_data_analysis}: the Glasgow respiratory disease data are available through the \texttt{R} package \texttt{CARBayes} \citep{CARBayes2013}, and the California lung cancer data were extracted from the SEER$^*$Stat database using the SEER$^*$Stat software \citep{seer}. Scripts for reproducing the empirical analyses are available in the code repository listed in the Software statement.

\section*{Software}

Code to reproduce all results reported in the manuscript and Appendices is available at \url{https://github.com/lucaaiello/ABI_boundary_detection/tree/main}. The repository includes Python notebooks and scripts for simulated-data generation, amortized posterior training, posterior sampling, posterior predictive checks, ablation analyses, runtime summaries, tables, and figures, as well as \texttt{R} and \texttt{Rcpp} code for the MCMC-DAGAR and \texttt{CARBayes} benchmark analyses.

\bibliographystyle{apalike}
\bibliography{reference}

@article{datta2019spatial,
  title={{Spatial disease mapping using directed acyclic graph auto-regressive (DAGAR) models}},
  author={Datta, Abhirup and Banerjee, Sudipto and Hodges, James S and Gao, Leiwen},
  journal={Bayesian Analysis},
  volume={14},
  number={4},
  pages={1221},
  year={2019},
  publisher={NIH Public Access}
}

@article{gao2023spatial,
  title={Spatial difference boundary detection for multiple outcomes using {B}ayesian disease mapping},
  author={Gao, Leiwen and Banerjee, Sudipto and Ritz, Beate},
  journal={Biostatistics},
  volume={24},
  number={4},
  pages={922--944},
  year={2023},
  publisher={Oxford University Press}
}

@article{lee2012boundary,
  title={Boundary detection in disease mapping studies},
  author={Lee, Duncan and Mitchell, Richard},
  journal={Biostatistics},
  volume={13},
  number={3},
  pages={415--426},
  year={2012},
  publisher={Oxford University Press}
}

@article{li2015bayesian,
  title={{B}ayesian models for detecting difference boundaries in areal data},
  author={Li, Pei and Banerjee, Sudipto and Hanson, Timothy A and McBean, Alexander M},
  journal={Statistica Sinica},
  volume={25},
  number={1},
  pages={385},
  year={2015},
  publisher={NIH Public Access}
}

@article{lu2007bayesian,
  title={{B}ayesian areal wombling via adjacency modeling},
  author={Lu, Haolan and Reilly, Cavan S and Banerjee, Sudipto and Carlin, Bradley P},
  journal={Environmental and Ecological Statistics},
  volume={14},
  pages={433--452},
  year={2007},
  publisher={Springer}
}

@article{ma2007bayesian,
  title={{B}ayesian Multivariate Areal Wombling for Multiple Disease Boundary Analysis},
  author={Ma, Haijun and Carlin, Bradley P},
  journal={Bayesian Analysis},
  volume={2},
  number={2},
  pages={281--302},
  year={2007}
}

@article{ma2010hierarchical,
  title={Hierarchical and joint site-edge methods for Medicare hospice service region boundary analysis},
  author={Ma, Haijun and Carlin, Bradley P and Banerjee, Sudipto},
  journal={Biometrics},
  volume={66},
  number={2},
  pages={355--364},
  year={2010},
  publisher={Wiley Online Library}
}

@book{koch2005cartographies,
title={Cartographies of Disease: Maps, Mapping, and Medicine},
author={Koch, Tom},
year={2005},
publisher={Esri Press Redlands, CA}
}

@book{lawson2016handbook,
	title={Handbook of Spatial Epidemiology},
	author={Lawson, Andrew, B. and Banerjee, Sudipto and Haining, Robert and Ugarte, Maria, D.},
	year={2016},
	publisher={CRC press, Boca Raton, FL}
}

@article{lu2005bayesian,
	title={{B}ayesian Areal Wombling for Geographical Boundary Analysis},
	author={Lu, Haolan and Carlin, Bradley P},
	journal={Geographical Analysis},
	volume={37},
	number={3},
	pages={265--285},
	year={2005},
	publisher={Wiley Online Library}
}

@misc{seer, 
	title={{SEER*Stat software}}, 
	author={{National Cancer Institute}}, 
	url={https://seer.cancer.gov/seerstat/}, 
	journal={SEER}, 
	year={2019}, 
	month={Aug}
}

@article{zhang2025jasa, 
    title={Bayesian Geostatistics Using Predictive Stacking}, 
    author={Zhang, Lu and Tang, Wenpin and Banerjee, Sudipto},
    journal={Journal of the American Statistical Association}, 
    publisher={Informa UK Limited}, 
    year={2025},
    doi={10.1080/01621459.2025.2566449},
    url={https://doi.org/10.1080/01621459.2025.2566449},
    volume={(In press)},
    page={}
}

@article{besag1991bayesian,
  title={Bayesian image restoration, with two applications in spatial statistics},
  author={Besag, Julian and York, Jeremy and Molli{\'e}, Annie},
  journal={Annals of the Institute of Statistical Mathematics},
  volume={43},
  number={1},
  pages={1--20},
  year={1991},
  publisher={Springer}
}

@article{radev2020bayesflow,
  title={Bayes{F}low: Learning complex stochastic models with invertible neural networks},
  author={Radev, Stefan T and Mertens, Ulf K and Voss, Andreas and Ardizzone, Lynton and K{\"o}the, Ullrich},
  journal={IEEE transactions on neural networks and learning systems},
  volume={33},
  number={4},
  pages={1452--1466},
  year={2020},
  publisher={IEEE}
}

@article{zammit2025neural,
  title={Neural methods for amortized inference},
  author={Zammit-Mangion, Andrew and Sainsbury-Dale, Matthew and Huser, Rapha{\"e}l},
  journal={Annual Review of Statistics and Its Application},
  volume={12},
  number={1},
  pages={311--335},
  year={2025},
  publisher={Annual Reviews}
}

@article{sainsbury2024likelihood,
  title={Likelihood-free parameter estimation with neural {B}ayes estimators},
  author={Sainsbury-Dale, Matthew and Zammit-Mangion, Andrew and Huser, Rapha{\"e}l},
  journal={The American Statistician},
  volume={78},
  number={1},
  pages={1--14},
  year={2024},
  publisher={Taylor \& Francis}
}

@article{aiello2023detecting,
  title={Detecting spatial health disparities using disease maps},
  author={Aiello, Luca and Banerjee, Sudipto},
  journal={arXiv preprint arXiv:2309.02086},
  year={2023}
}

@inproceedings{rezende2015variational,
  title     = {Variational Inference with Normalizing Flows},
  author    = {Rezende, Danilo Jimenez and Mohamed, Shakir},
  booktitle = {Proceedings of the 32nd International Conference on Machine Learning},
  series    = {Proceedings of Machine Learning Research},
  volume    = {37},
  pages     = {1530--1538},
  year      = {2015}
}

@inproceedings{papamakarios2017masked,
  title     = {Masked Autoregressive Flow for Density Estimation},
  author    = {Papamakarios, George and Pavlakou, Theo and Murray, Iain},
  booktitle = {Advances in Neural Information Processing Systems},
  volume    = {30},
  year      = {2017}
}

@inproceedings{durkan2019neural,
  title     = {Neural Spline Flows},
  author    = {Durkan, Conor and Bekasov, Artur and Murray, Iain and Papamakarios, George},
  booktitle = {Advances in Neural Information Processing Systems},
  volume    = {32},
  year      = {2019}
}

@article{papamakarios2021normalizing,
  title   = {Normalizing Flows for Probabilistic Modeling and Inference},
  author  = {Papamakarios, George and Nalisnick, Eric and Rezende, Danilo Jimenez and Mohamed, Shakir and Lakshminarayanan, Balaji},
  journal = {Journal of Machine Learning Research},
  volume  = {22},
  number  = {57},
  pages   = {1--64},
  year    = {2021}
}

@inproceedings{lee2019set,
  title     = {Set {T}ransformer: A Framework for Attention-based Permutation-Invariant Neural Networks},
  author    = {Lee, Juho and Lee, Yoonho and Kim, Jungtaek and Kosiorek, Adam and Choi, Seungjin and Teh, Yee Whye},
  booktitle = {Proceedings of the 36th International Conference on Machine Learning},
  series    = {Proceedings of Machine Learning Research},
  volume    = {97},
  pages     = {3744--3753},
  year      = {2019}
}

@book{lawson2018bayesian,
  title={Bayesian disease mapping: hierarchical modeling in spatial epidemiology},
  author={Lawson, Andrew B},
  year={2018},
  publisher={Chapman and Hall/CRC}
}

@article{wakefield2007disease,
  title={Disease mapping and spatial regression with count data},
  author={Wakefield, Jon},
  journal={Biostatistics},
  volume={8},
  number={2},
  pages={158--183},
  year={2007},
  publisher={Oxford University Press}
}

@incollection{leroux2000estimation,
  title={Estimation of disease rates in small areas: a new mixed model for spatial dependence},
  author={Leroux, Brian G and Lei, Xingye and Breslow, Norman},
  booktitle={Statistical models in epidemiology, the environment, and clinical trials},
  pages={179--191},
  year={2000},
  publisher={Springer}
}

@article{lee2014bayesian,
  title   = {A {B}ayesian Localized Conditional Autoregressive Model for Estimating the Health Effects of Air Pollution},
  author  = {Lee, Duncan and Mitchell, Richard},
  journal = {Biometrics},
  volume  = {70},
  number  = {2},
  pages   = {419--429},
  year    = {2014},
  doi     = {10.1111/biom.12156}
}

@article{rushworth2017adaptive,
  title={An adaptive spatiotemporal smoothing model for estimating trends and step changes in disease risk},
  author={Rushworth, Alastair and Lee, Duncan and Sarran, Christophe},
  journal={Journal of the Royal Statistical Society Series C: Applied Statistics},
  volume={66},
  number={1},
  pages={141--157},
  year={2017},
  publisher={Oxford University Press}
}

@article{lee2021improved,
  title   = {Improved Inference for Areal Unit Count Data Using Graph-Based Optimisation},
  author  = {Lee, Duncan and Meeks, Kitty and Pettersson, William},
  journal = {Statistics and Computing},
  volume  = {31},
  pages   = {51},
  year    = {2021},
  doi     = {10.1007/s11222-021-10025-7}
}

@article{wu2025assessing,
  title={Assessing spatial disparities: a {B}ayesian linear regression approach},
  author={Wu, Kyle and Banerjee, Sudipto},
  journal={Biostatistics},
  volume={26},
  number={1},
  pages={kxaf048},
  year={2025},
  publisher={Oxford University Press}
}

@Article{CARBayes2013,
    author = {Duncan Lee},
    title = {{CARBayes}: An {R} Package for {B}ayesian Spatial Modeling
      with Conditional Autoregressive Priors},
    year = {2013},
    journal = {{Journal of Statistical Software}},
    doi = {10.18637/jss.v055.i13},
    url =
      {https://www.jstatsoft.org/htaccess.php?volume=55&type=i&issue=13},
    pages = {1--24},
    volume = {55},
    number = {13},
  }

@article{besag1974spatial,
  title={Spatial interaction and the statistical analysis of lattice systems},
  author={Besag, Julian},
  journal={Journal of the Royal Statistical Society: Series B (Methodological)},
  volume={36},
  number={2},
  pages={192--225},
  year={1974},
  publisher={Wiley Online Library}
}

@inproceedings{zaheer2017deep,
  title     = {Deep {S}ets},
  author    = {Zaheer, Manzil and Kottur, Satwik and Ravanbakhsh, Siamak and Poczos, Barnabas and Salakhutdinov, Ruslan and Smola, Alexander},
  booktitle = {Advances in Neural Information Processing Systems},
  volume    = {30},
  year      = {2017}
}

@article{cook2006validation,
  title={Validation of software for {B}ayesian models using posterior quantiles},
  author={Cook, Samantha R and Gelman, Andrew and Rubin, Donald B},
  journal={Journal of Computational and Graphical Statistics},
  volume={15},
  number={3},
  pages={675--692},
  year={2006},
  publisher={Taylor \& Francis}
}

@article{sainsbury2025neural,
  title={Neural {B}ayes estimators for irregular spatial data using graph neural networks},
  author={Sainsbury-Dale, Matthew and Zammit-Mangion, Andrew and Richards, Jordan and Huser, Rapha{\"e}l},
  journal={Journal of Computational and Graphical Statistics},
  volume={34},
  number={3},
  pages={1153--1168},
  year={2025},
  publisher={Taylor \& Francis}
}

@article{wikle2023statistical,
  title={Statistical deep learning for spatial and spatiotemporal data},
  author={Wikle, Christopher K and Zammit-Mangion, Andrew},
  journal={Annual Review of Statistics and Its Application},
  volume={10},
  number={1},
  pages={247--270},
  year={2023},
  publisher={Annual Reviews}
}

@misc{navottEtAl2026,
      title={DeepRV: Accelerating Spatiotemporal Inference with Pre-trained Neural Priors}, 
      author={Jhonathan Navott and Daniel Jenson and Seth Flaxman and Elizaveta Semenova},
      year={2026},
      eprint={2503.21473},
      archivePrefix={arXiv},
      primaryClass={stat.ML},
      url={https://arxiv.org/abs/2503.21473}, 
}

@article{presicce_bayesian_2024,
	title = {Bayesian Transfer Learning for Artificially Intelligent Geospatial Systems: {A} Predictive Stacking Approach},
	doi = {10.48550/arXiv.2410.09504},
	journal = {arXiv preprint},
	author = {Presicce, Luca and Banerjee, Sudipto},
	year = {2024},
	note = {arXiv:2410.09504 [stat.ME]},
}

@article{panEtAl2025ba,
author = {Soumyakanti Pan and Lu Zhang and Jonathan R. Bradley and Sudipto Banerjee},
title = {{Bayesian Inference for Spatial-Temporal Non-Gaussian Data Using Predictive Stacking}},
journal = {Bayesian Analysis},
publisher = {International Society for Bayesian Analysis},
volume = {(In press)},
pages = {},
year = {2025},
doi = {10.1214/25-BA1582},
URL = {https://doi.org/10.1214/25-BA1582}
}

@book{banerjee2025book,
    author = {Sudipto Banerjee and Alan E. Gelfand and Brad P. Carlin},
    title = {Hierarchical Modeling and Analysis for Spatial Data },
    publisher = {Chapman and Hall/CRC},
    edition = {3rd},
    year = {2025},
    doi = {10.1201/9781003401728},
    url = {https://doi.org/10.1201/9781003401728}
}

\newpage

\appendix

% Number appendix figures, tables, and equations by appendix section:
% Figure A1, Table B2, Equation (A3), etc.
\numberwithin{equation}{section}
\numberwithin{figure}{section}
\numberwithin{table}{section}

\renewcommand{\theequation}{\thesection\arabic{equation}}
\renewcommand{\thefigure}{\thesection\arabic{figure}}
\renewcommand{\thetable}{\thesection\arabic{table}}

\section{Observed-data representation}
\label{supp:observed_data_representation}

This section gives the full observed-data representation supplied to the amortized posterior approximator. The representation is constructed from the observed counts $y_i$, offsets $e_i$, covariate values $x_i$, and the observed adjacency matrix $\bs{A}=(a_{ij})$. The latent filtered graph $\bs{A}^{\ast}$ and latent spatial effects $\bs{w}$ are not used as inputs, since they are unavailable in empirical applications. Section~\ref*{sec:summary_network} of the main text summarizes the inferential role of the main groups of summaries; here we give their full construction.

Let $x_i$ denote the covariate used to define edge dissimilarity. Before constructing edge-specific summaries, the covariate is standardized as $\tilde{x}_i=(x_i-\bar x)/(s_x)$, where $\bar x$ and $s_x$ are the sample mean and sample standard deviation of the raw covariate. For neighboring areas, define $z_{ij}=|\tilde x_i-\tilde x_j|$, and $\widetilde z_{ij}=z_{ij}/Z_{0.5}$, where $Z_{0.5}$ is the median of $z_{ij}$ over observed neighboring pairs. The graph-specific scaling factor used in the boundary model is $M=\log 2/Z_{0.5}$.

We use the offset-adjusted residual proxy
\begin{equation*}
    r_i=\log(y_i+0.5)-\log e_i.
\end{equation*}
The baseline node-level summaries are
\begin{equation*}
    \tilde{x}_i, y_i, e_i, \log(1+y_i), \log e_i, r_i.    
\end{equation*}

Let
\begin{equation*}
    \mathcal{N}(i)=\{j:a_{ij}=1\},
    \qquad
    d_i=\sum_{j=1}^N a_{ij}.
\end{equation*}
For $d_i>0$, define
\begin{equation*}
    \bar r_i=\frac{1}{d_i}\sum_{j\in\mathcal{N}(i)} r_j,
    \qquad
    \delta_i=\frac{1}{d_i}\sum_{j\in\mathcal{N}(i)} |r_i-r_j|.
\end{equation*}
We also compute local dissimilarity summaries
\begin{equation*}
    \bar z_i=\frac{1}{d_i}\sum_{j\in\mathcal{N}(i)} \widetilde z_{ij},
    \qquad
    z_i^{\max}=\max_{j\in\mathcal{N}(i)} \widetilde z_{ij}.
\end{equation*}
When $d_i=0$, neighborhood summaries are set to zero.

To capture information about boundary formation, we partition the neighborhood of each node according to standardized covariate dissimilarity:
\begin{align*}
    \mathcal{N}_L(i)&=\{j\in\mathcal{N}(i):\widetilde z_{ij}\leq 0.75\},\\
    \mathcal{N}_M(i)&=\{j\in\mathcal{N}(i):0.75<\widetilde z_{ij}\leq 1.25\},\\
    \mathcal{N}_H(i)&=\{j\in\mathcal{N}(i):\widetilde z_{ij}>1.25\}.
\end{align*}
Let $d_i^{(L)}$, $d_i^{(M)}$, and $d_i^{(H)}$ denote the corresponding cardinalities. For each bin $B\in\{L,M,H\}$, we compute
\begin{equation*}
    \bar r_i^{(B)}
    =
    \frac{1}{d_i^{(B)}}\sum_{j\in\mathcal{N}_B(i)} r_j,
    \qquad
    \delta_i^{(B)}
    =
    \frac{1}{d_i^{(B)}}\sum_{j\in\mathcal{N}_B(i)} |r_i-r_j|,
    \qquad
    p_i^{(B)}=\frac{d_i^{(B)}}{d_i}.
\end{equation*}
The summaries are set to zero whenever the relevant denominator is zero.

Let $\mathcal{E}={(i,j):i<j,\ a_{ij}=1}$ denote the observed edge set. For each edge, define $\Delta_{ij}=|r_i-r_j|$, $C_{ij}=(r_i-\bar r_{\cdot})(r_j-\bar r_{\cdot})$, and $\bar r_{\cdot}=\frac{1}{N}\sum_{i=1}^N r_i$. Partition $\mathcal{E}$ into $\mathcal{E}_L,\mathcal{E}_M,\mathcal{E}_H$ according to the same thresholds on $\widetilde z_{ij}$. Let $\Delta_B$ and $C_B$ denote the average of $\Delta_{ij}$ and $C_{ij}$, respectively, over edges in bin $B$. We then compute
\begin{equation*}
G_{\Delta}=\Delta_H-\Delta_L,
\qquad
G_C=C_L-C_H.
\end{equation*}
We also compute the slope
\begin{equation*}
B_{\Delta}=
\frac{\sum_{(i,j)\in\mathcal{E}}(\widetilde z_{ij}-\bar z_{\mathcal{E}})(\Delta_{ij}-\bar\Delta_{\mathcal{E}})}
{\sum_{(i,j)\in\mathcal{E}}(\widetilde z_{ij}-\bar z_{\mathcal{E}})^2},
\end{equation*}
where $\bar z_{\mathcal{E}}=|\mathcal{E}|^{-1}\sum_{(i,j)\in\mathcal{E}}\widetilde z_{ij}$ and $\bar\Delta_{\mathcal{E}}=|\mathcal{E}|^{-1}\sum_{(i,j)\in\mathcal{E}}\Delta_{ij}$. If the denominator in $B_{\Delta}$ is zero, we set $B_{\Delta}=0$.

Let
\begin{equation*}
    s_r^2=\frac{1}{N}\sum_{i=1}^N (r_i-\bar r_{\cdot})^2.    
\end{equation*}
When $s_r^2=0$, the standardized autocorrelation summaries are set to zero. At the node level, we include
\begin{equation*}
I_i=\frac{(r_i-\bar r_{\cdot})\bar r_i}{s_r^2},
\qquad
V_i=\frac{(r_i-\bar r_i)^2}{s_r^2}.
\end{equation*}
Here $I_i$ is a local Moran-type measure of agreement between node $i$ and its neighborhood, while $V_i$ measures local semivariance.

At the graph level, we compute
\begin{equation*}
C_{\mathrm{edge}}
=
\frac{1}{|\mathcal{E}|s_r^2}
\sum_{(i,j)\in\mathcal{E}} C_{ij},
\qquad
C_{\mathrm{lag}}
=
\operatorname{corr}{(r_i,\bar r_i):i=1,\ldots,N},
\end{equation*}
\begin{equation*}
B_{\mathrm{lag}}
=
\frac{N^{-1}\sum_{i=1}^N (r_i-\bar r_{\cdot})(\bar r_i-\bar r_{\mathcal N})}{s_r^2},
\qquad
\bar r_{\mathcal N}=\frac{1}{N}\sum_{i=1}^N \bar r_i,
\end{equation*}
and
\begin{equation*}
\Gamma_{\mathrm{edge}}
=
\frac{1}{2|\mathcal{E}|}
\sum_{(i,j)\in\mathcal{E}}(r_i-r_j)^2.
\end{equation*}

The graph-level summaries are replicated across nodes and appended to each node-specific feature vector. The resulting input vector is
\begin{align*}
\bs{s}_i=\Big(
&\tilde{x}_i,\;
y_i,\;
e_i,\;
\log(1+y_i),\;
\log e_i,\;
r_i,\;
d_i,\;
\bar r_i,\;
\delta_i,\;
\bar r_i^{(L)},\;
\bar r_i^{(M)},\;
\bar r_i^{(H)},\;
\delta_i^{(L)},\;
\delta_i^{(M)},\;
\delta_i^{(H)},\\
&p_i^{(L)},\;
p_i^{(M)},\;
p_i^{(H)},\;
\bar z_i,\;
z_i^{\max},\;
I_i,\;
V_i,\;
M,\;
B_{\Delta},\;
G_{\Delta},\;
G_C,\;
C_{\mathrm{edge}},\;
C_{\mathrm{lag}},\;
B_{\mathrm{lag}},\;
\Gamma_{\mathrm{edge}}
\Big).
\end{align*}

\section{Additional simulation diagnostics}
\label{sec:supp_simulation}

This section provides supplementary diagnostic plots and distributional summaries for recovery, calibration, posterior boundary probabilities, posterior predictive adequacy, the ABI-DAGAR versus MCMC-DAGAR benchmark, the ablation study, and computation. Unless otherwise stated, results are based on the same 200 held-out simulated datasets used in the main text, for which the number of areas ranged from 40 to 299. The ablation study uses a separate validation design described in Section~\ref{sec:supp_ablation}.

\subsection{Additional recovery and calibration diagnostics}
\label{sec:supp_sim_recovery_calibration}

\begin{figure}[t]
\centering
\includegraphics[width=0.75\textwidth]{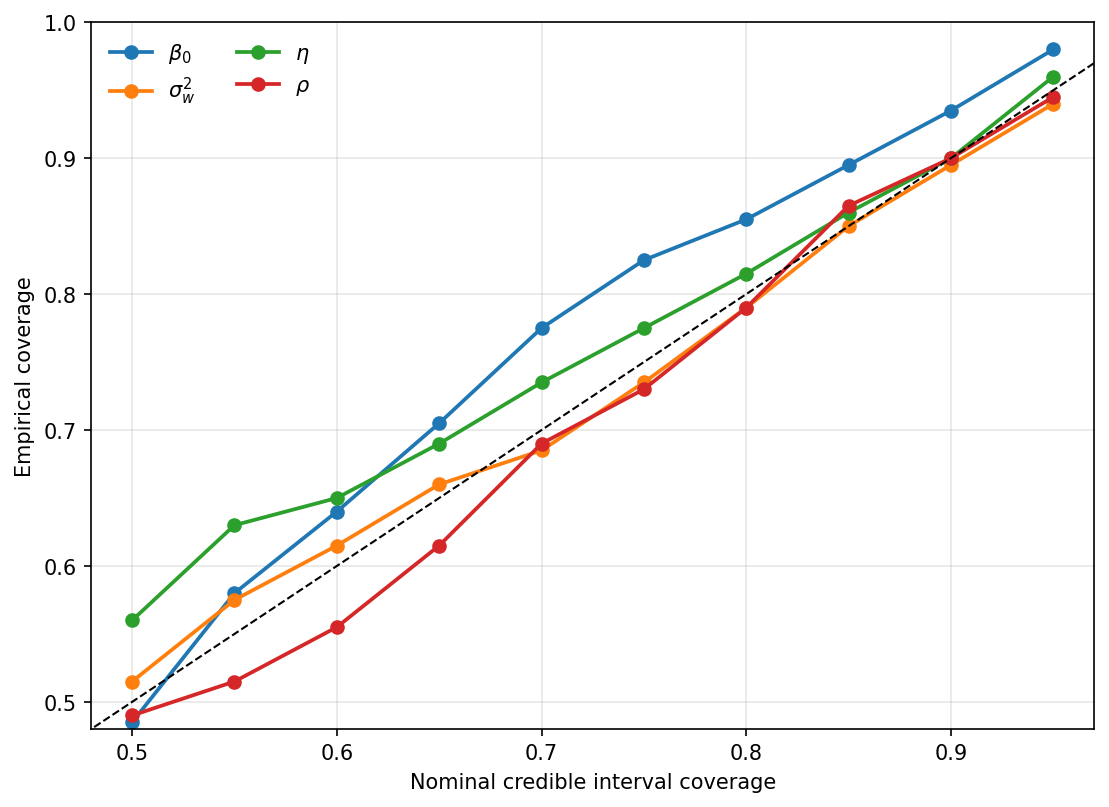}
\caption{Empirical coverage vs nominal coverage for the amortized posterior approximation.}
\label{fig:sim_coverage_sup}
\end{figure}

This subsection complements the SBC rank histograms in Fig.~\ref{fig:sbc_hist} of the main text. Fig.~\ref{fig:sim_coverage_sup} reports empirical coverage against nominal coverage. The curves remain close to the diagonal for all parameters, indicating generally well-calibrated posterior intervals, with only mild conservatism for $\beta_0$.

\begin{figure}[t]
\centering
\includegraphics[width=\textwidth]{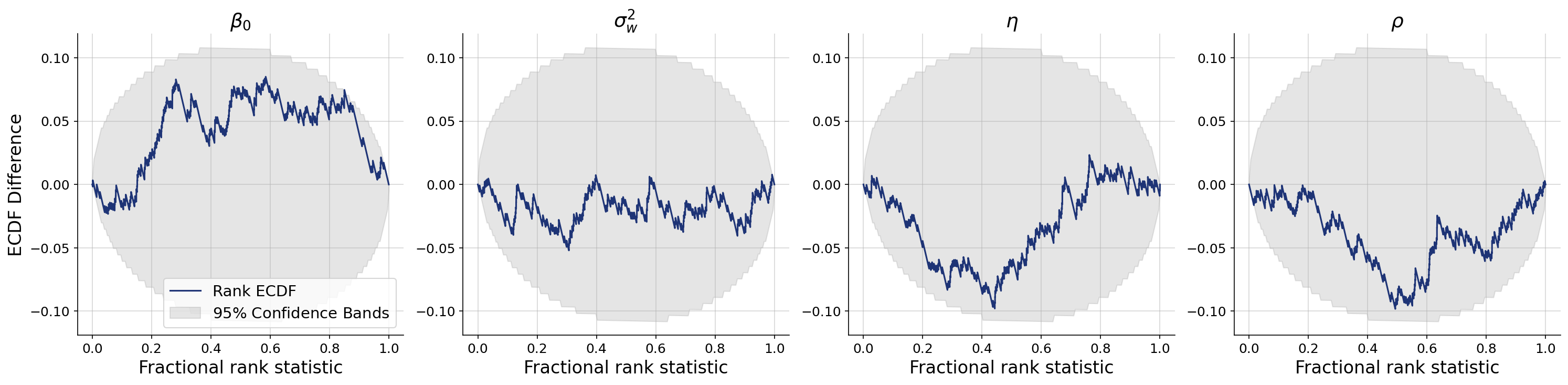}
\caption{Simulation-based calibration ECDF-difference plots for the four model parameters.}
\label{fig:sbc_ecdf_sup}
\end{figure}

Fig.~\ref{fig:sbc_ecdf_sup} shows SBC ECDF-difference plots. The curves remain close to zero, suggesting no substantial systematic distortion beyond the mild irregularities already visible in the rank histograms.

\begin{figure}[t]
\centering
\includegraphics[width=\textwidth]{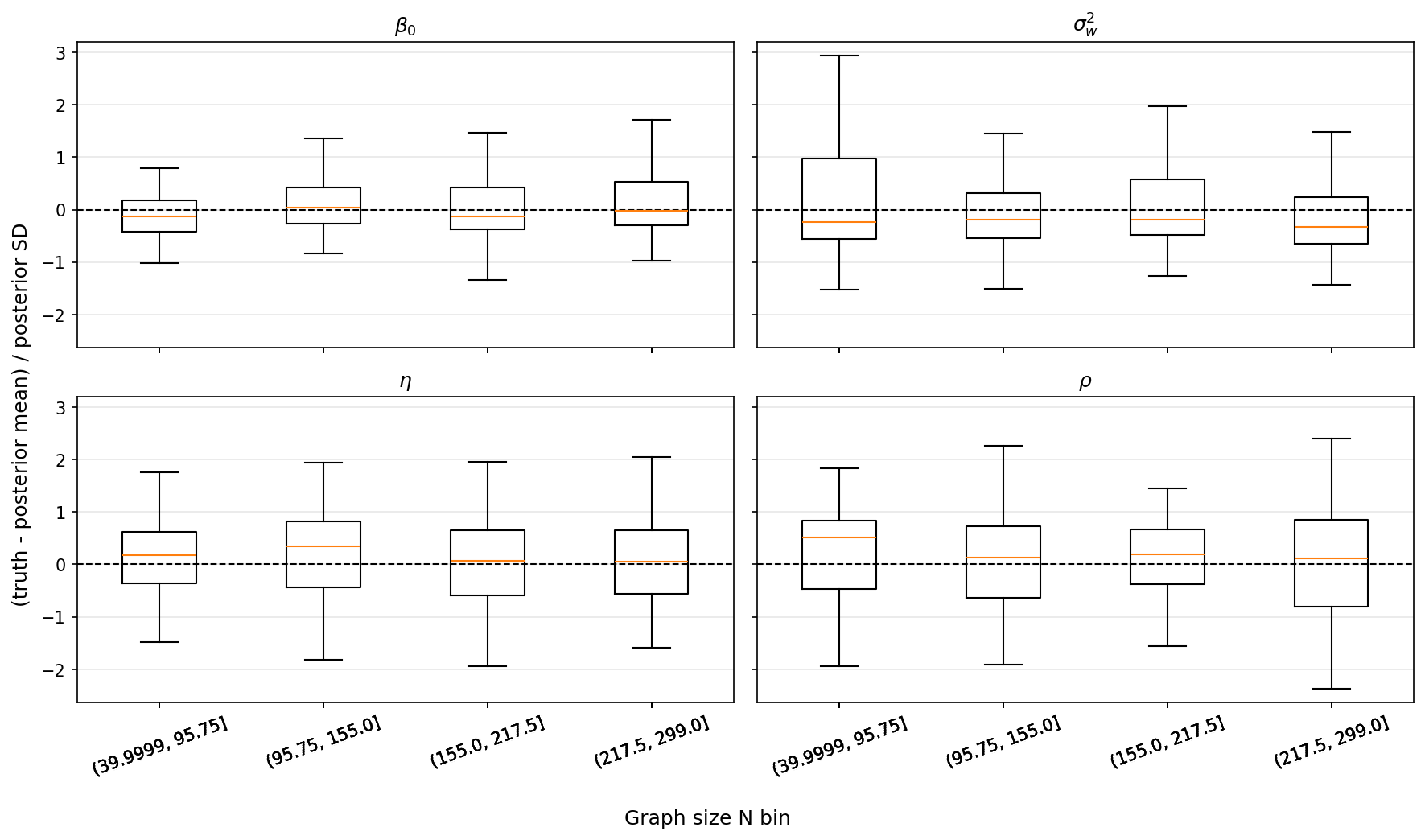}
\caption{Posterior $z$-scores, stratified by graph-size bin.}
\label{fig:graph_size_z_sup}
\end{figure}

Fig.~\ref{fig:graph_size_z_sup} stratifies posterior $z$-scores, i.e., $(\theta_{\mathrm{true}}-\mathbb{E}(\theta\mid\mathcal{D}))/\operatorname{sd}(\theta\mid\mathcal{D})$, by graph-size bin, showing boxplot regarding. The boxplots do not suggest a pronounced graph-size-dependent shift or rescaling failure, indicating that posterior centering and uncertainty remain reasonably stable across graph sizes.

\begin{figure}[t]
\centering
\includegraphics[width=\textwidth]{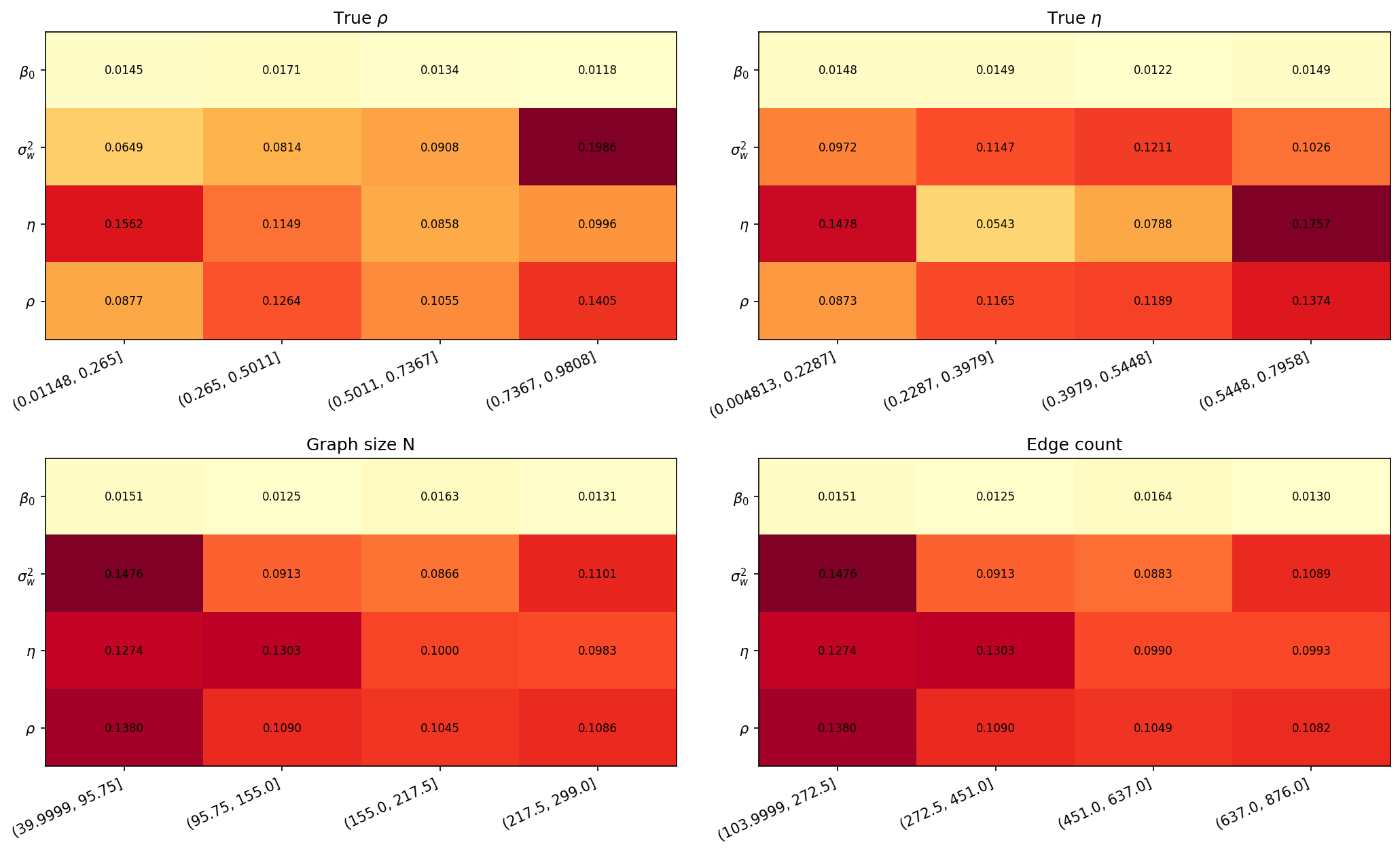}
\caption{Mean absolute recovery error by true $\rho$, true $\eta$, graph size, and edge count.}
\label{fig:regime_error_sup}
\end{figure}

Fig.~\ref{fig:regime_error_sup} reports mean absolute recovery error stratified by true $\rho$, true $\eta$, graph size, and edge count. Recovery of $\beta_0$ remains uniformly accurate, whereas $\sigma_w^2$, $\eta$, and $\rho$ show greater heterogeneity, as expected for parameters governing latent dependence and boundary formation. Errors increase gradually in more challenging parts of the design space, especially larger or denser graphs and settings with stronger spatial structure, with no narrow failure regime apparent.

\subsection{Additional boundary-probability diagnostics}
\label{sec:supp_sim_boundary}

For an observed edge $(i,j)$, the posterior boundary probability, i.e. $p_{ij}=\Pr\!\left(z_{ij}\eta > \log 2 \mid \bs{y}\right)$, is estimated from the amortized posterior draws. This subsection evaluates these edge-level probabilities beyond any single thresholding rule.

Of the 200 held-out datasets, 161 contained both boundary and non-boundary edges, while 39 contained no true boundaries. Dataset-level sensitivity is undefined when no true boundaries are present, and AUROC and average precision are not informative in datasets containing only non-boundary edges. These metrics were therefore summarized over the 161 datasets containing true boundaries. Brier scores, boundary-count summaries, selected-boundary counts, and specificity remained defined and were summarized over all 200 datasets. The number of datasets contributing to each metric is reported explicitly in Table~\ref{tab:sim_boundary_sup}.

Dataset-level AUROC and average precision remained high, with means of 0.996 and 0.988, respectively, while Brier scores were low. The FDR-controlling rule was highly conservative: its median sensitivity was zero, and 120 of the 161 datasets containing true boundaries had zero sensitivity under this rule. Its specificity was nevertheless extremely high, with median specificity equal to one.

The median-probability rule provided a less conservative operating point. Its mean sensitivity was 0.719 and its median sensitivity was 0.857 across datasets containing true boundaries, while mean specificity remained 0.962 across all datasets. Only 7 of the 161 datasets containing true boundaries had zero sensitivity under the median-probability rule. These results indicate that the median-probability rule recovers substantially more true boundaries than the FDR-controlling rule while retaining strong false-positive control.

\begin{table}[t]
\centering
\caption{Additional boundary-probability summaries on held-out simulated datasets. Sensitivity, AUROC, and average precision exclude the 39 datasets containing no true boundaries.}
\label{tab:sim_boundary_sup}
\resizebox{\textwidth}{!}{%
\begin{tabular}{lccccc}
\hline
\textbf{Metric} & $\bs{n}$ & \textbf{Mean} & \textbf{SD} & \textbf{Median} & \textbf{IQR} \\
\hline
Dataset AUROC & 161 & 0.996 & 0.005 & 0.997 & 0.007 \\
Dataset AP & 161 & 0.988 & 0.012 & 0.990 & 0.020 \\
Dataset Brier score & 200 & 0.061 & 0.048 & 0.050 & 0.047 \\
FDR sensitivity & 161 & 0.169 & 0.329 & 0.000 & 0.004 \\
FDR specificity & 200 & 0.995 & 0.016 & 1.000 & 0.000 \\
Median-probability selected boundaries & 200 & 80.740 & 80.344 & 56.500 & 93.000 \\
True boundaries & 200 & 91.240 & 93.093 & 65.500 & 142.000 \\
Median-probability sensitivity & 161 & 0.719 & 0.322 & 0.857 & 0.537 \\
Median-probability specificity & 200 & 0.962 & 0.059 & 0.988 & 0.050 \\
\hline
\end{tabular}%
}
\end{table}

\begin{figure}[t]
\centering
\includegraphics[width=\textwidth]{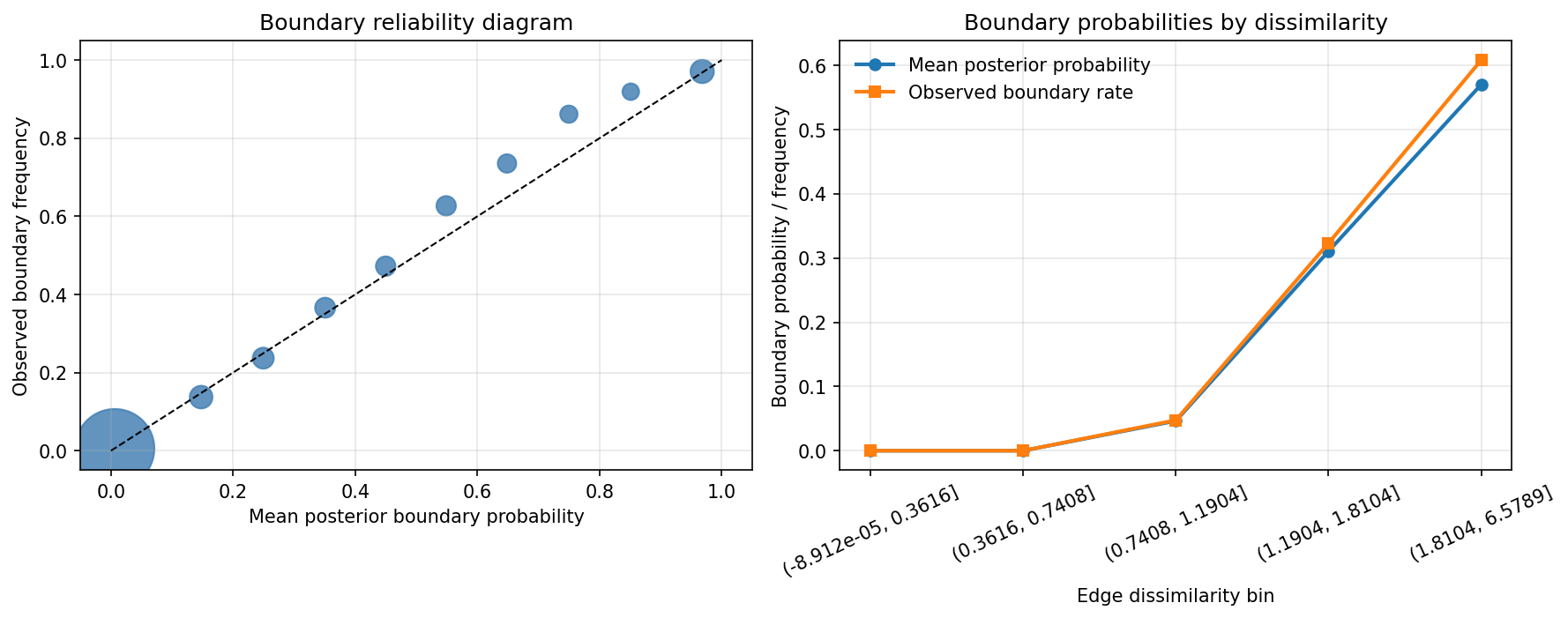}
\caption{Boundary-probability diagnostics. Left: reliability diagram for posterior boundary probabilities, with point diameter proportional to the number of edges in each probability bin. Right: posterior and empirical boundary frequencies by dissimilarity bin.}
\label{fig:boundary_prob_sup}
\end{figure}

Fig.~\ref{fig:boundary_prob_sup} provides two complementary diagnostics. The reliability diagram shows that binned posterior boundary probabilities remain reasonably close to empirical boundary frequencies over the main mass of the distribution. The dissimilarity-binned panel shows close agreement between mean posterior boundary probability and empirical boundary frequency, indicating that the amortized posterior reproduces the covariate-driven relationship between edge dissimilarity and boundary formation.

The threshold-based operating characteristics for the FDR-controlling rule and median-probability rule are shown in Figs.~\ref*{fig:boundary_curve} and~\ref*{fig:boundary_mpm_hist} of the main text. These figures follow the same eligibility convention as Table~\ref{tab:sim_boundary_sup}: sensitivity excludes datasets containing no true boundaries, whereas specificity includes all held-out datasets.

\subsection{Additional posterior predictive diagnostics}
\label{sec:supp_sim_ppc}

Posterior predictive checks were conducted on 40 representative held-out datasets spanning the graph-size range, using 100 posterior predictive replicates per dataset. For a summary statistic $T(\bs{y})$, we report posterior predictive $p$-values, i.e., $p_{\mathrm{ppc}} = \Pr\!\bigl(T(\bs{y}^{\mathrm{rep}})\ge T(\bs{y}) \mid \mathcal{D}\bigr)$, estimated from posterior predictive replicates.

Table~\ref{tab:sim_ppc_sup} summarizes the diagnostics. Node-level count coverage was close to nominal, with observed counts falling inside the 95\% posterior predictive interval for a mean of 0.940 of areas per dataset. The Moran-type posterior predictive $p$-values were centered near 0.5, with mean 0.486, indicating no systematic mismatch in residual spatial dependence.

For edge-level diagnostics, we considered the residual log-risk contrast
\begin{equation*}
\Delta_{ij}
=
\left|
\bigl[\log(y_i+0.5)-\log(e_i)\bigr]
-
\bigl[\log(y_j+0.5)-\log(e_j)\bigr]
\right|
\end{equation*}
for adjacent pairs $(i,j)$, averaged within low-, medium-, and high-dissimilarity bins. The corresponding mean posterior predictive $p$-values were 0.536, 0.489, and 0.561, respectively, suggesting adequate reproduction of local edge contrasts across the dissimilarity spectrum.

\begin{table}[t]
\centering
\caption{Additional posterior predictive summaries on held-out simulated datasets.}
\label{tab:sim_ppc_sup}
\begin{tabular}{lcccc}
\hline
\textbf{Metric} & \textbf{Mean} & \textbf{SD} & \textbf{Median} & \textbf{IQR} \\
\hline
Node-level count coverage (95\% PPC interval) & 0.940 & 0.019 & 0.938 & 0.025 \\
Moran-type posterior predictive $p$-value & 0.486 & 0.204 & 0.520 & 0.240 \\
Low-dissimilarity edge-contrast $p$-value & 0.536 & 0.263 & 0.575 & 0.423 \\
Mid-dissimilarity edge-contrast $p$-value & 0.489 & 0.268 & 0.430 & 0.450 \\
High-dissimilarity edge-contrast $p$-value & 0.561 & 0.245 & 0.600 & 0.345 \\
\hline
\end{tabular}
\end{table}

\begin{figure}[t]
\centering
\includegraphics[width=\textwidth]{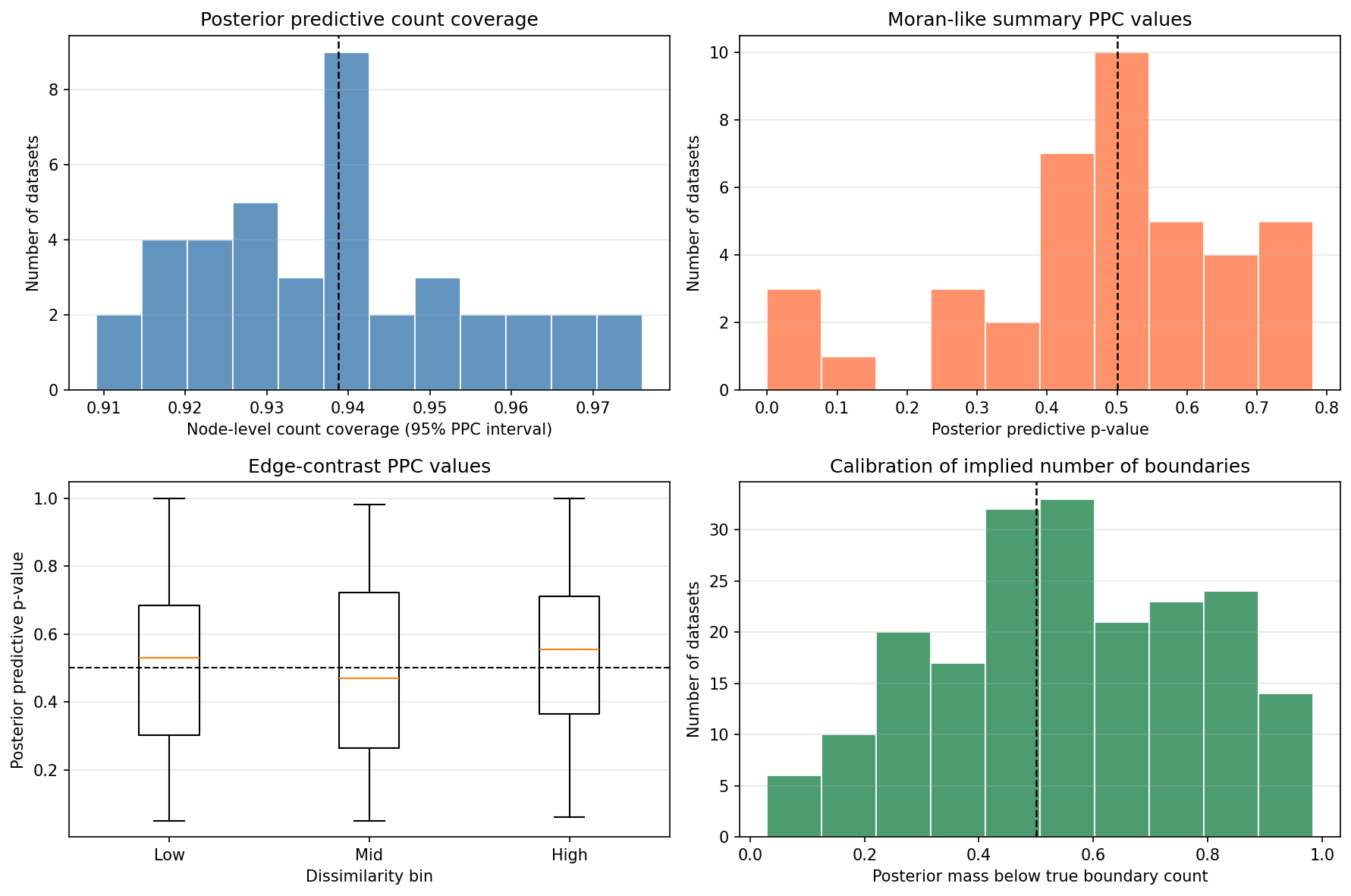}
\caption{Posterior predictive diagnostics for counts, Moran-type summaries, edge contrasts, and the implied number of boundaries.}
\label{fig:ppc_sup}
\end{figure}

Fig.~\ref{fig:ppc_sup} visualizes marginal count coverage, Moran-type residual spatial dependence, edge contrasts by dissimilarity bin, and the implied number of boundaries. Together, these summaries indicate that the amortized posterior predictive distribution reproduces key features of the data-generating model on held-out graphs.

\subsection{Additional MCMC-DAGAR benchmark diagnostics}
\label{sec:supp_sim_abi_mcmc}

This subsection provides additional details for the benchmark comparing ABI-DAGAR with the model-matched MCMC-DAGAR implementation on 100 held-out simulated datasets. The purpose is to assess whether the amortized posterior approximation reproduces the output of a dataset-specific Bayesian sampler under the same thresholded Poisson-DAGAR model.

At the parameter level, agreement is strongest for $\beta_0$, for which the mean absolute difference in posterior means is 0.010 and the posterior-mean correlation is 1.000. Agreement remains strong, though weaker, for the spatial parameters: the mean absolute differences are 0.096 for $\sigma_w^2$, 0.066 for $\eta$, and 0.068 for $\rho$, with corresponding posterior-mean correlations 0.899, 0.846, and 0.927. The corresponding parameter-level graphical comparison is reported in Fig.~\ref*{fig:sim_abi_mcmc_param_main} of the main manuscript.

\begin{figure}[t]
\centering
\includegraphics[width=\textwidth]{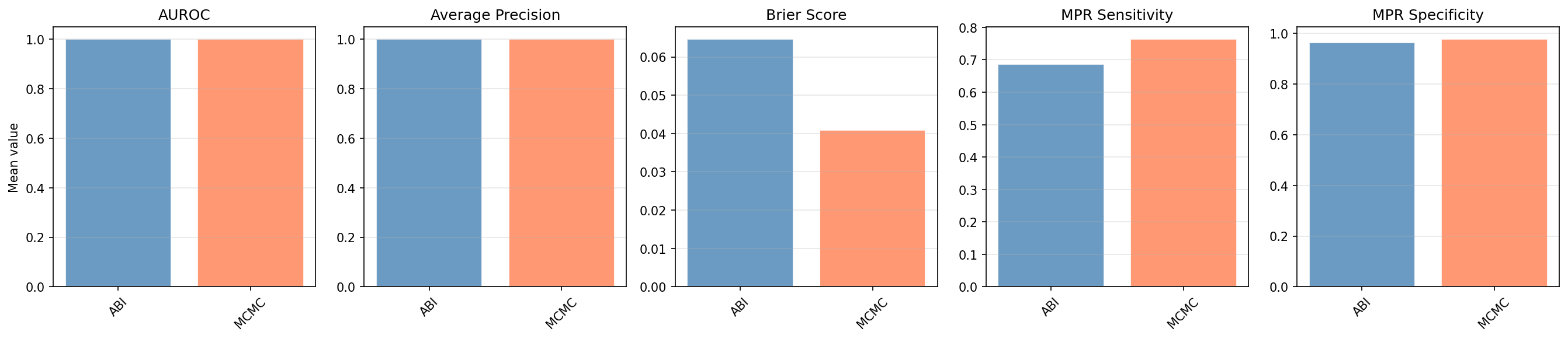}
\includegraphics[width=\textwidth]{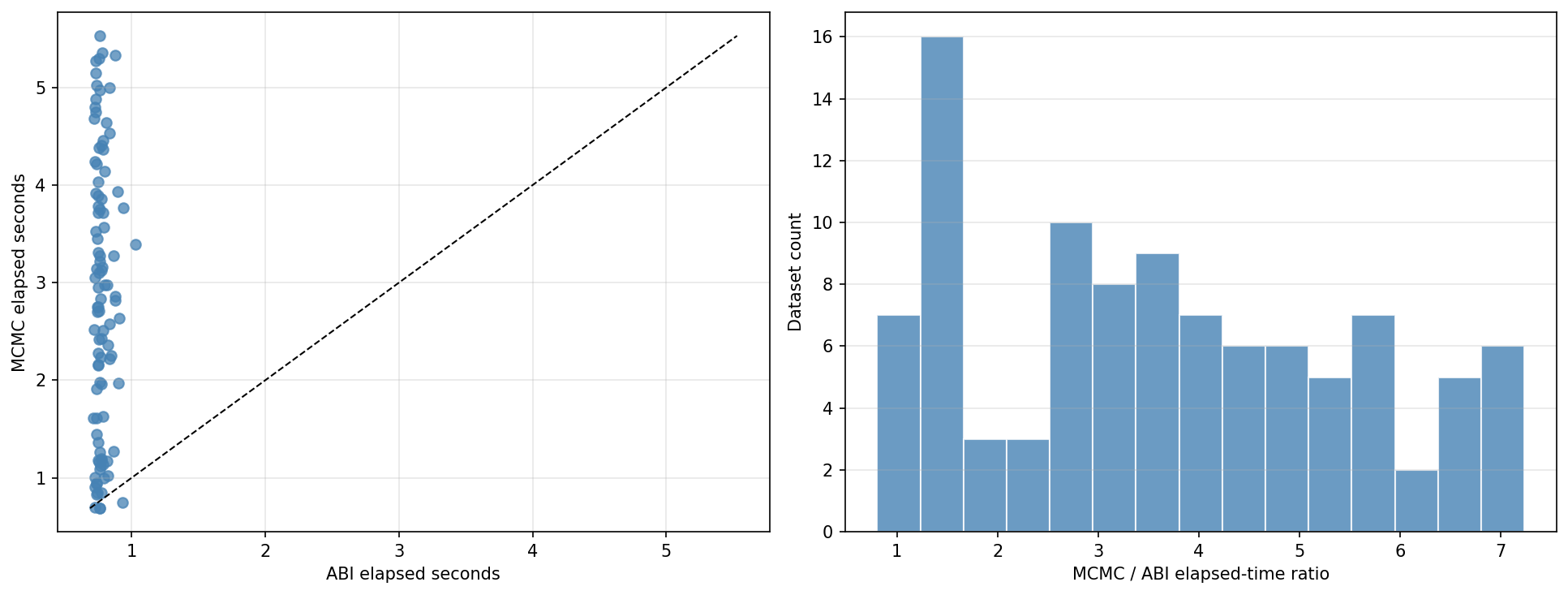}
\caption{Additional ABI-DAGAR and MCMC-DAGAR benchmark diagnostics. Top: dataset-level boundary-probability and decision metrics. Bottom: runtime comparison across datasets, including elapsed time and the distribution of MCMC-DAGAR/ABI-DAGAR runtime ratios.}
\label{fig:sim_abi_mcmc_boundary_runtime_sup}
\end{figure}

At the edge level, posterior boundary probabilities from ABI-DAGAR and MCMC-DAGAR have mean correlation 0.920 and mean absolute difference 0.066. Under the median-probability rule, ABI-DAGAR selects 94.48 boundaries on average, compared with 94.47 under MCMC-DAGAR, with a mean of 79.10 shared selected edges and mean Jaccard overlap 0.593. Fig.~\ref{fig:sim_abi_mcmc_boundary_runtime_sup} shows that both methods achieve very strong boundary discrimination, while MCMC-DAGAR attains a lower mean Brier score, 0.041 versus 0.065, and slightly higher mean median-probability sensitivity and specificity, 0.764 and 0.977 versus 0.686 and 0.962 under ABI-DAGAR.

Computationally, ABI-DAGAR requires 0.78 seconds per dataset on average to generate 10{,}000 posterior draws, compared with 2.84 seconds for MCMC-DAGAR. Including the one-time training cost of 5 hours and 45 minutes, however, the break-even point occurs only after roughly $1.0\times 10^4$ datasets. These results support interpreting ABI-DAGAR as a reusable posterior approximation for repeated deployment, rather than as an immediate end-to-end computational saving in a moderate-sized benchmark experiment.

\subsection{Additional ablation-study details}
\label{sec:supp_ablation}

This subsection provides additional details for the ablation study of the summary representation used by the amortized posterior approximator. The baseline model used the complete summary design, including core observation features, graph-topology features, dissimilarity-based features, local spatial features, and global graph features. The ablated models removed one block at a time, and an additional reduced model used only the core observation features. For each summary representation, we retrained a separate amortized posterior approximator from scratch using the same network architecture and training protocol. All runs were evaluated on the same 4050 held-out datasets, with 50 datasets at each graph size \(N=40,\ldots,120\).

\begin{figure}[t]
\centering
\includegraphics[width=\textwidth]{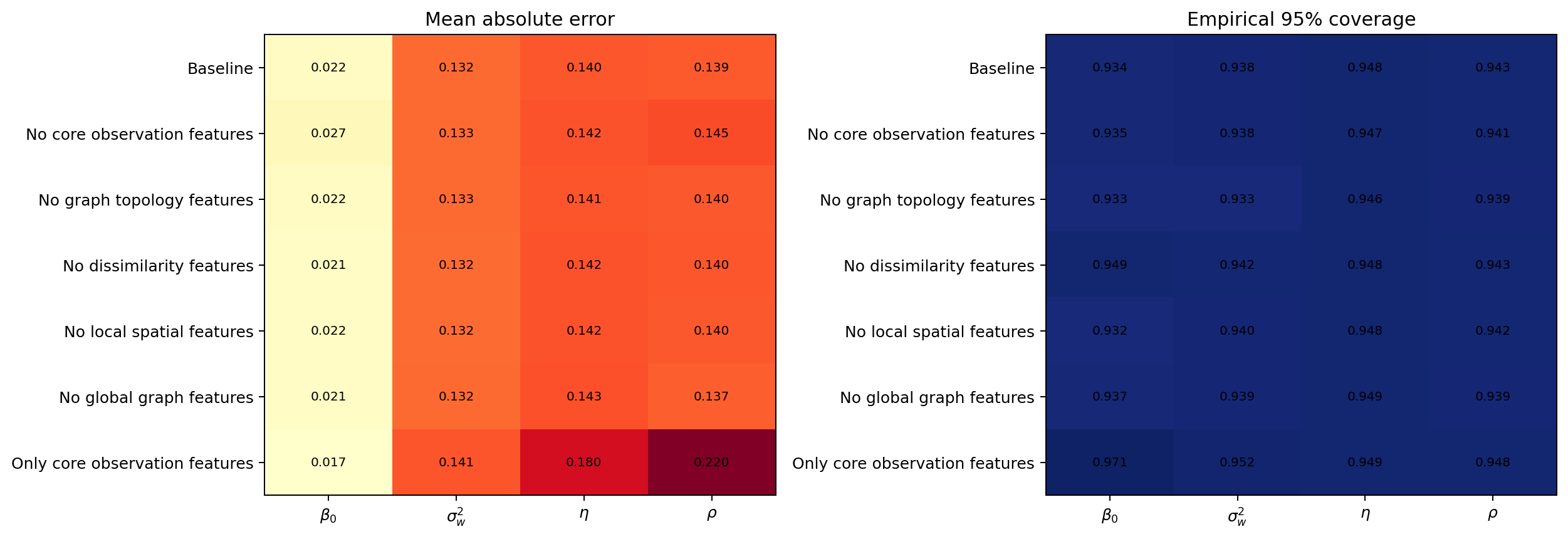}
\caption{Parameter-specific recovery metrics for the ablation study. Rows correspond to summary representations and columns summarize recovery for $\beta_0$, $\sigma_w^2$, $\eta$, and $\rho$.}
\label{fig:ablation_recovery_heatmaps}
\end{figure}

The full representation gave balanced recovery across parameters, with average MAE across parameters 0.1084, average RMSE 0.1431, empirical 95\% coverage 0.9406, and mean correlation 0.7878. Recovery was strongest for $\beta_0$, with MAE 0.0223 and correlation 0.9986. The boundary parameter $\eta$ was the most difficult component, with MAE 0.1403 and correlation 0.5692, but the baseline gave the strongest $\eta$ recovery among the compared representations. Fig.~\ref{fig:ablation_recovery_heatmaps} summarizes the parameter-specific patterns.

The core-observation-only representation shows that marginal information alone is insufficient for the full spatial boundary-detection task. It recovered $\beta_0$ very well, with MAE 0.0171 and correlation 0.9990, but recovery deteriorated for parameters depending on spatial and boundary structure: the correlation dropped to 0.1717 for $\eta$ and 0.4643 for $\rho$. Removing the core observation block produced the largest degradation among the one-block ablations and worsened optimization, with final validation loss increasing from 0.6699 under the full representation to 1.0242.

Boundary-detection diagnostics show the same hierarchy. Under the full representation, pooled posterior boundary probabilities achieved AUROC 0.9441, AP 0.7723, and Brier score 0.0738. The core-observation-only model showed the clearest deterioration, with AUROC 0.9013, AP 0.5905, and Brier score 0.0965. By contrast, the one-block non-core ablations remained close to the baseline across pooled AUROC, AP, and Brier score, suggesting that these blocks contribute mainly through refinement and robustness.

\begin{figure}[t]
\centering
\includegraphics[width=\textwidth]{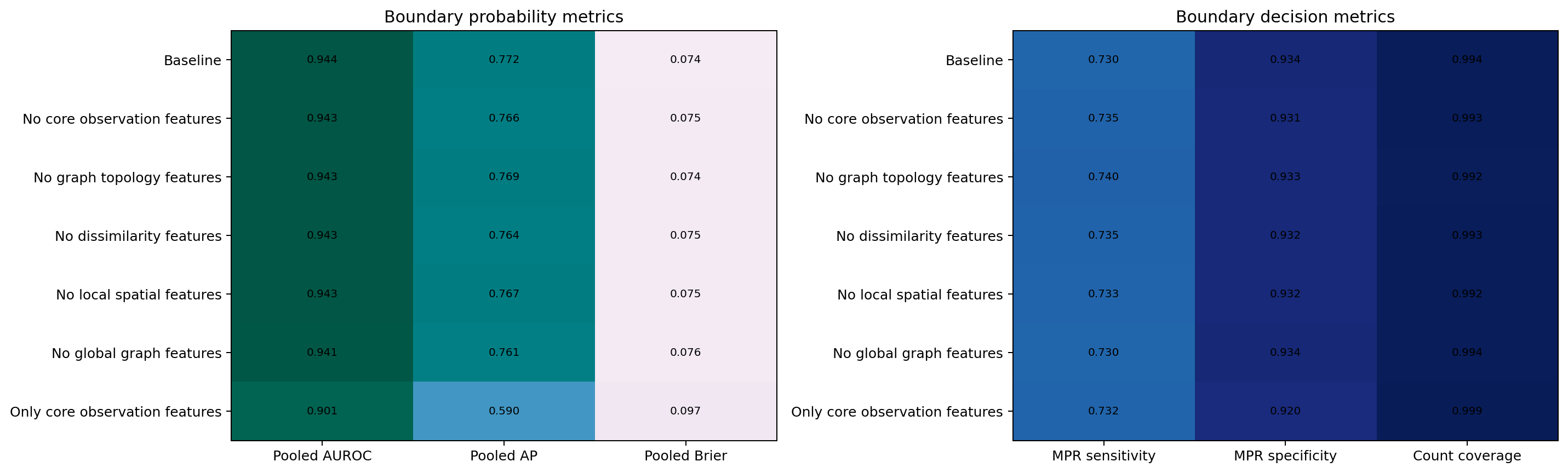}
\caption{Boundary-detection diagnostics across summary representations. The left panel reports pooled boundary-probability metrics, and the right panel reports median-probability decision metrics and boundary-count coverage.}
\label{fig:ablation_boundary_heatmaps}
\end{figure}

\begin{figure}[t]
\centering
\includegraphics[width=\textwidth]{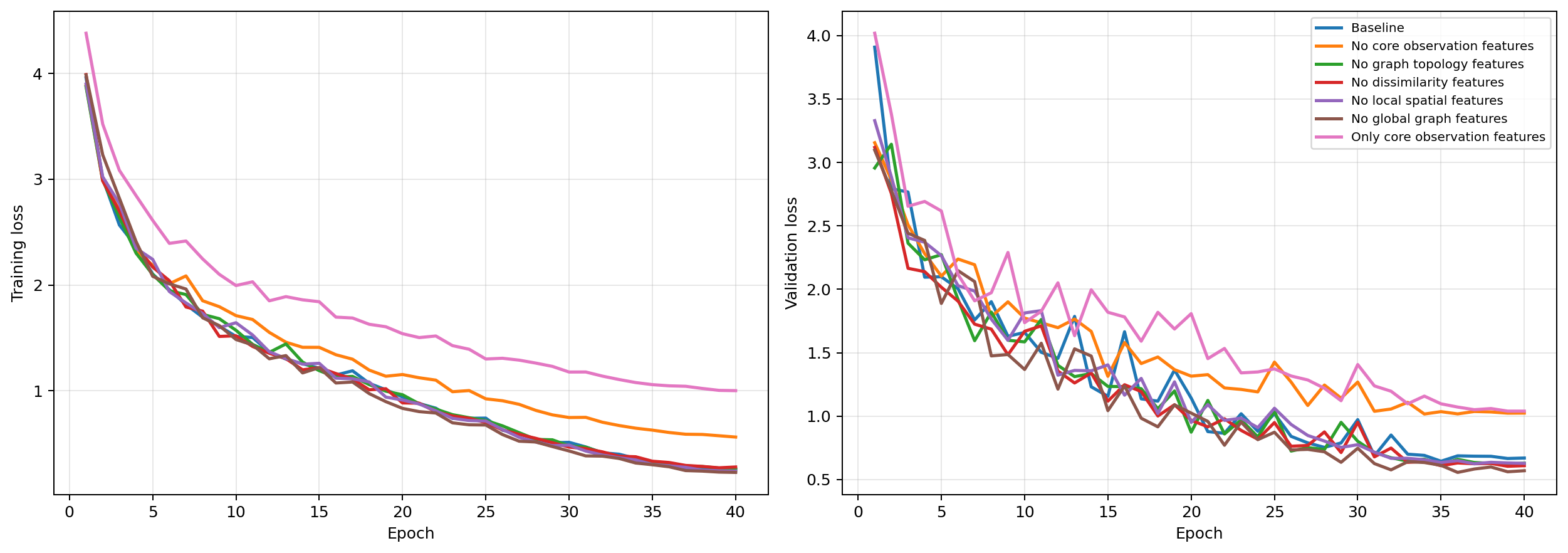}
\caption{Training and validation loss trajectories for the baseline and ablated summary representations.}
\label{fig:ablation_training_histories}
\end{figure}

Fig.~\ref{fig:ablation_training_histories} reports training and validation loss trajectories. The core-only representation had much larger final training and validation losses, 1.0009 and 1.0395, respectively, while the no-core representation also showed worse optimization behavior, with final losses 0.5619 and 1.0242. The one-block non-core ablations were closer to the baseline. These diagnostics support retaining the full summary design as a stable and interpretable basis for learning posterior inference across heterogeneous areal graphs.

\subsection{Additional computational details}
\label{sec:supp_sim_computation}

Table~\ref{tab:sim_computation} summarizes the computational details of the simulation study. The one-time training stage required 5 hours and 45 minutes on an Intel(R) Core(TM) i7-10750H CPU @ 2.60GHz. Validation on the 200 held-out datasets, with 10{,}000 posterior draws per dataset, required 153.59 seconds in total, corresponding to 0.77 seconds per dataset, with a memory footprint of 1.168 GB. The posterior predictive diagnostic block required an additional 23.52 seconds. These timings characterize the deployment profile of the validated amortized posterior approximation after the one-time training stage.

\begin{table}[t]
\centering
\caption{Computational details for the simulation study.}
\label{tab:sim_computation}
\begin{tabular}{ll}
\hline
\textbf{Quantity} & \textbf{Value} \\
\hline
Hardware & Intel(R) Core(TM) i7-10750H CPU @ 2.60GHz \\
Training epochs & 100 \\
Batch size & 64 \\
Batches per epoch & 200 \\
Total training simulations & 1,280,000 \\
Training time & 5 hours and 45 minutes \\
Held-out validation datasets & 200 \\
Posterior draws per dataset & 10,000 \\
Validation sampling time (total) & 153.59 seconds \\
Validation sampling time (per dataset) & 0.77 seconds \\
Validation memory & 1.168 GB \\
PPC datasets / PPC draws per dataset & 40 / 100 \\
PPC runtime & 23.52 seconds \\
\hline
\end{tabular}
\end{table}

\section{Additional real data analysis results}
\label{sec:additional_details}

This section reports supplementary diagnostics for the Glasgow and California applications, including posterior predictive checks, fitted-risk comparisons, edge-level boundary diagnostics, a model-matched MCMC-DAGAR comparison, runtime details, and implementation details. Posterior summaries were based on 100{,}000 amortized posterior draws, and posterior predictive diagnostics used 1{,}000 replicates per dataset.

\subsection{Posterior predictive diagnostics}
\label{sec:additional_details_ppc}

\begin{figure}[t]
    \centering
    \includegraphics[width=0.495\linewidth]{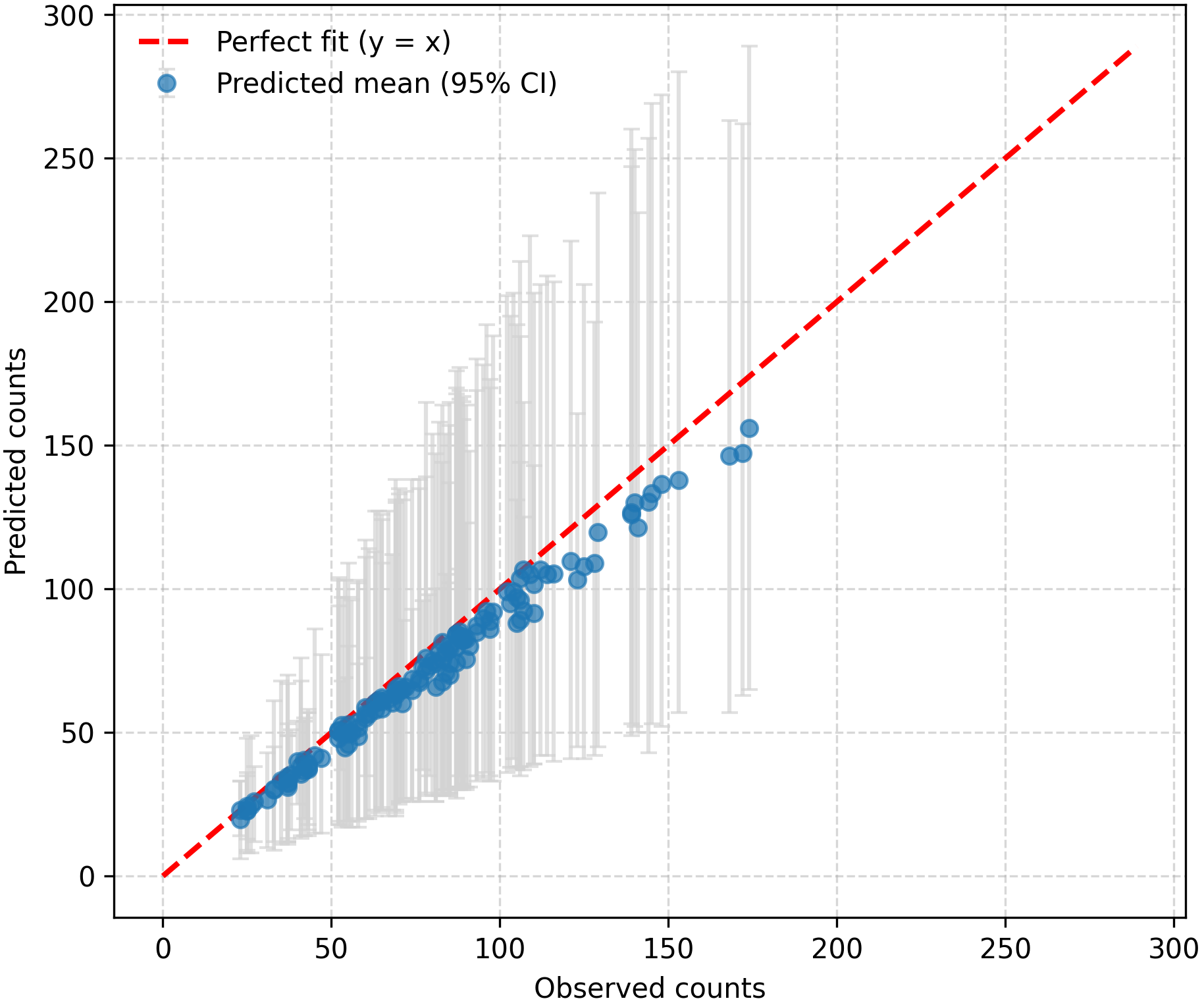}
    \includegraphics[width=0.495\linewidth]{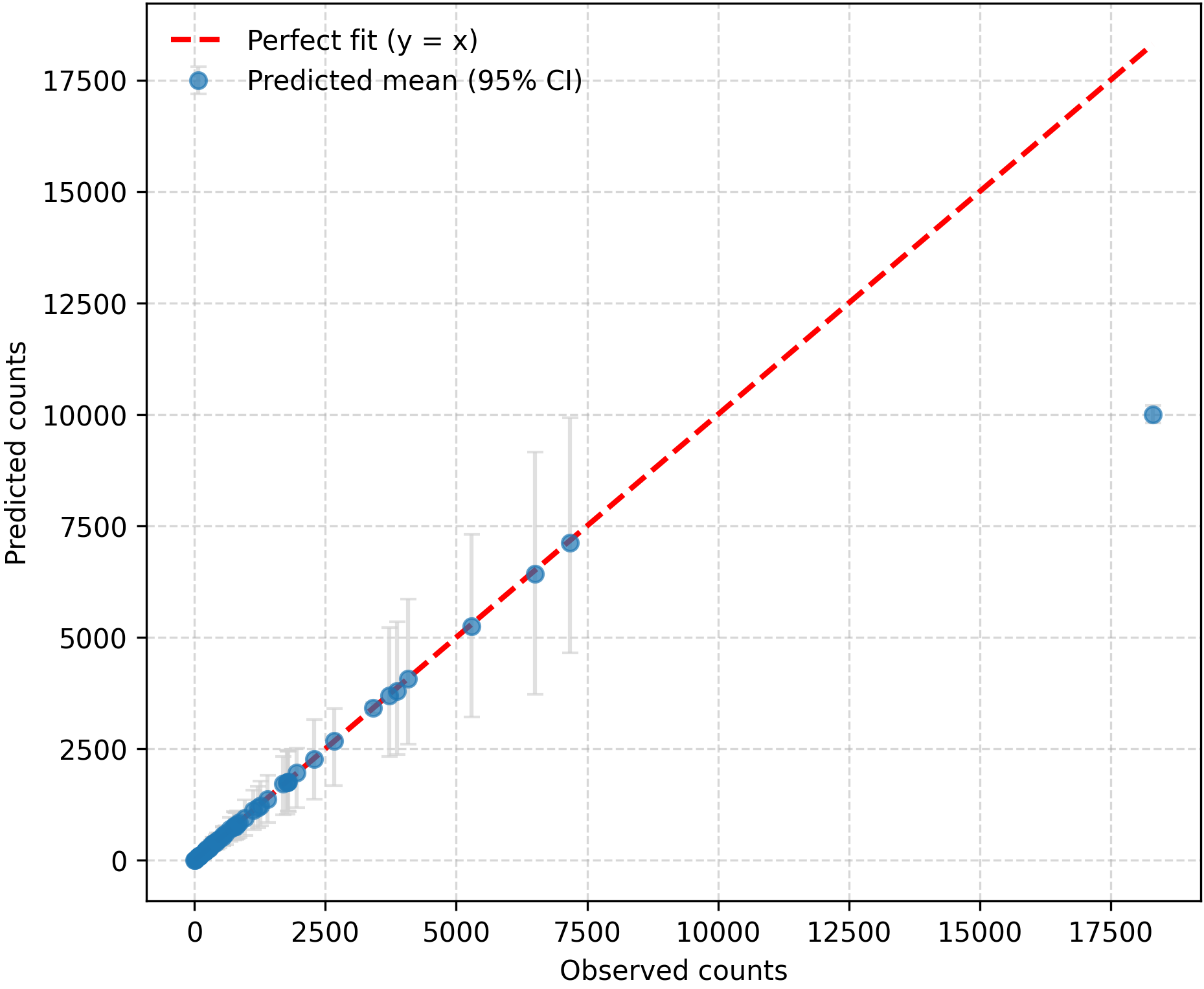}
    \caption{Posterior predictive checks for the real-data applications. Left: Glasgow. Right: California.}
    \label{fig:realdata_ppc}
\end{figure}

Fig.~\ref{fig:realdata_ppc} reports posterior predictive checks for the two applications. Observed counts are plotted against posterior predictive means with pointwise 95\% posterior predictive intervals. These plots are intended as graphical checks of predictive adequacy rather than formal goodness-of-fit tests.

In both applications, posterior predictive means track the observed counts reasonably well over most of the range, with wider intervals for larger counts. The Glasgow application shows greater spread and more visible departures among the largest observations, whereas the California application lies closer to the 45-degree line for most counties. Overall, the diagnostics suggest that the amortized posterior predictive distribution preserves the main count structure in both datasets, while reflecting greater residual heterogeneity in Glasgow.

\subsection{Additional fitted-risk comparison}
\label{sec:additional_details_risk}

\begin{figure}[t]
    \centering
    \includegraphics[width=0.495\linewidth]{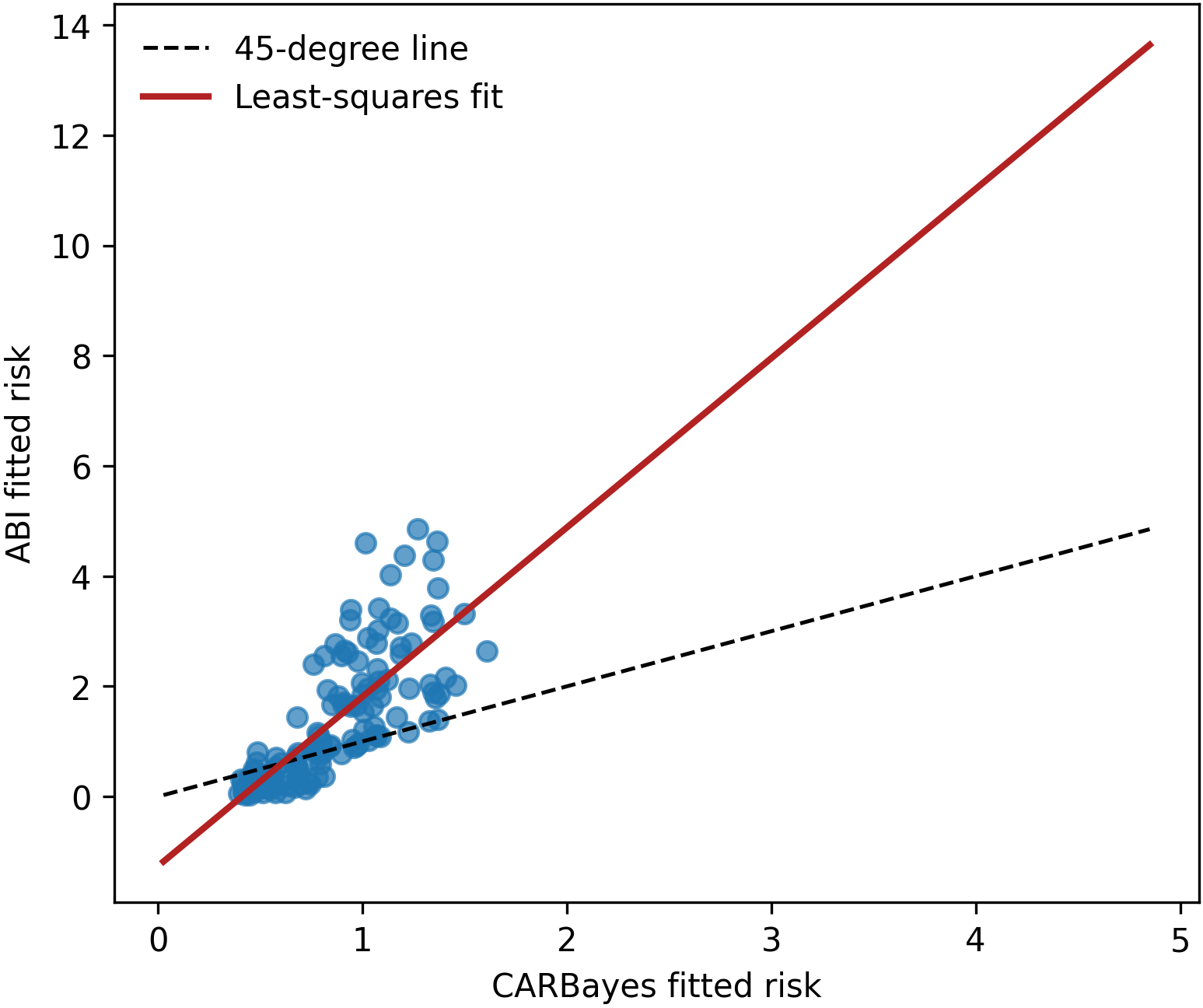}
    \includegraphics[width=0.495\linewidth]{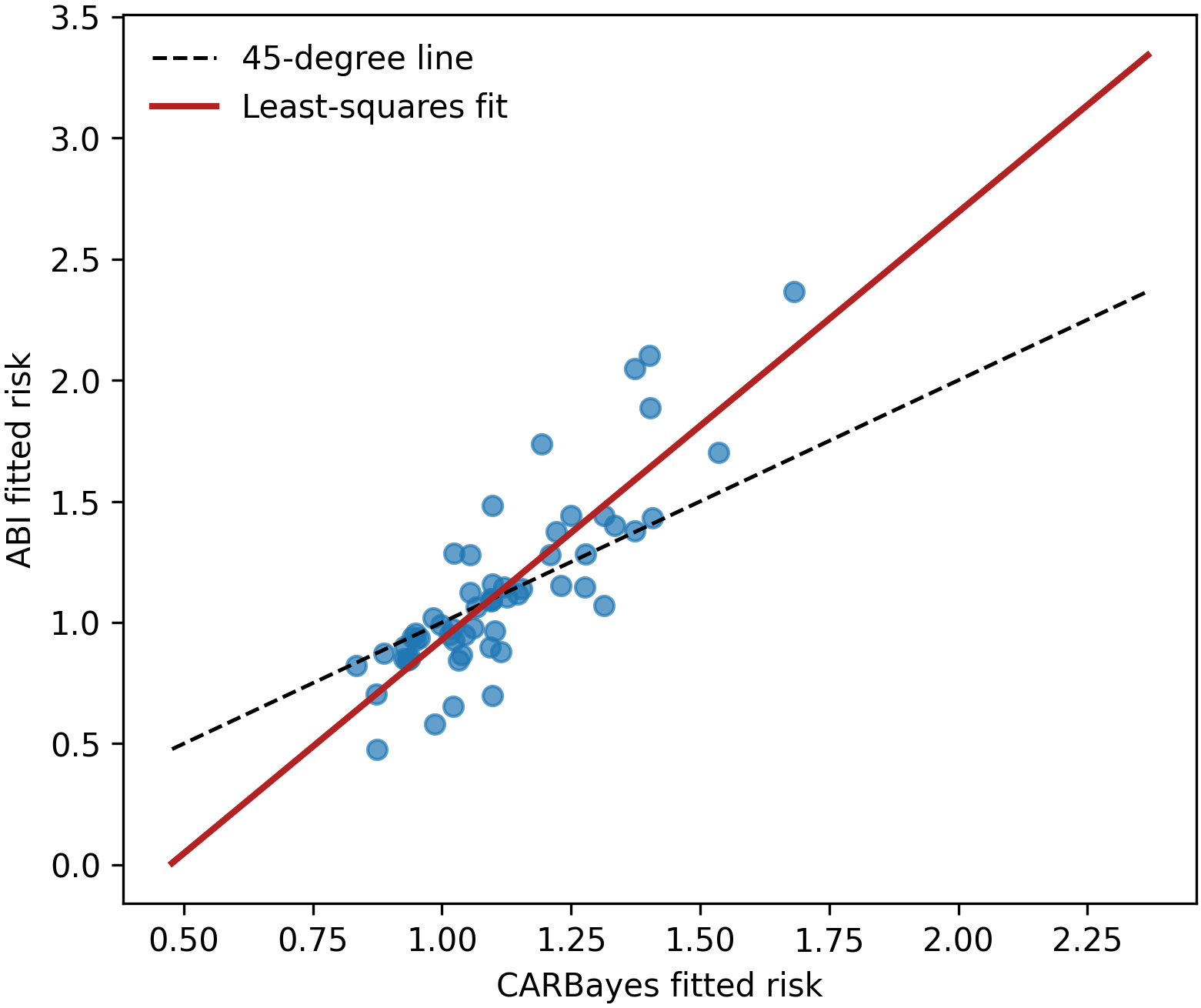}
    \caption{Fitted-risk comparison between ABI-DAGAR and \texttt{CARBayes}. Left: Glasgow. Right: California. }
    \label{fig:risk_comparison}
\end{figure}

Fig.~\ref{fig:risk_comparison} compares fitted risks from ABI-DAGAR and \texttt{CARBayes}. This comparison is a secondary deployment diagnostic: exact agreement is not expected because the two approaches use different latent spatial priors, and the ABI implementation targets posterior inference for $(\beta_0,\sigma_w^2,\eta,\rho)$ rather than direct posterior sampling of the latent field $\bs{w}$.

For ABI-DAGAR, fitted risks are reconstructed as follows. For each posterior draw of $(\beta_0,\eta,\rho)$, we construct the filtered graph implied by $\eta$, form an empirical log-residual field from the observed data, smooth this field using the corresponding DAGAR operator, and compute the implied fitted risks. The ABI fitted risk in Fig.~\ref{fig:risk_comparison} is the posterior average of these reconstructed fitted risks.

Table~\ref{tab:realdata_risk_extra} reports the corresponding summaries. Fitted-risk correlations are positive in both applications and stronger in California than in Glasgow. The slopes exceeding one indicate that the reconstructed ABI-DAGAR fitted-risk surface varies more strongly over space than the \texttt{CARBayes} surface, especially in Glasgow. Given the reconstructed nature of the ABI-DAGAR fitted risks, these discrepancies should not be interpreted as direct failures of posterior recovery; the edge-level boundary comparison remains the more direct benchmark for the scientific target of interest.

\begin{table}[t]
\centering
\caption{Additional fitted-risk comparison metrics between ABI-DAGAR and \texttt{CARBayes}.}
\label{tab:realdata_risk_extra}
\begin{tabular}{lccccc}
\hline
\textbf{Dataset} & \textbf{Correlation} & \textbf{MAE} & \textbf{RMSE} & \textbf{ABI on CB slope} & \textbf{ABI on CB intercept} \\
\hline
Glasgow & 0.767 & 0.737 & 1.110 & 3.072 & $-1.261$ \\
California & 0.843 & 0.155 & 0.239 & 1.765 & $-0.836$ \\
\hline
\end{tabular}
\end{table}

\subsection{Additional boundary-probability diagnostics}
\label{sec:additional_details_edge}

\begin{figure}[t]
    \centering
    \includegraphics[width=0.495\linewidth]{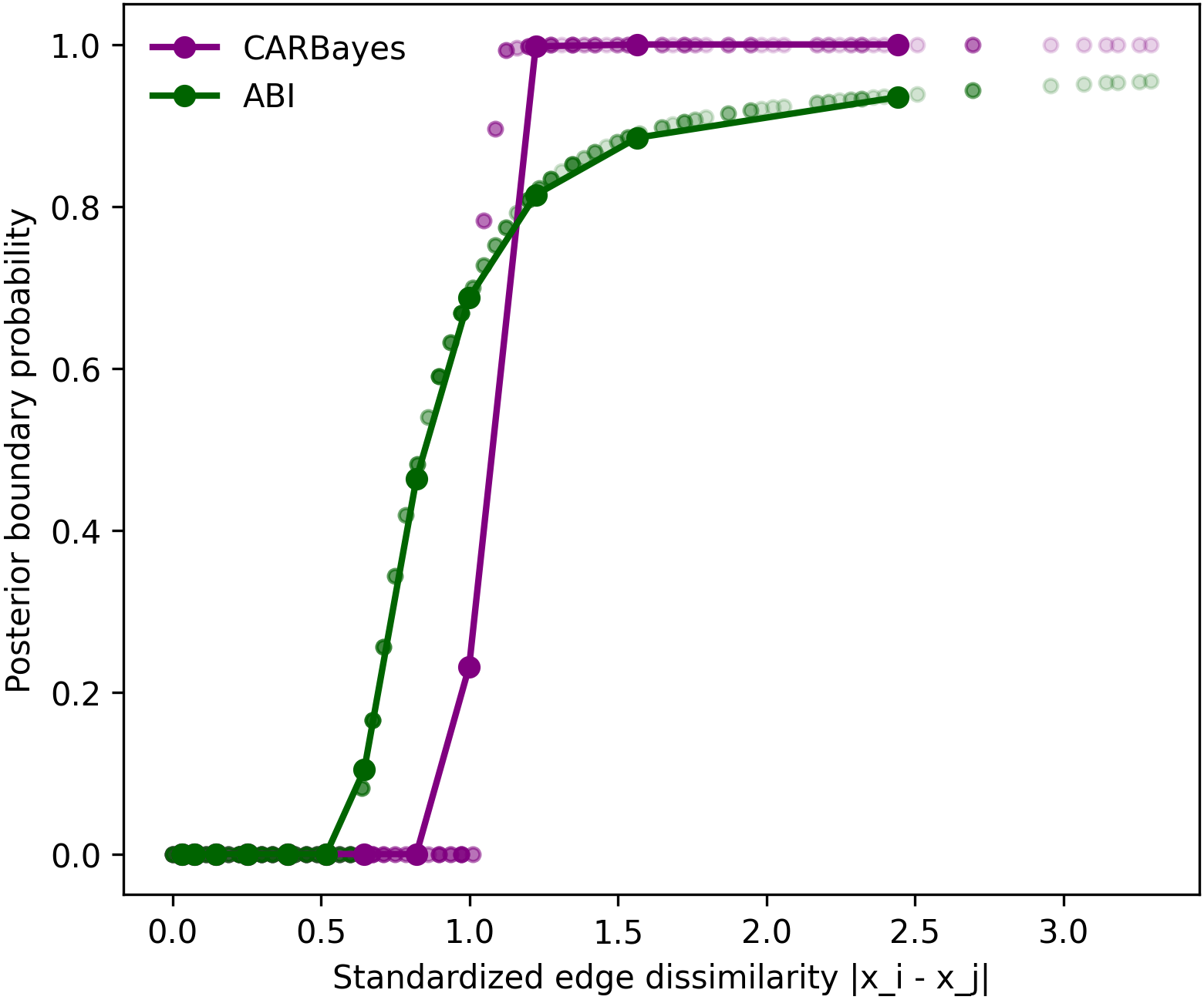}
    \includegraphics[width=0.495\linewidth]{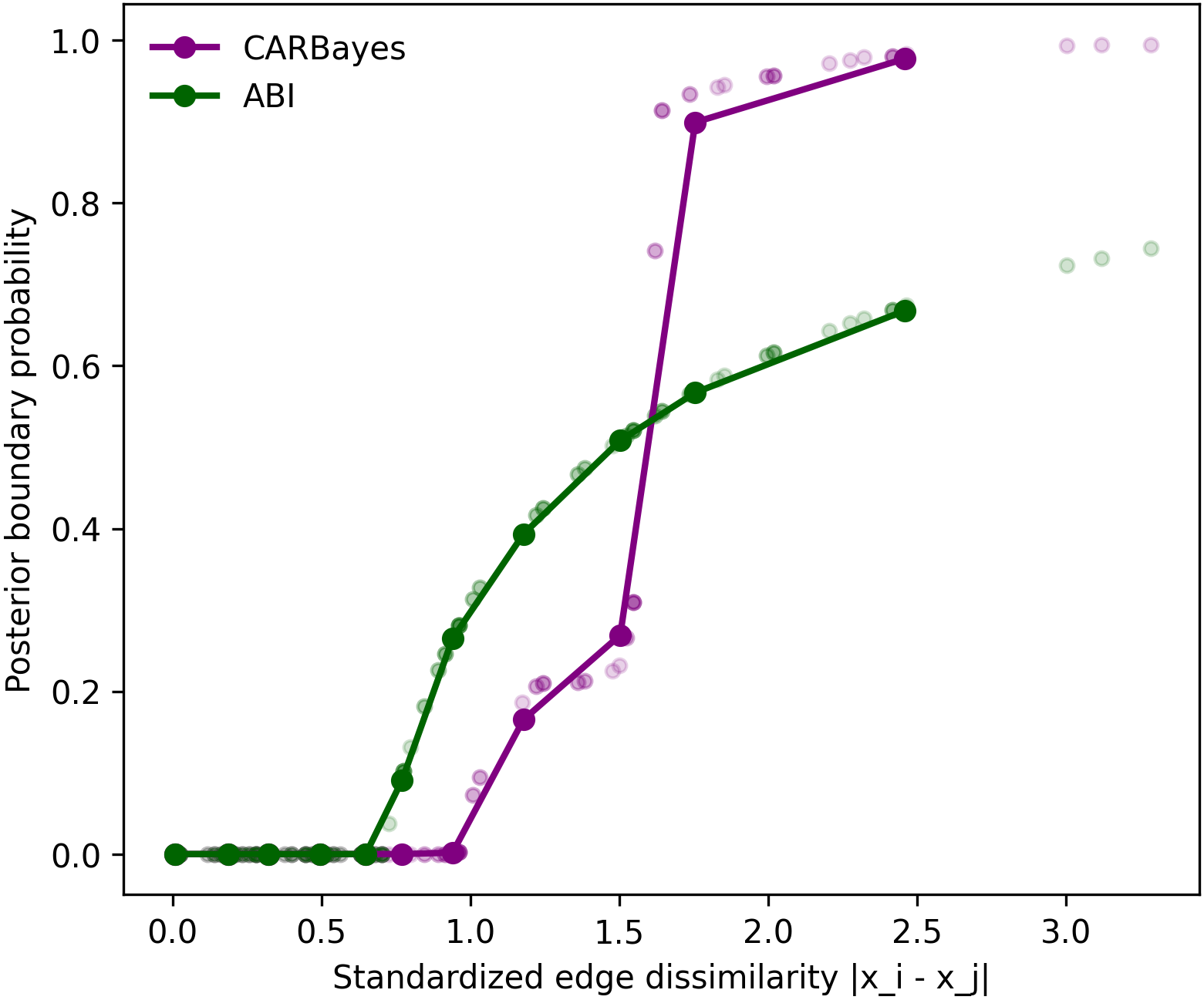}
    \caption{Posterior boundary probability versus standardized edge dissimilarity. Left: Glasgow. Right: California.}
    \label{fig:edge_probability_vs_dissimilarity}
\end{figure}

Fig.~\ref{fig:edge_probability_vs_dissimilarity} plots posterior boundary probability against standardized edge dissimilarity for ABI-DAGAR and \texttt{CARBayes}. These structural diagnostics assess whether posterior boundary probabilities increase with dissimilarity, as expected under the localized-smoothing mechanism. The increasing pattern is evident in both datasets, and the two methods track each other closely, with ABI-DAGAR slightly more elevated at larger dissimilarities.

Table~\ref{tab:realdata_edge_extra} summarizes the edge-level comparison. Posterior boundary probabilities are highly correlated across methods, and the mean absolute differences are modest. The Jaccard indices of the median-probability boundary sets, 0.739 in Glasgow and 0.727 in California, indicate substantial overlap even though ABI-DAGAR selects additional edges.

\begin{table}[t]
\centering
\caption{Additional boundary-probability comparison metrics between ABI-DAGAR and \texttt{CARBayes}. Here ABI-$z_{ij}$ and CB-$z_{ij}$ denote the correlation between posterior boundary probability and standardized edge dissimilarity under ABI-DAGAR and \texttt{CARBayes}, respectively.}
\label{tab:realdata_edge_extra}
\resizebox{\textwidth}{!}{%
\begin{tabular}{lcccccccc}
\hline
\textbf{Dataset} & \textbf{Prob. corr.} & \textbf{Prob. MAE} & \textbf{ABI-}$\bs{z}$ & \textbf{CB-}$\bs{z}$ & \textbf{ABI} & \textbf{CB} & \textbf{Shared} & \textbf{Jaccard} \\
\hline
Glasgow & 0.862 & 0.123 & 0.893 & 0.825 & 134 & 99 & 99 & 0.739 \\
California & 0.863 & 0.124 & 0.945 & 0.868 & 33 & 24 & 24 & 0.727 \\
\hline
\end{tabular}%
}
\end{table}

\subsection{Additional comparison with a model-matched MCMC-DAGAR benchmark}
\label{sec:additional_details_mcmc_dagar}

To complement the external \texttt{CARBayes} benchmark in the main text, we also compared ABI-DAGAR with a model-matched MCMC implementation of the same DAGAR boundary model, denoted MCMC-DAGAR. This comparison separates model-class differences from amortization error: \texttt{CARBayes} provides an established localized-CAR benchmark, whereas MCMC-DAGAR targets the same thresholded Poisson-DAGAR model as ABI-DAGAR.

Table~\ref{tab:mcmc_dagar_parameter_comparison} reports posterior medians and 95\% credible intervals. Compared with \texttt{CARBayes}, the MCMC-DAGAR benchmark already yields broader uncertainty for several spatial parameters, indicating that part of the wider ABI-DAGAR intervals in the main text reflects the DAGAR model class rather than amortization alone. Relative to MCMC-DAGAR, ABI-DAGAR posterior centers are broadly similar, while the amortized intervals remain more diffuse for some latent-structure parameters, especially $\eta$. This supports the interpretation used in the main text: ABI-DAGAR recovers the main posterior location but provides conservative scalar uncertainty for the most weakly identified spatial components.

\begin{table}[t]
\centering
\caption{Posterior medians with 95\% credible intervals for ABI-DAGAR and the model-matched MCMC-DAGAR benchmark.}
\label{tab:mcmc_dagar_parameter_comparison}
\begin{tabular}{lccc}
\hline
\textbf{Dataset} & \textbf{Parameter} & \textbf{ABI-DAGAR} & \textbf{MCMC-DAGAR} \\
\hline
Glasgow & $\beta_0$    & $-0.239$ $(-0.309,\,-0.171)$ & $-0.220$ $(-0.241,\,-0.199)$  \\
Glasgow & $\sigma_w^2$ & $0.338$ $(0.078,\,1.093)$    & $0.209$ $(0.104,\,0.801)$     \\
Glasgow & $\eta$       & $0.832$ $(0.116,\,1.137)$    & $0.672$ $(0.561,\,0.688)$     \\
Glasgow & $\rho$       & $0.878$ $(0.453,\,0.975)$    & $0.760$ $(0.513,\,0.938)$     \\
\hline
California & $\beta_0$    & $0.086$ $(0.015,\,0.151)$ & $0.091$ $(0.073,\,0.109)$  \\
California & $\sigma_w^2$ & $0.107$ $(0.011,\,0.834)$ & $0.030$ $(0.016,\,0.108)$  \\
California & $\eta$       & $0.467$ $(0.017,\,0.964)$ & $0.439$ $(0.258,\,0.725)$  \\
California & $\rho$       & $0.817$ $(0.051,\,0.989)$ & $0.570$ $(0.263,\,0.879)$  \\
\hline
\end{tabular}
\end{table}

The edge-level conclusions are also consistent under the model-matched benchmark. Table~\ref{tab:mcmc_dagar_agreement} shows high correlations between ABI-DAGAR and MCMC-DAGAR posterior boundary probabilities in both datasets. Under the median-probability rule, ABI-DAGAR selects somewhat more boundaries, but all MCMC-DAGAR-selected boundaries are recovered by ABI-DAGAR in both applications. The Jaccard overlap is 0.739 in Glasgow and 0.750 in California, and fitted-risk correlations are positive in both datasets.

\begin{table}[t]
\centering
\caption{Agreement between ABI-DAGAR and the model-matched MCMC-DAGAR benchmark.}
%``ABI in MCMC'' denotes the percentage of ABI-DAGAR selected boundaries also selected by MCMC-DAGAR; ``MCMC in ABI'' denotes the percentage of MCMC-DAGAR-selected boundaries also selected by ABI-DAGAR.
\label{tab:mcmc_dagar_agreement}
\resizebox{\textwidth}{!}{%
\begin{tabular}{lcccccccc}
\hline
\textbf{Dataset} & \textbf{Risk corr.} & \textbf{Bound. corr.} & \textbf{ABI sel.} & \textbf{MCMC sel.} & \textbf{Shared} & \textbf{ABI in MCMC} & \textbf{MCMC in ABI} & \textbf{Jaccard} \\
\hline
Glasgow & 0.775 & 0.863 & 134 & 99 & 99 & 73.9 \% & 100.0 \% & 0.739 \\
California & 0.859 & 0.845 & 32 & 24 & 24 & 75.0 \% & 100.0 \%& 0.750 \\
\hline
\end{tabular}%
}
\end{table}

\begin{figure}[t]
\centering
\includegraphics[width=0.495\textwidth]{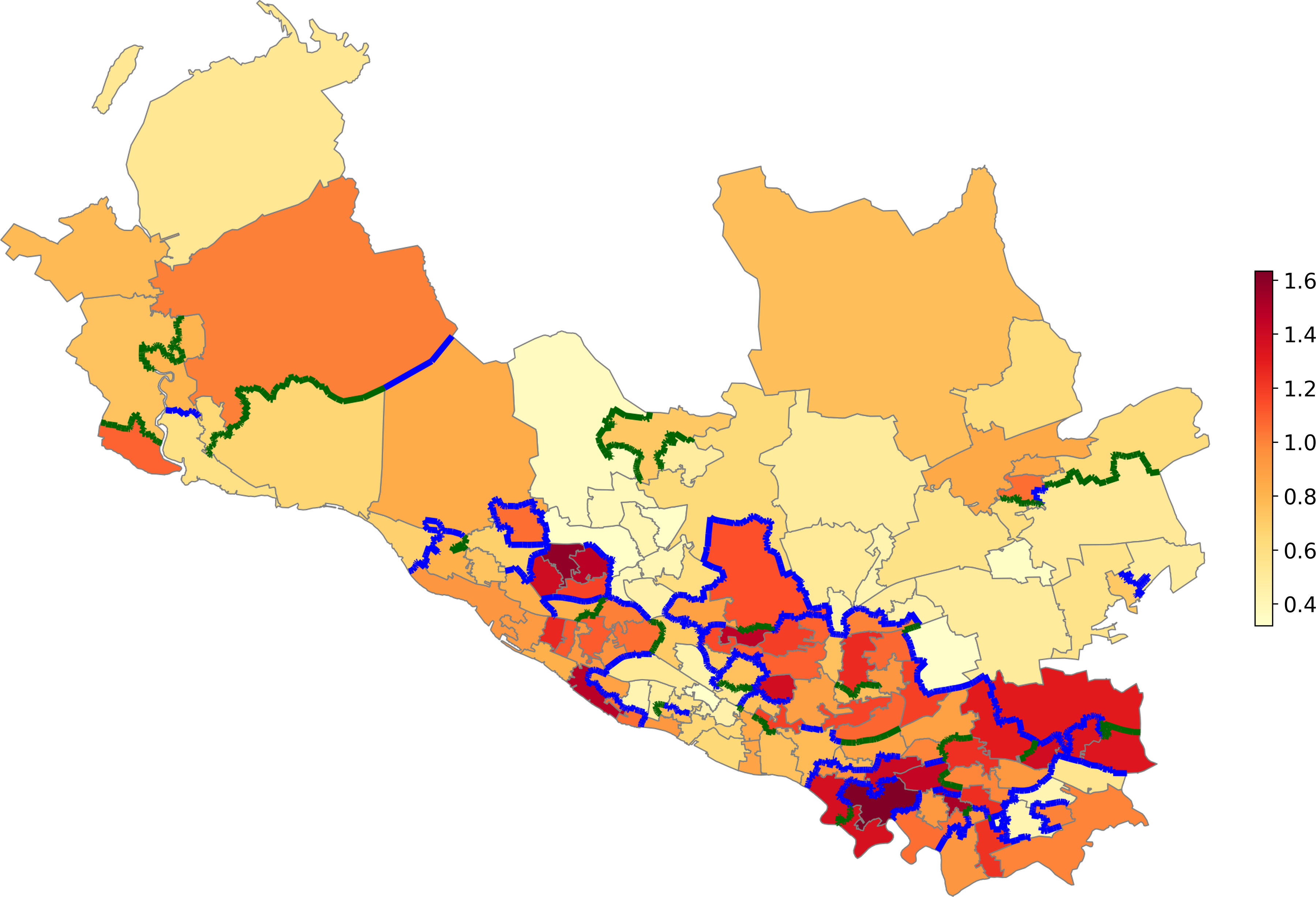}
\includegraphics[width=0.495\textwidth]{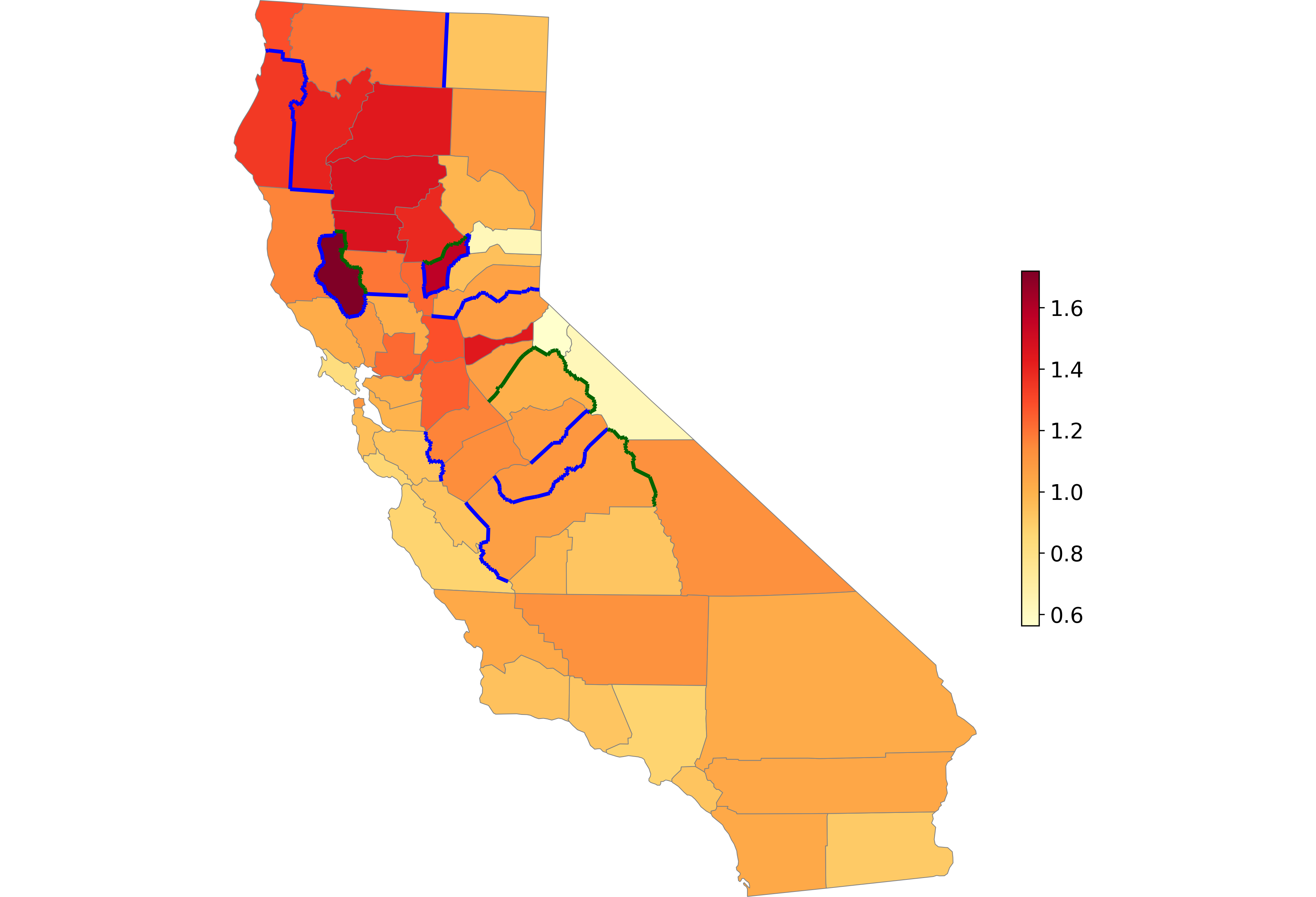}
\caption{Glasgow (left) and California (right) boundary agreement between ABI-DAGAR and the model-matched MCMC-DAGAR benchmark. Blue indicates boundaries selected by both methods, and green ABI-DAGAR-only edges.}
\label{fig:mcmc_dagar_boundary_agreement}
\end{figure}

Fig.~\ref{fig:mcmc_dagar_boundary_agreement} maps the selected boundary sets. In both datasets, the additional ABI-DAGAR boundaries appear mainly as local expansions around the same dominant discontinuity structure, rather than as qualitatively different boundary systems. Thus, although ABI-DAGAR remains more diffuse for some scalar spatial parameters, the posterior boundary evidence is closely aligned with the model-matched MCMC-DAGAR benchmark.

This model-matched benchmark clarifies the real-data uncertainty comparison. The narrower \texttt{CARBayes} intervals in the main text should not be viewed as a direct uncertainty gold standard for ABI-DAGAR, because \texttt{CARBayes} and DAGAR-based inference use different latent spatial priors and dependence parameterizations. The remaining ABI-versus-MCMC-DAGAR gap is more specifically an amortization effect: the learned posterior approximation is conservative for some weakly identified latent-structure parameters, while preserving posterior centers and edge-level boundary conclusions.

\subsection{Additional real-data runtime comparison}
\label{sec:additional_details_runtime}

Table~\ref{tab:realdata_runtime_extra} reports wall-clock timings for posterior sampling in the two real-data applications. The \texttt{CARBayes} posterior summaries and boundary comparisons reported in the main manuscript were based on 300{,}000 MCMC iterations, 100{,}000 burn-in iterations, and thinning by 20, yielding 10{,}000 retained draws. Under this configuration, \texttt{CARBayes} required 615 seconds for Glasgow and 395 seconds for California. Drawing 10{,}000 samples from the trained ABI-DAGAR posterior approximation required approximately 1 second per dataset.

\begin{table}[t]
\centering
\caption{Real-data posterior sampling runtimes for ABI-DAGAR and \texttt{CARBayes}.}
\label{tab:realdata_runtime_extra}
\begin{tabular}{llccc}
\hline
\textbf{Dataset} & \textbf{Method} & \textbf{Retained draws} & \textbf{Runtime} \\
\hline
Glasgow & ABI-DAGAR & 10{,}000 & $\approx 1$ s  \\
Glasgow & \texttt{CARBayes}, 300k iter., 100k burn-in, thin 20 & 10{,}000 & 615 s \\
Glasgow & \texttt{CARBayes}, 20k iter., 10k burn-in, thin 1 & 10{,}000 & 45 s \\
\hline
California & ABI-DAGAR & 10{,}000 & $\approx 1$ s \\
California & \texttt{CARBayes}, 300k iter., 100k burn-in, thin 20 & 10{,}000 & 395 s \\
California & \texttt{CARBayes}, 20k iter., 10k burn-in, thin 1 & 10{,}000 & 30 s \\
\hline
\end{tabular}
\end{table}

We also recorded a shorter \texttt{CARBayes} configuration with 20{,}000 iterations, 10{,}000 burn-in iterations, and no thinning, which also yields 10{,}000 retained draws. These shorter runs required 45 seconds for Glasgow and 30 seconds for California, but were used only to contextualize computational cost; the posterior summaries and boundary comparisons reported in the main text are based on the longer 300{,}000-iteration configuration. These timings should therefore be interpreted as deployment costs for already specified analyses, not as substitutes for statistical validation and benchmark comparison.

Including the one-time ABI-DAGAR training cost of 5 hours and 45 minutes, the break-even point relative to the main \texttt{CARBayes} configuration is approximately 34 Glasgow-sized analyses or 53 California-sized analyses. Under the shorter \texttt{CARBayes} configuration, the corresponding break-even points are approximately 471 and 714 analyses. Thus, the computational advantage of amortization is most relevant for repeated deployment across many related maps, outcomes, or sensitivity analyses.

\subsection{Neural-network implementation and reproducibility details}
\label{sec:supp_network_reproducibility}

The amortized posterior approximation was implemented in Python using BayesFlow with Keras on the TensorFlow backend. The training notebook sets \texttt{KERAS\_BACKEND=tensorflow}. The recorded environment used Python 3.10.19, BayesFlow 2.0.8, Keras 3.12.1, and TensorFlow 2.21.0. Randomness was controlled through \texttt{np.random.seed(123)} in the notebook; no separate TensorFlow or Keras seeds were specified.

The summary network was a BayesFlow \texttt{SetTransformer} with a 32-dimensional output representation, and the inference network was a BayesFlow \texttt{CouplingFlow} with spline transforms. Both inference variables and summary variables were standardized within the BayesFlow workflow. Training was performed online for 100 epochs with batch size 64 and 200 batches per epoch, corresponding to $1{,}280{,}000$ simulated datasets. No user-defined callbacks, early stopping, or custom learning-rate schedule were specified, and BayesFlow's default online-training optimizer was used. Table~\ref{tab:supp_network_impl} summarizes the implementation details.
\begin{table}[t!]
\centering
%\small
\caption{High-level neural-network and training configuration for the ABI-DAGAR implementation.}
\label{tab:supp_network_impl}
\begin{tabular}{ll}
\hline
\textbf{Component} & \textbf{Specification} \\
\hline
Software stack & Python 3.10.19;  BayesFlow 2.0.8; \\
               & Keras 3.12.1; TensorFlow 2.21.0 \\
Backend & \texttt{KERAS\_BACKEND=tensorflow} \\
Random seed & \texttt{np.random.seed(123)} \\
Summary network & BayesFlow \texttt{SetTransformer} \\
Summary output dimension & 32 \\
Inference network & BayesFlow \texttt{CouplingFlow} \\
Transform type & Spline \\
Training mode & Online simulation \\
Epochs & 100 \\
Batch size & 64 \\
Batches per epoch & 200 \\
Total simulated datasets seen in training & $1{,}280{,}000$ \\
Hardware & Intel(R) Core(TM) i7-10750H CPU @ 2.60GHz \\
Training time & 5 hours and 45 minutes \\
\hline
\end{tabular}
\end{table}

\end{document}